\documentclass[aps,twocolumn,superscriptaddress]{revtex4}
\usepackage{dsfont,amsthm,amsbsy}
\usepackage{amssymb}
\usepackage{amsmath}
\usepackage{bbm}
\usepackage{graphicx}
\usepackage{epstopdf}
\usepackage{subfigure}
\usepackage{natbib}
\usepackage{epsfig}
\usepackage{amsfonts}
\usepackage{mathrsfs}
\usepackage{sidecap}
\usepackage{lipsum}
\usepackage{xcolor}
\usepackage[toc,page,title,titletoc,header]{appendix}
\usepackage[colorlinks,linkcolor=blue,citecolor=blue,anchorcolor=blue, urlcolor=blue]{hyperref}
\usepackage{hyperref}
\usepackage{resizegather}
\usepackage{float}
\usepackage{mathbbol}
\usepackage[normalem]{ulem}
\usepackage{cancel}
\usepackage{upgreek}
\usepackage{graphics,dcolumn,bm}
\usepackage{rotate,color}
\usepackage{times}
\newcommand\redsout{\bgroup\markoverwith{\textcolor{red}{\rule[0.5ex]{2pt}{0.4pt}}}\ULon}

\newcommand{\bl}{\begin{aligned}}
\newcommand{\el}{\end{aligned}}
\def\be{\begin{equation}}
\def\ee{\end{equation}}

\def\bi{\begin{itemize}}
\def\ei{\end{itemize}}
\def\bn{\begin{enumerate}}
\def\en{\end{enumerate}}
\def\bea{\begin{eqnarray}}
\def\eea{\end{eqnarray}}
\def\no{\nonumber}
\def\ba{\begin{array}}
\def\ea{\end{array}}
\def\bd{\begin{displaymath}}
\def\ed{\end{displaymath}}

\begin{document}
\title{Out-of-time-order correlations and Floquet dynamical quantum phase transition}
\author{Sara Zamani}
\email[]{sara.zamani@iasbs.ac.ir}
\affiliation{Department of Physics, Institute for Advanced Studies in Basic Sciences (IASBS), Zanjan 45137-66731, Iran}

\author{R. Jafari}
\email[]{jafari@iasbs.ac.ir, rohollah.jafari@gmail.com}
\affiliation{Department of Physics, Institute for Advanced Studies in Basic Sciences (IASBS), Zanjan 45137-66731, Iran}
\affiliation{School of physics, Institute for Research in Fundamental Sciences (IPM), P.O. Box 19395-5531, Tehran, Iran}
\affiliation{Department of Physics, University of Gothenburg, SE 412 96 Gothenburg, Sweden}

\author{A. Langari}
\email[]{langari@sharif.edu}
\affiliation{Department of Physics, Sharif University of Technology, P.O.Box 11155-9161, Tehran, Iran}

\begin{abstract}
Out-of-time-order correlators (OTOCs) progressively play an important role in different fields of
physics, particularly in the non-equilibrium quantum many-body systems.
In this paper, we show that OTOCs can be used to prob the Floquet dynamical quantum phase transitions (FDQPTs).
We investigate the OTOCs of two exactly solvable Floquet spin models,  namely: Floquet XY chain and synchronized Floquet
XY model. We show that the border of driven frequency range, over which the Floquet XY model shows FDQPT,
signals by the global minimum of the infinite-temperature time averaged OTOC.
Moreover, our results manifest that OTOCs decay algebraically in the long time, for which the decay exponent in the FDQPT region is different from that of in the region where the system does not show FDQPTs.
In addition, for the synchronized Floquet XY model, where FDQPT occurs at any driven frequency depending on the initial condition at infinite or finite temperature, the imaginary part of the OTOCs become zero whenever the system shows FDQPT.
\end{abstract}

\pacs{}

\maketitle

\section{Introduction}

Recently, out-of-time-order correlation (OTOC) has gained much attention in the physics
community across many different fields, due to its feasibility in experiments~\cite{nie2020experimental,swingle2016measuring,Zhu2016,Kaufman794,garttner2017measuring,li2017measuring,lewis2019unifying,wei2018exploring} and also its richness in theoretical physics~\cite{Roberts2016,Luitz2017,Iyoda2018,Keyserlingk2018,Niknam2020,Alavirad2019,hosur2016chaos,
Swingle2017,schleier2017probing,Pappalardi2018,Klug2018,Khemani2018,Alavirad2019,Maldacena2016,shenker2014black,kitaev2018soft,fan2017out,larkin1969quasiclassical}. Recent progress in the experimental detection of quantum correlations and in quantum control techniques applied to systems as photons, molecules, and atoms, made it possible to direct
observation of an OTOC in nuclear magnetic resonance quantum simulator~\cite{li2017measuring,wei2018exploring}
and trapped ions quantum magnets~\cite{garttner2017measuring}.

The OTOC was first introduced by Larkin and Ovchinnikov in the context of superconductivity~\cite{larkin1969quasiclassical}. Lately, it has been revitalized, because it propounds an interesting and different insight into physical systems~\cite{kitaev2018soft}. Some of the most important results involve the dynamics of quantum systems~\cite{Roberts2016,Luitz2017,Iyoda2018,Keyserlingk2018,Niknam2020,Alavirad2019} such as quantum information scrambling~\cite{hosur2016chaos,Swingle2017,schleier2017probing,Pappalardi2018,Klug2018,
Khemani2018,Alavirad2019,Maldacena2016,shenker2014black} and quantum entanglement~\cite{fan2017out,hosur2016chaos}.
The decay of OTOC is closely related to the delocalization of information and implies the information-theoretic definition of scrambling. Scrambling is a process by which the information stored in local degrees of freedom spreads over the many-body degrees of freedom of a quantum system, becoming inaccessible to local probes and apparently lost. A connection between the OTOC and the growth of entanglement entropy, at the infinite temperature, in quantum many-body systems has also been discovered quite recently~\cite{fan2017out,hosur2016chaos}.

In addition, the study of OTOC has renovated the interest in the correspondence between classical and quantum chaos~\cite{Kukuljan2017,hashimoto2017out,Rozenbaum2017,Rozenbaum2019,Mata2018,Herrera2018,Carlos2019} with some analytical advances in the field of high-energy physics, mostly regarding the black hole information problem~\cite{maldacena2017diving} and the Sachdev-Ye-Kitaev model~\cite{Maldacena2016PRD}. OTOCs have been also developed into condensed-matter systems~\cite{patel2017quantum,dora2017out,shen2017out,heyl2018detecting,sahu2019scrambling}
as well as in statistical physics~\cite{campisi2017thermodynamics,chenu2018quantum}. For instance,  OTOC has been analyzed in conformal field theories~\cite{patel2017quantum}, fermionic models with critical Fermi surface~\cite{Patel1844}, weakly diffusive metals\cite{patel2017quantum}, Luttinger liquids~\cite{dora2017out}, hardcore boson model~\cite{lin2018out1}, random field XX spin chain~\cite{Riddell2019}, symmetric Kitaev chain~\cite{mcginley2018slow} and the $O(N)$ model~\cite{Chowdhury2017PRD}.
Besides, it has been shown that OTOC equals the thermal average of the Loschmidt echo~\cite{yan2020information} and theoretically proposed that OTOC can be used as an order parameter to dynamically detect ergodic-nonergodic transitions\cite{Buijsman2017,ray2018signature}, many-body localization transition~\cite{Maldacena2016,huang2017out}, excited-state quantum phase transition (ESQPT)~\cite{wang2019probing,lobez2016entropy}, equilibrium quantum phase transitions (EQPTs)~\cite{daug2019detection,heyl2018detecting,wei2019dynamical} and quench dynamical quantum phase transitions (DQPTs) in many-body systems~\cite{nie2020experimental,heyl2018detecting}.

Despite considerable studies on OTOCs in a wide variety of quantum systems, comparatively little attention has been focussed on the Floquet systems. In the present work, we study OTOCs in two Floquet spin systems, where both models show FDQPTs. To the best of our knowledge, such contributions have not been explored in previous works and would bring several new realizations to this subject.
In the first model, the Floquet XY model in which  FDQPT occurs at any temperature within a finite range of driving frequency,  where we show that
the border of the driven frequency window are captured by the global minimum of the infinite-temperature time averaged OTOCs.
In other words, the time averaged OTOC can be used as an order parameter to detect the range of driven frequency over which FDQPTs occur.
Moreover, the long time behavior of OTOCs represents power law decay with an exponent, which is different in the FDQPTs and no-FDQPTs regimes.
Furthermore, in the synchronized Floquet XY model, in which the FDQPTs occur for any driving frequency
at infinite or finite temperature, the imaginary part of OTOCs composed with local and nonlocal operators becomes zero when FDQPTs are present.

The paper is organized as follows: In the next section we define the OTOCs and some background materials.
In section \ref{DPT}, we review the notion of dynamical phase transition and its features. Section \ref{FXYM} is dedicated to
introducing the Floquet XY Hamiltonian, its FDQPT features and discussing the OTOC behavior in the model.
In section \ref{FSXYM} we first introduce the synchronized Floquet XY model and its FDQPTs properties and then we study the OTOCs characteristics.

\section{OTOCs\label{OTOC}}
Consider a system with a Hamiltonian $H$, an initial state $|\psi\rangle$, and two local operators $W_{i}$ and $V_{i+r}$ ,
on sites $i$ and $i+r$ of the system. The spreading of the operator $W_{i}$ with time can be probed through the expectation value of the squared module of a commutator with a second operator $V_{i+r}$,

\begin{equation}
\label{eq1}
C_{i, r}(t)=\frac{1}{2}\langle\Big[W_{i}(t),V_{i+r}(0)\Big]^\dagger \Big[W_{i}(t),V_{i+r}(0)\Big]\rangle,
\end{equation}
%
where $W_{i}(t)\equiv e^{iHt}W_{i}(0)e^{-iHt}$ is the Heisenberg evolution of the operator $W_{i}$, and $\langle {\cal O}\rangle= {\rm Tr}(e^{-\beta H}{\cal O} )/{\rm Tr}(e^{-\beta H})$ denotes averaging over the thermal ensemble with $\beta=1/T$  is the inverse temperature while setting the Boltzmann constant $K_{B}$ to unity.

We consider a translational invariant system such that Eq.(\ref{eq1}) depends only on the distance between two operators.
Assuming operators $W_{i}$ and $V_{i+r}$ are both Hermitian and unitary, one can show that
$C_{r}(t) \equiv C_{i, r}(t)=1-{\rm Re}(F_{r}(t))$, in which $F_{r}(t)=\langle W_{i}(t)V_{i+r}(0)W_{i}(t)V_{i+r}(0) \rangle$ dubbed OTOC for its unconventional time ordering~\cite{bao2020out,lin2018out}. From the operator delocalization assessment point of view, OTOC characterizes the spreading behavior of information. Vanishing $C_{r}(t)$ (or large $F_{r}(t)$) indicates that no information has traveled from site $i$ to $i+r$ at time $t$.

In addition, $C_{r}(t)$ characterizes the quantum chaos via an exponential growth bounded by a thermal Lyapunov exponent. In classical physics, a hallmark of chaos is that a small difference in the initial condition results in an exponential deviation of the trajectory i.e., $e^{\lambda_{L} t}$ where $\lambda_{L}$ is the Lyapunov exponent (butterfly effect). The OTOC could be considered as the overlap of two states $W_{i}(t)V_{i+r}(0)|\psi\rangle$ and $V_{i+r}(0)W_{i}(t)|\psi\rangle$, where $V_{i+r}(0)$  acts in different ways to affect the growth of the time-evolved operator $W_{i}(t)$. In other words, $C_{r}(t)$ explicitly exhibits the difference in the outcome when the order of two operations $V_{i+r}(0)$ and $W_{i}(t)$ is exchanged~\cite{shen2017out,garttner2017measuring,swingle2018unscrambling}. The exponential deviation of normalized OTOC from unity, i.e., $F_{r}(t)\sim 1-\text{\#}e^{\lambda_{L} t}$ diagnoses the chaos and the so-called "butterfly effect" in a quantum many-body system. Unlike classical systems where the Lyapunov exponent $\lambda_{L}$ is unbounded, in quantum systems it is bounded by $2\pi/\beta$ (assuming $\hbar=1$)~\cite{Maldacena2016}.
Those systems which saturate the aforementioned bound are called fast scramblers, with examples including black holes~\cite{lashkari2013towards,sekino2008fast}, fermionic models with critical Fermi surface~\cite{Patel1844}, weakly diffusive metals~\cite{patel2017quantum}, and the $O(N)$ model~\cite{Chowdhury2017PRD}. However, some systems, do not show such exponential growth, for example Luttinger liquids \cite{dora2017out} and many-body localized systems~\cite{Swingle2017,slagle2017out,huang2017out}, and hence characterized as less chaotic or as slow scramblers. These many-body quantum systems include rich information to connect thermalization and information scrambling, and may also be related to the study of hiding information behind black hole horizon.

\subsection{OTOC in the one dimensional spin $1/2$ exactly solvable models}

In the one dimensional spin $1/2$ models, which are exactly solvable by means of Jordan-Wigner transformation~\cite{LIEB1961407,Barouch1970,Barouch1971,Eriksson2009,titvinidze2003,Jafari2011,Jafari2012,Mahdavifar2017},
the operators $W$ and $V$ are replaced by single-site Pauli matrices $\sigma^{\alpha}, (\alpha=\{x,y,z\}$) and consequently the OTOC is given by

\begin{equation}
\label{eq2}
F^{\mu,\nu}_{r}(t)=\langle \sigma_{r}^{\mu}(t) \sigma_{0}^{\nu} \sigma_{r}^{\mu}(t) \sigma_{0}^{\nu} \rangle,
\end{equation}
where, $\mu,\nu=\{x,y,z\}$ and $\sigma^{\alpha}(t)=e^{iHt}\sigma^{\alpha}e^{-iHt}$. Since the models is exactly solvable by means of Jordan-Wigner transformation, it is convenient to express Pauli matrices by fermionic operators,

\begin{eqnarray}
\label{eq3}
\sigma_{m}^{x}&&=\sigma_{m}^{+}+\sigma_{m}^{-}\\
\nonumber
&&=\Pi_{l<m}(1-2c_{l}^{\dagger}c_{l})(c_{m}^{\dagger}+c_{m})=\Pi_{l<m} A_l B_l A_m\\
\nonumber
\sigma_{m}^{y}&&=-i~(\sigma_{m}^{+}-\sigma_{m}^{-})\\
\nonumber
&&=-i~\Pi_{l<m}(1-2c_{l}^{\dagger}c_{l})(c_{m}^{\dagger}-c_{m})=-i~\Pi_{l<m} A_l B_l B_m\\
\nonumber
\sigma_{m}^{z}&&=2c_{m}^{\dagger}c_{m}-1=-A_m B_m,
\end{eqnarray}

where $A_m=c_{m}^{\dagger}+c_{m}$, $B_m=c_{m}^{\dagger}-c_{m}$, and $c^{\dagger}_{m}$ ($c_{m}$) is the fermion creation (annihilation) operator.

In terms of Jordan-Wigner fermions, some spin operators are local and some become nonlocal. Local operator i.e., $\sigma_{m}^{z}$ is consisted of fermions only located at site $m$, while $\sigma_{m}^{x,y}$ are nonlocal according to their connections with all fermions before site $m$. It has been shown that the relation of two-point correlations and OTOCs of local operators is different from nonlocal ones~\cite{bao2020out,lin2018out,chapman2018classical,rossini2010long,sachdev2007quantum}. All OTOCs can be expressed in terms of thermal average of $A_m$ and $B_m$ sequences. So, we need to calculate the expectation values of long sequences of $A_m$ and $B_m$ fermion operators, which can be turn into the sum of all possible products of two-point correlation functions, using the Wick's theorem. It should be noted that conservation of the fermion parity via operators is necessary for using free-fermion calculations and Wick’s theorem~\cite{bao2020out,lin2018out,mccoy1971statistical}. The basic time dependent correlation functions, which should be calculated, are $\langle A_p(t) A_q\rangle$, $\langle A_p(t) B_q\rangle$,
$\langle B_p(t) A_q\rangle$ and $\langle B_p(t) B_q\rangle$.
Using the Fourier transformations,  the mentioned correlators are expressed as

{\small
\begin{eqnarray}
\nonumber
\langle A_p(t) A_q \rangle &=& \frac{1}{N} \sum_{k} e^{ik(p-q)} \langle U_{k}^{\dagger}(t)(c_{k}^{\dagger}+c_{-k})U_{k}(t)(c_{-k}^{\dagger}+c_{k})\rangle,\\
\nonumber
\langle A_p(t) B_q \rangle &=& \frac{1}{N} \sum_{k} e^{ik(p-q)} \langle U_{k}^{\dagger}(t)(c_{k}^{\dagger}+c_{-k})U_{k}(t)(c_{-k}^{\dagger}-c_{k})\rangle,\\
\nonumber
\langle B_p(t) A_q \rangle &=& \frac{1}{N} \sum_{k} e^{ik(p-q)} \langle U_{k}^{\dagger}(t)(c_{k}^{\dagger}-c_{-k})U_{k}(t)(c_{-k}^{\dagger}+c_{k})\rangle,\\
\nonumber
\langle B_p(t) B_q \rangle &=& \frac{1}{N} \sum_{k} e^{ik(p-q)} \langle U_{k}^{\dagger}(t)(c_{k}^{\dagger}-c_{-k})U_{k}(t)(c_{-k}^{\dagger}-c_{k})\rangle,\\
\label{eq4}
\end{eqnarray}
}
where $N+1$ is the size of the system and $p, q$ denote the position of operators in the spin chain.

\begin{figure*}
\begin{minipage}{\linewidth}
\centerline{\includegraphics[width=0.34\linewidth]{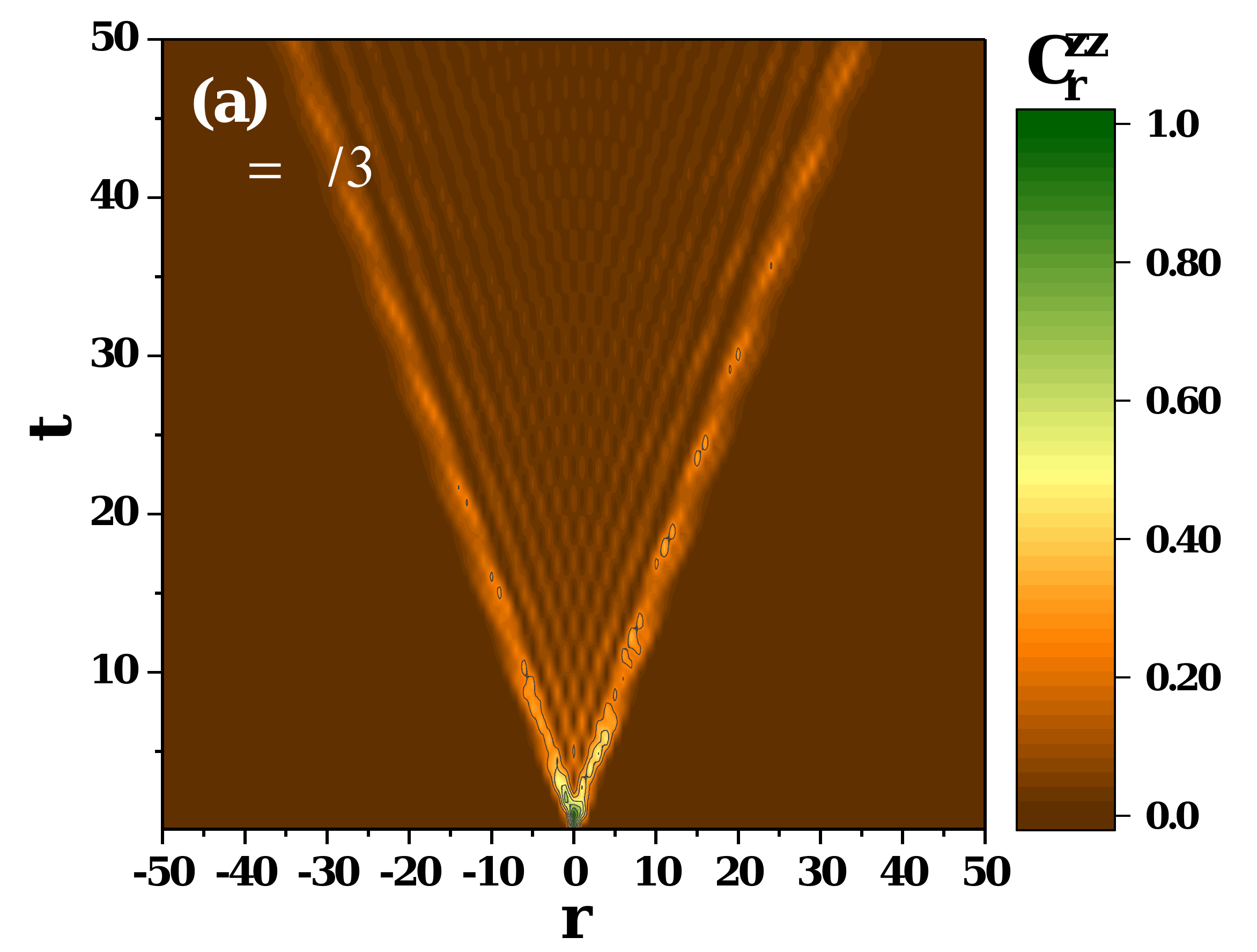}
\includegraphics[width=0.33\linewidth]{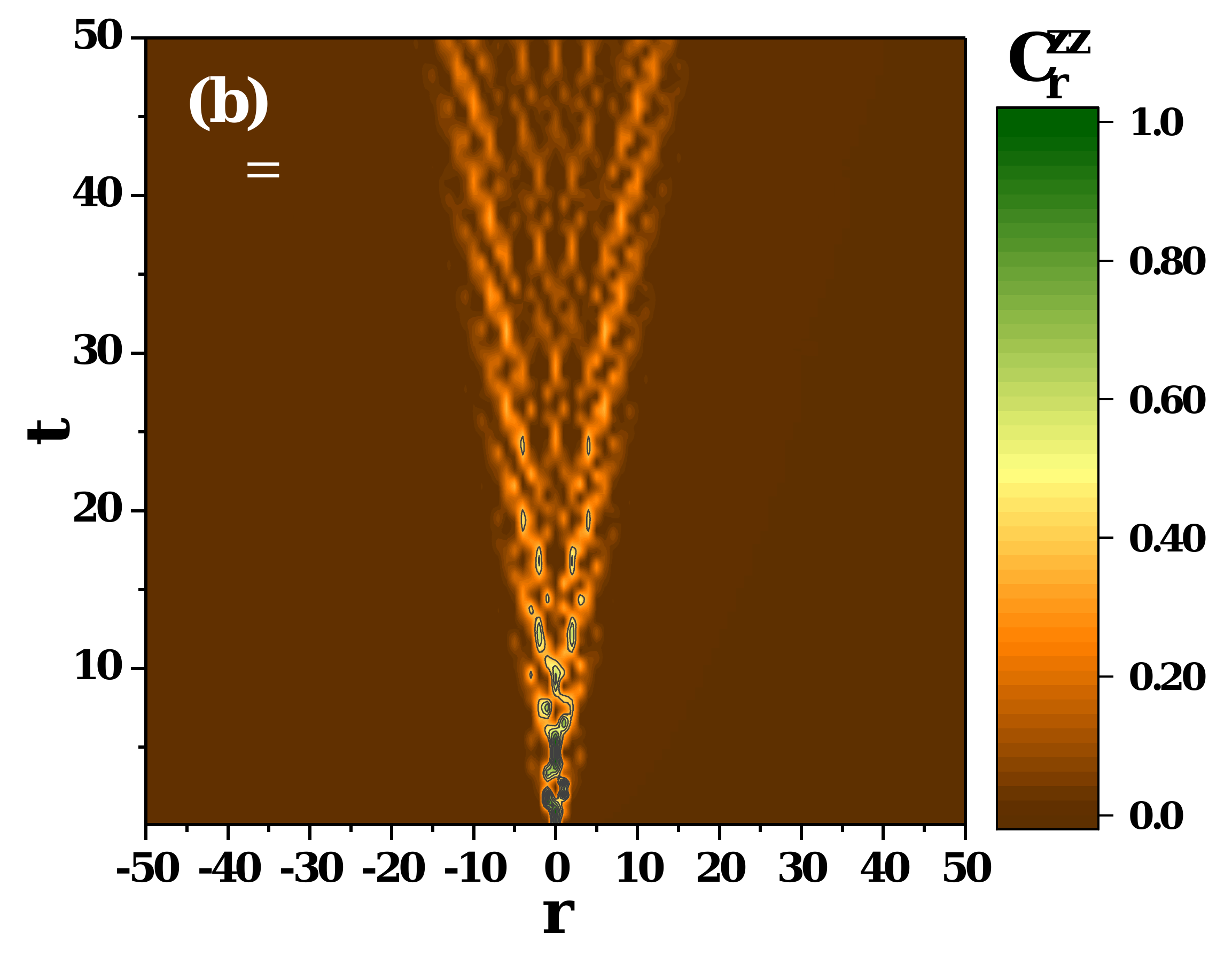}
\includegraphics[width=0.33\linewidth]{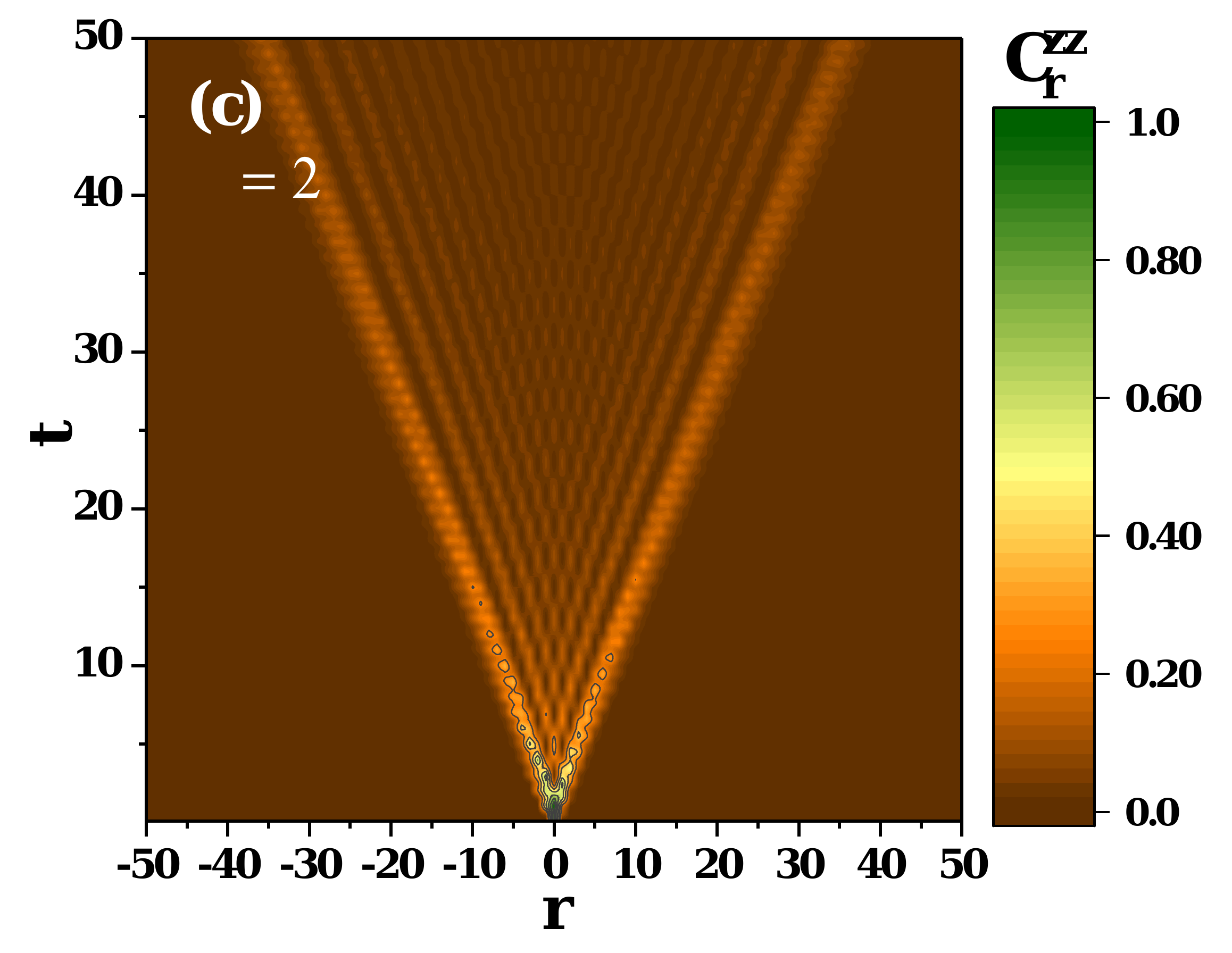}}
\centering
\end{minipage}
\begin{minipage}{\linewidth}
\centerline{\includegraphics[width=0.33\linewidth]{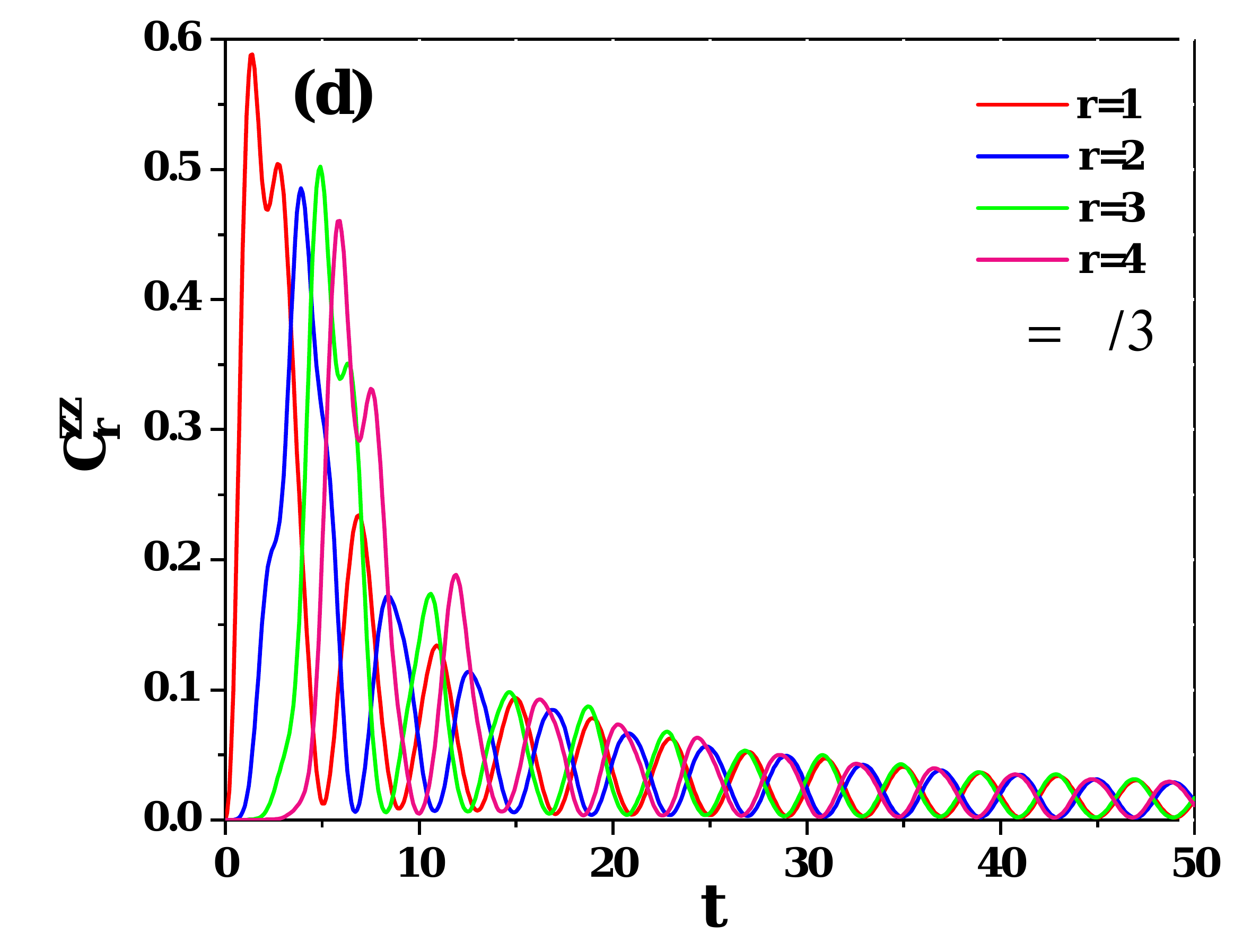}
\includegraphics[width=0.33\linewidth]{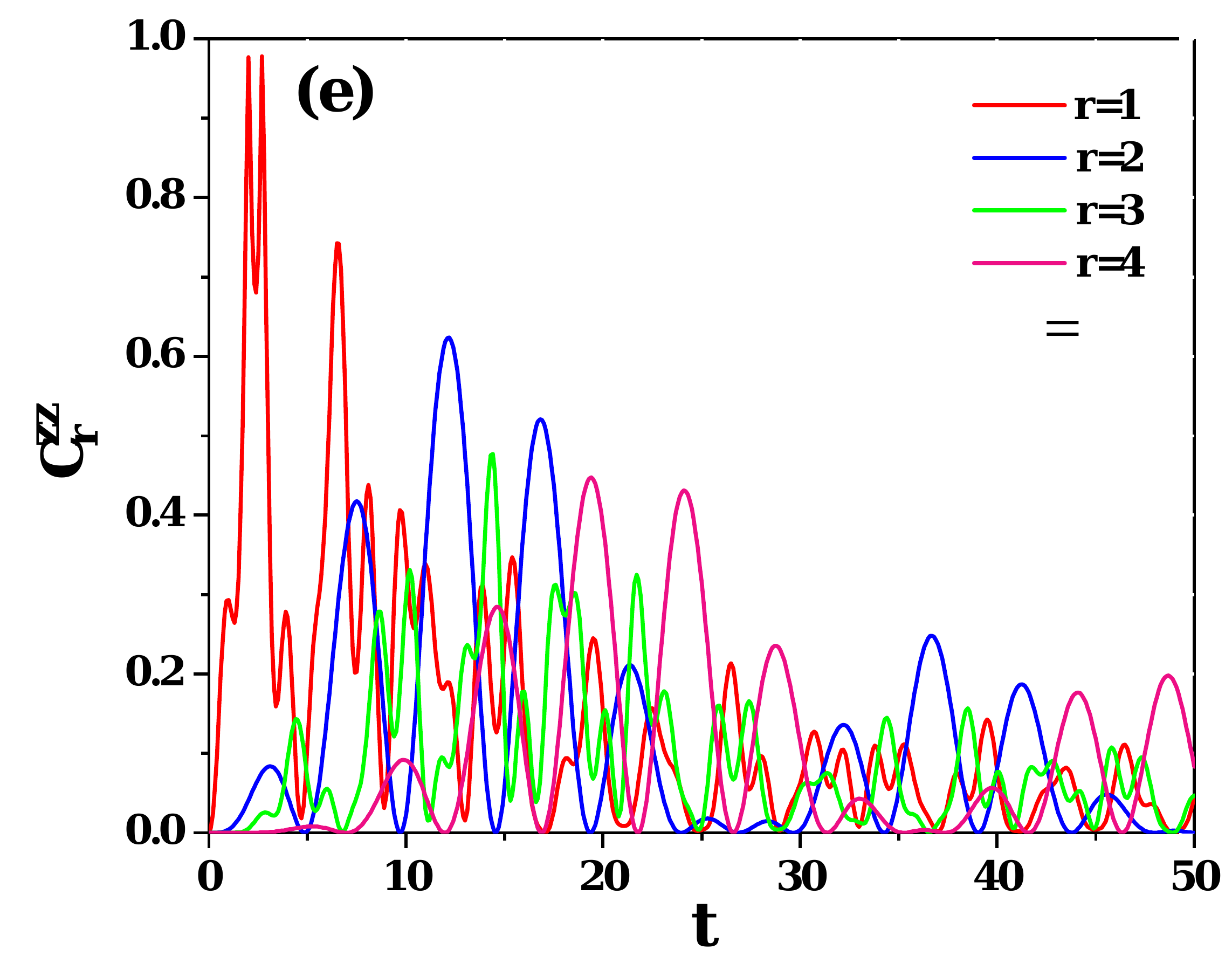}
\includegraphics[width=0.33\linewidth]{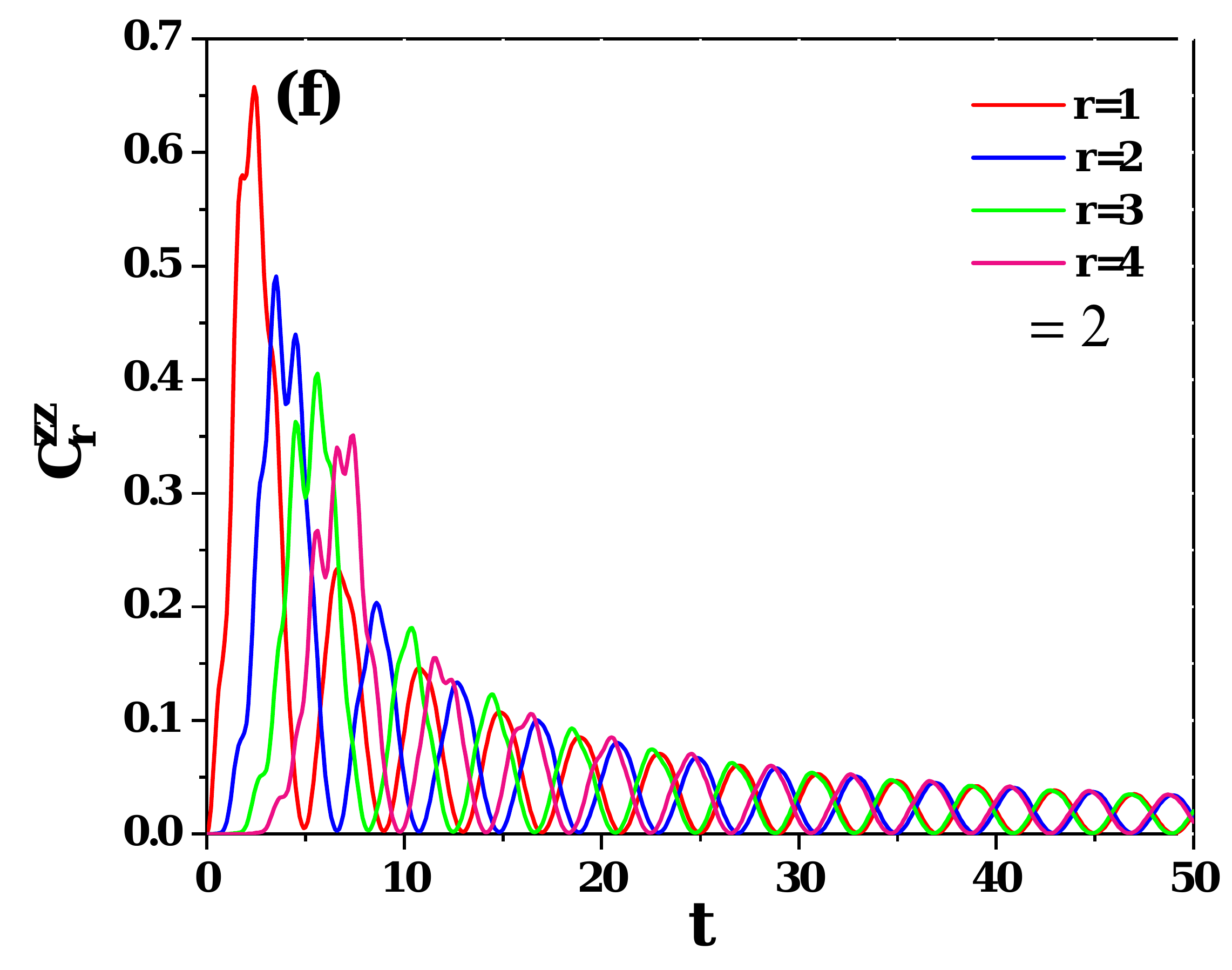}}
\centering
\end{minipage}
\caption{(Color online)
Density plot of  $C^{zz}_{r}(t)$ versus separation, $r$,  and time, $t$ for
(a) $\omega=\pi/3$, (b) $\omega=\pi$, (c) $\omega=2 \pi$. The numerical simulation of Floquet $XY$ model
is done for system size $N = 100$, inverse temperature $\beta=0$ and $J=0.25\pi, h=0.5\pi,
\gamma=0.5$,  which shows FDQPT for
$\pi/2<\omega<3\pi/2$.  $C^{zz}_{r}(t)$ is plotted at fixed separations, $r=1, 2, 3, 4$ versus time and different
driving frequency (d) $\omega=\pi/3$, (e) $\omega=\pi$, (f) $\omega=2 \pi$.
}
\label{Fig1}
\end{figure*}
%
%
\begin{figure}
\centerline{\includegraphics[width=\columnwidth]{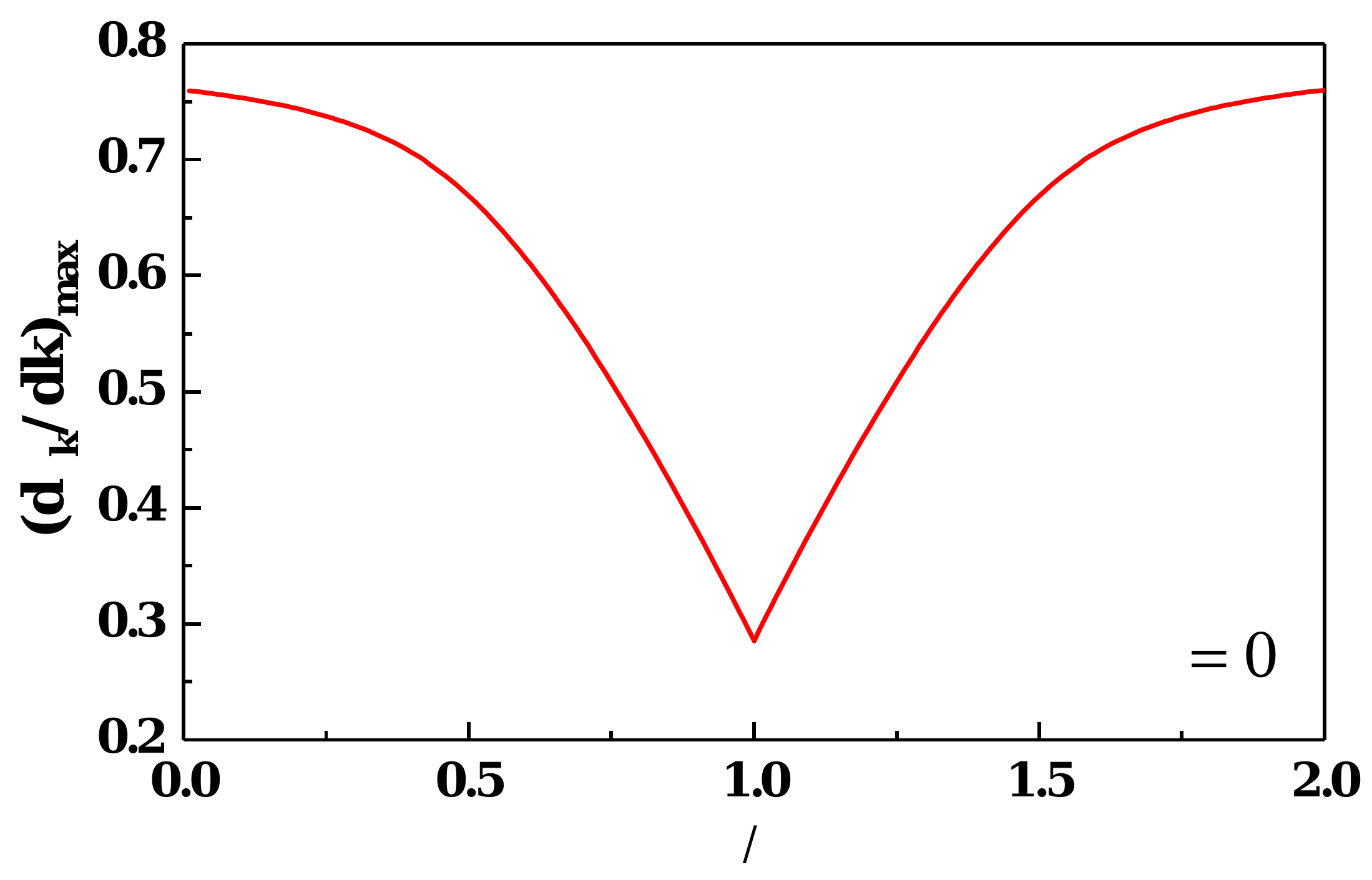}}
\caption{The maximum quasiparticle group velocity of time-independent Floquet Hamiltonian
Eq. (\ref{eq:B7}) versus $\omega/\pi$.
The maximum group velocity reaches a minimum at the middle of driven frequency
window, where Floquet dynamical phase transition occurs.}
\label{Fig2}
\end{figure}
%
\subsection{OTOC of local operators}
As mentioned, OTOCs characterize the delocalization of operators, and study of local operators plays a key role in this context.
By means of Jordan-Wigner transformation (Eq. (\ref{eq3})),
OTOC of local operators, $F^{zz}_{r}(t)$, is given by
%
\begin{equation}
\label{eq5}
F^{zz}_{r}(t)=\langle A_{r}(t)B_{r}(t)A_{0}B_{0}A_{r}(t)B_{r}(t)A_{0}B_{0} \rangle,
\end{equation}
%
for the exactly solvable spin-$1/2$ chain.
In the thermodynamic limit, the above relation could be computed using the Wick's theorem.
In this calculation, $\langle A_p(t) A_q\rangle$ and $\langle B_p(t) B_q\rangle$  terms do not
vanish and we must consider combination of all two-point correlation functions constructed with
$A$ and $B$ operators. We can simplify calculate Eq. (\ref{eq5}) using Pfaffian method~\cite{cayley1852},
which can be expressed in terms of skew-symmetric matrix $\Phi$.
%
\begin{equation}
F^{zz}_{r}(t)=\pm Pf(\Phi_{zz})=\pm\sqrt{Det(\Phi_{zz})}
\label{eq6},
\end{equation}
%
where $\Phi_{zz}$ is constructed from two point correlation functions (Eq. (\ref{eq4})), $(\Phi_{zz})_{ij}=\langle X_{i}X_{j} \rangle$, where $X_i$ is the $i$-th element inside thermal average expression $F^{zz}_{r}(t)$ (Eq. (\ref{eq5})).

\subsection{OTOC of nonlocal operators}
As mentioned before, the dynamical correlation functions of nonlocal operators are qualitatively distinct from local ones~\cite{sachdev2007quantum,essler2016quench,rossini2010long,bao2020out}. For two-point correlation functions at non-zero temperature, the time dependent decaying of nonlocal operators, which is exponential, is more similar to thermal behavior, in comparison with the behavior of local operators, which is power-law. There are three different types of nonlocal OTOCs corresponding to various combinations of local and nonlocal operators. It should be mentioned that the operators $\sigma_{\mu}^{x}$ and $\sigma_{\mu}^{y}$ change the fermion parity.
%
\begin{figure*}
\begin{minipage}{\linewidth}
\centerline{\includegraphics[width=0.35\linewidth]{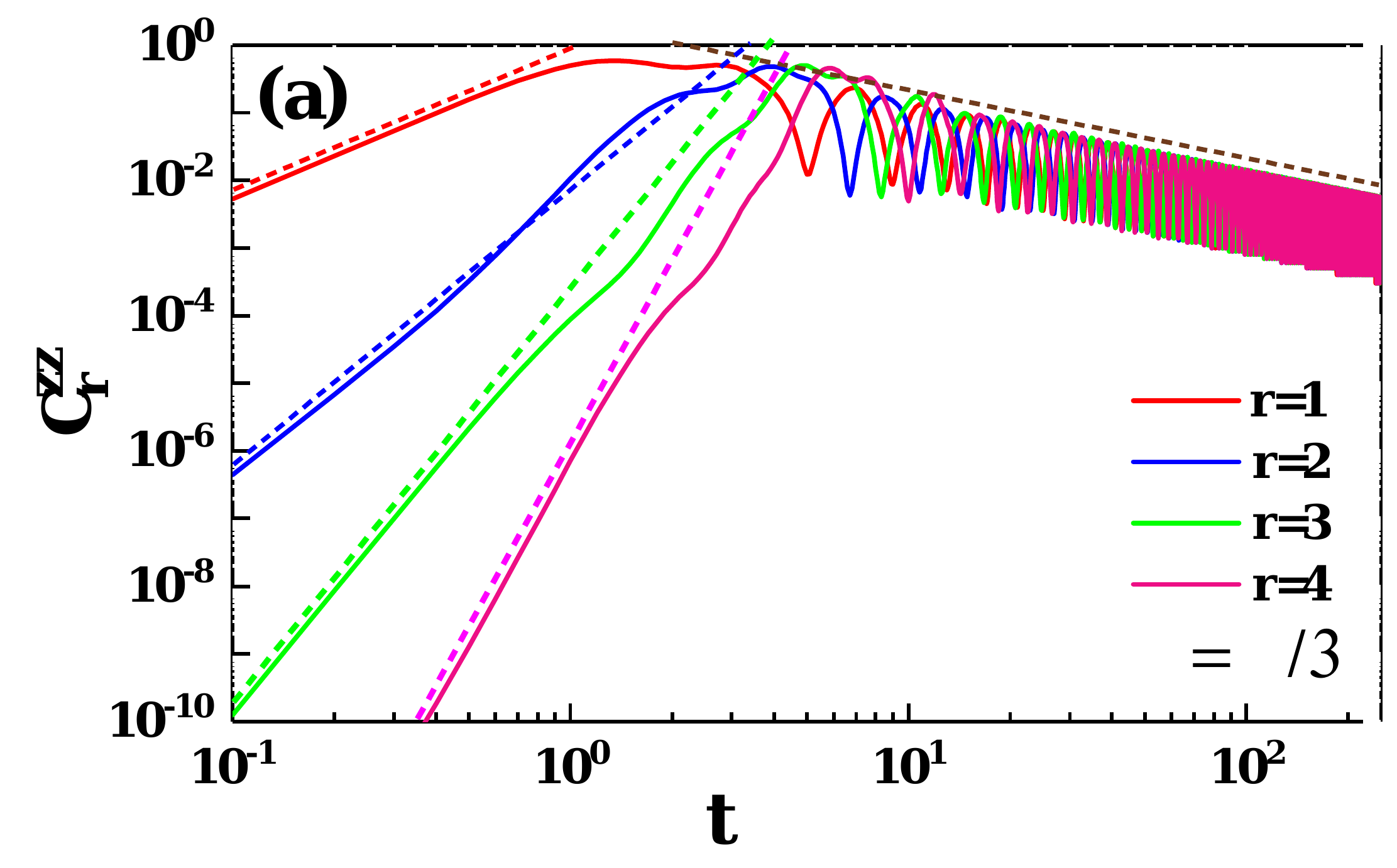}
\includegraphics[width=0.33\linewidth]{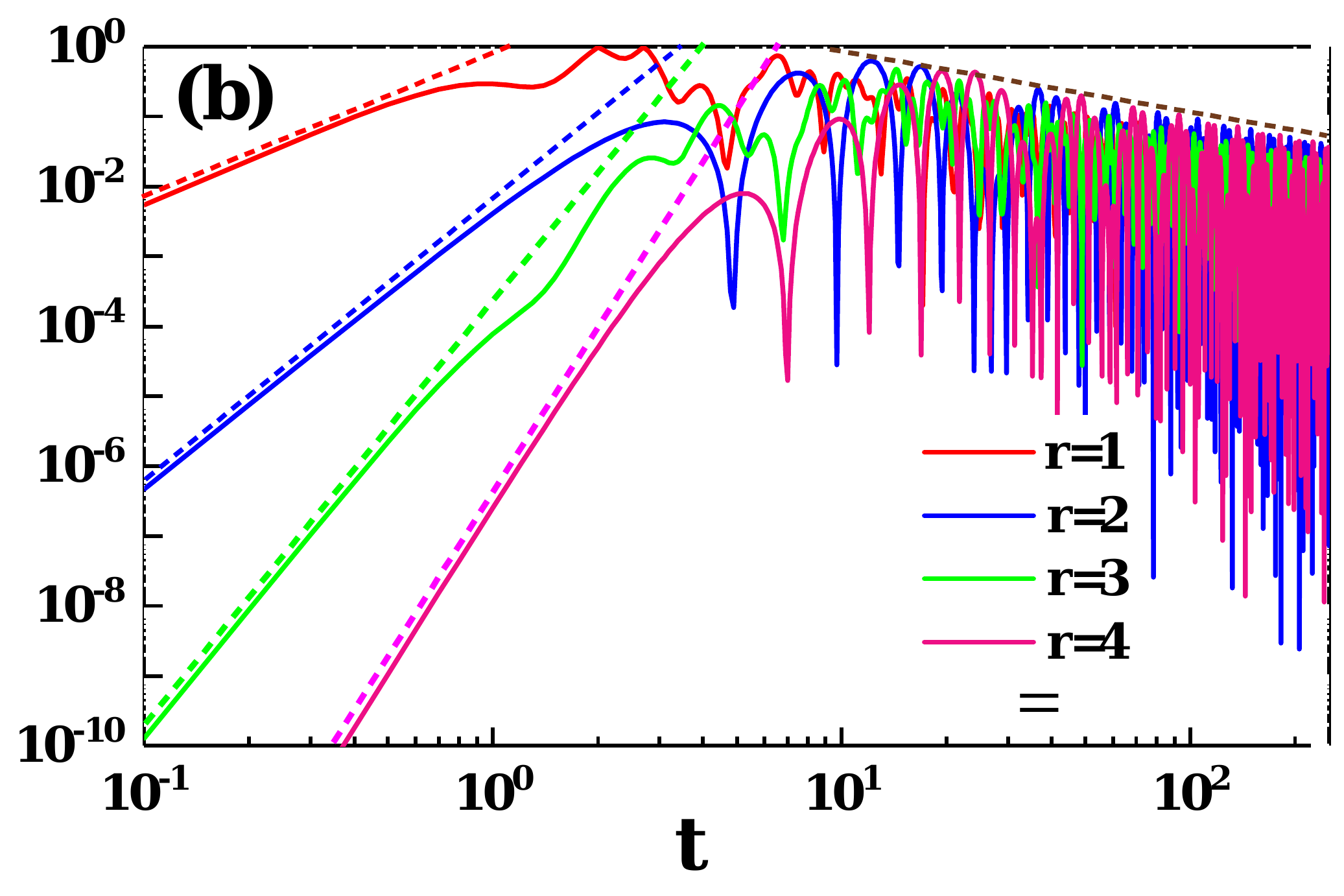}
\includegraphics[width=0.33\linewidth]{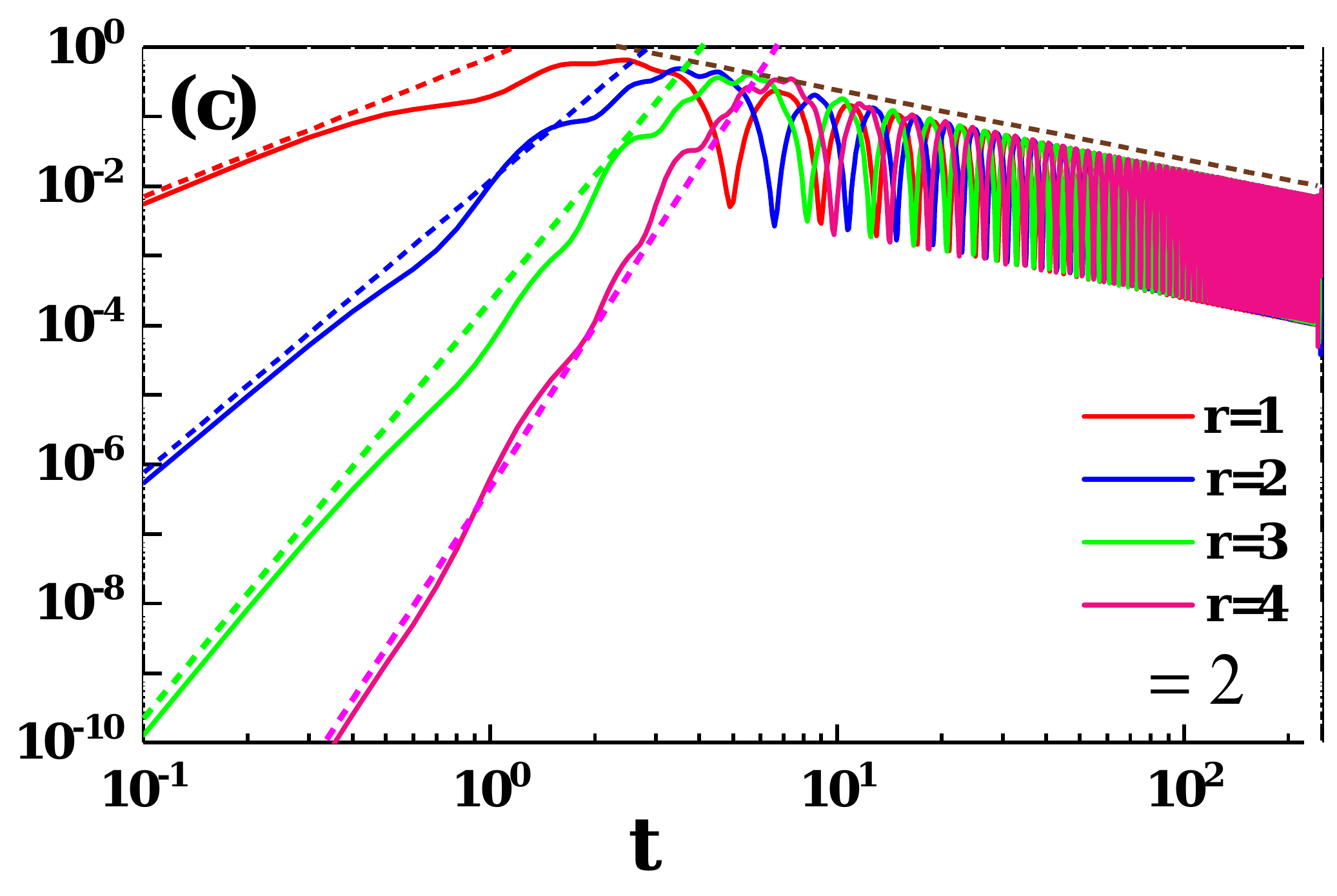}}
\centering
\end{minipage}
\caption{(Color online)
The scaling behavior of $C^{zz}_{r}$ versus time at fixed separation $r=1, 2, 3, 4$ for the Floquet $XY$ model.
The dashed lines are used for power
law fitting, which show $t^{2r}$ power law
growth at early time, independent of the driving frequency,
(a) $\omega=\pi/3$, (b) $\omega=\pi$, (c) $\omega=2 \pi$. Besides, we see
$t^{-1}$ decaying behavior at long time, except for (b) which shows
$t^{-2.6}$.  The system size is $N = 200$,
inverse temperature $\beta=0$ and Hamiltonian parameters are
$J=0.25\pi, h=0.5\pi, \gamma=0.5$.}
\label{Fig3}
\end{figure*}
%
%
\begin{figure}
\centerline{\includegraphics[width=\columnwidth]{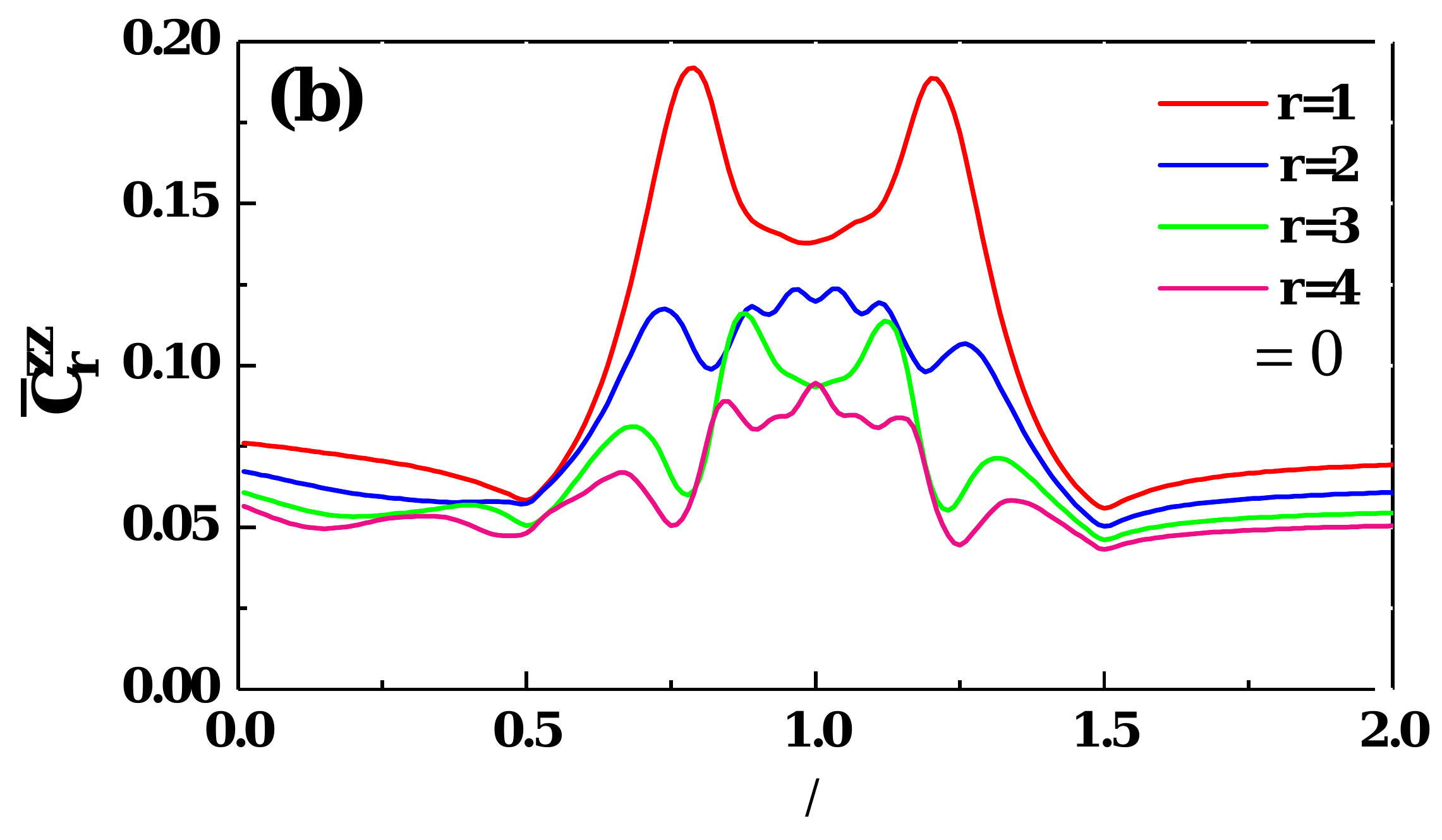}}
\caption{Time averaged of $C^{zz}_{r}$ versus
driving frequency, for several fixed separations of the Floquet $XY$ model,
in the presence of periodically time dependent Hamiltonian. System
size is $N = 100$, and Hamiltonian parameters are $J=0.25\pi, h=0.5\pi,
\gamma=0.5$.}
\label{Fig4}
\end{figure}
%
So their Heisenberg evolution can not be obtained simply from the free-fermion Heisenberg-evolved operator $A_{\mu}(t)$ and $B_{\mu}(t)$, because the Heisenberg evolution of the fermion operators are simple only
when the proposed Hamiltonian in free-fermion language is fixed over the full Fock space, including both parity sectors.
However, we can use the "double trick" to deal with this case~\cite{mccoy1971statistical,bao2020out}, by defining the following
quantity,
%
\begin{equation}
\label{eq7}
\Gamma^{\mu,\nu}_{r}(t)=\langle (\sigma_{\frac{N}{2}}^{\mu}(t) \sigma_{N-r}^{\mu}(t) \sigma_{0}^{\nu} \sigma_{\frac{N}{2}-r}^{\nu})^2 \rangle.
\end{equation}
%
Introducing the function $\Gamma^{\mu,\nu}_{r}(t)$ results in parity cancellation due to pairing of operators; and so one can simply use the Wick's theorem to expand the full function. For large size system and considering the mirror symmetry $F^{\mu,\nu}_{r}(t)=F^{\mu,\nu}_{-r}(t)$, we have~\cite{mccoy1971statistical,bao2020out},
%
\begin{eqnarray}
\label{eq8}
\Gamma^{\mu,\nu}_{r}(t)&&=\langle (\sigma_{\frac{N}{2}}^{\mu}(t)
\sigma_{\frac{N}{2}-r}^{\nu})^2 \rangle \langle (\sigma_{N-r}^{\mu}(t)
\sigma_{0}^{\nu})^2 \rangle
\nonumber\\
&&= F^{\mu,\nu}_{r}(t)F^{\mu,\nu}_{-r}(t)=(F^{\mu,\nu}_{r}(t))^2.
\end{eqnarray}
%
Therefore, to obtain $F^{xx}_{r}(t)$, $F^{xy}_{r}(t)$ and $F^{xz}_{r}(t)$, we need to calculate $\Gamma^{xx}_{r}(t)$, $\Gamma^{xy}_{r}(t)$ and $\Gamma^{xz}_{r}(t)$, respectively (see Appendix \ref{A}). Then, we make do similar procedure as of Sec. II.B using Pfaffian method of the appropriate antisymmetric matrices and finally obtain the OTOC of nonlocal operator as
%
\begin{equation}
\label{eq9}
F^{\mu,\nu}_{r}(t)=\pm \sqrt{Pf(\Phi_{\mu,\nu})}=\pm (Det(\Phi_{\mu,\nu}))^{\frac{1}{4}}
\end{equation}
%
In this paper, we will study both local and nonlocal OTOCs of two exactly solvable Floquet spin $1/2$ model to investigate the behaviour of OTOC and its ability to capture the FDQPT.

\section{Dynamical quantum phase transition\label{DPT}}
Recently, a new research area of quantum phase transition has been investigated in nonequilibrium quantum systems, called dynamical quantum phase transitions (DQPTs) as a counterpart of equilibrium thermal phase transitions~\cite{heyl2013dynamical,heyl2018dynamical}. DQPT represents a phase transitions between dynamically emerging quantum phases, that occurs during the nonequilibrium coherent quantum time evolution under quenching~\cite{Sadrzadeh2021,Uhrich2020,Halimeh2017,heyl2018dynamical,zvyagin2016dynamical} or time-periodic modulation of Hamiltonian~\cite{yang2019floquet,zamani2020floquet,jafari2021floquet,Asboth2012,Asboth2013,naji2021,Jafari2021c}. In DQPT the real time acts as a control parameter analogous to temperature in conventional equilibrium phase transitions.
The DQPT characterized by the nonanalytical behavior of dynamical free energy~\cite{heyl2013dynamical,heyl2018dynamical,jafari2019quench,jafari2019dynamical,
Jafari2017,Divakaran2013,Najafi2019,yan2020information,Zache2019,wang2019probing,Serbyn2017,jafari2021floquet,Yu2021} which is defined as
%
\bea
\no
g(t)=-\lim_{N\rightarrow\infty}\frac{1}{N}\ln|{\cal L}(t)|^{2}.
\eea
%
Here, $N$ is the system size and ${\cal L}(t)$ is the Loschmidt amplitude, which is given by ${\cal L}(t)=\langle\psi(0)|\psi(t)\rangle$,
%
%
where $|\psi(0)\rangle$ and $|\psi(t)\rangle$ are the initial state of system and its corresponding time evolved state at a later time $t$, respectively.

However, in experiments~\cite{flaschner2018observation,jurcevic2017direct}, to search the far-from-equilibrium theoretical concepts, the initial state in which system is prepared, is a mixed state. This motivates to propose the generalized  Loschmidt amplitude (GLA) for mixed thermal states, which perfectly replicate the nonanalyticities manifested in the pure state DQPTs~\cite{bhattacharya2017mixed,heyl2017dynamical}. The GLA for thermal mixed state is defined as follows
\begin{equation}
\nonumber
{\cal L}(t) =Tr \Big(\rho(0) U(t)\Big),
\end{equation}
where $\rho(0)$ is the mixed state density matrix at time $t=0$, and $U(t)$ is the time-evolution operator.
%
\begin{figure*}[t]
\begin{minipage}{\linewidth}
\centerline{\includegraphics[width=0.35\linewidth]{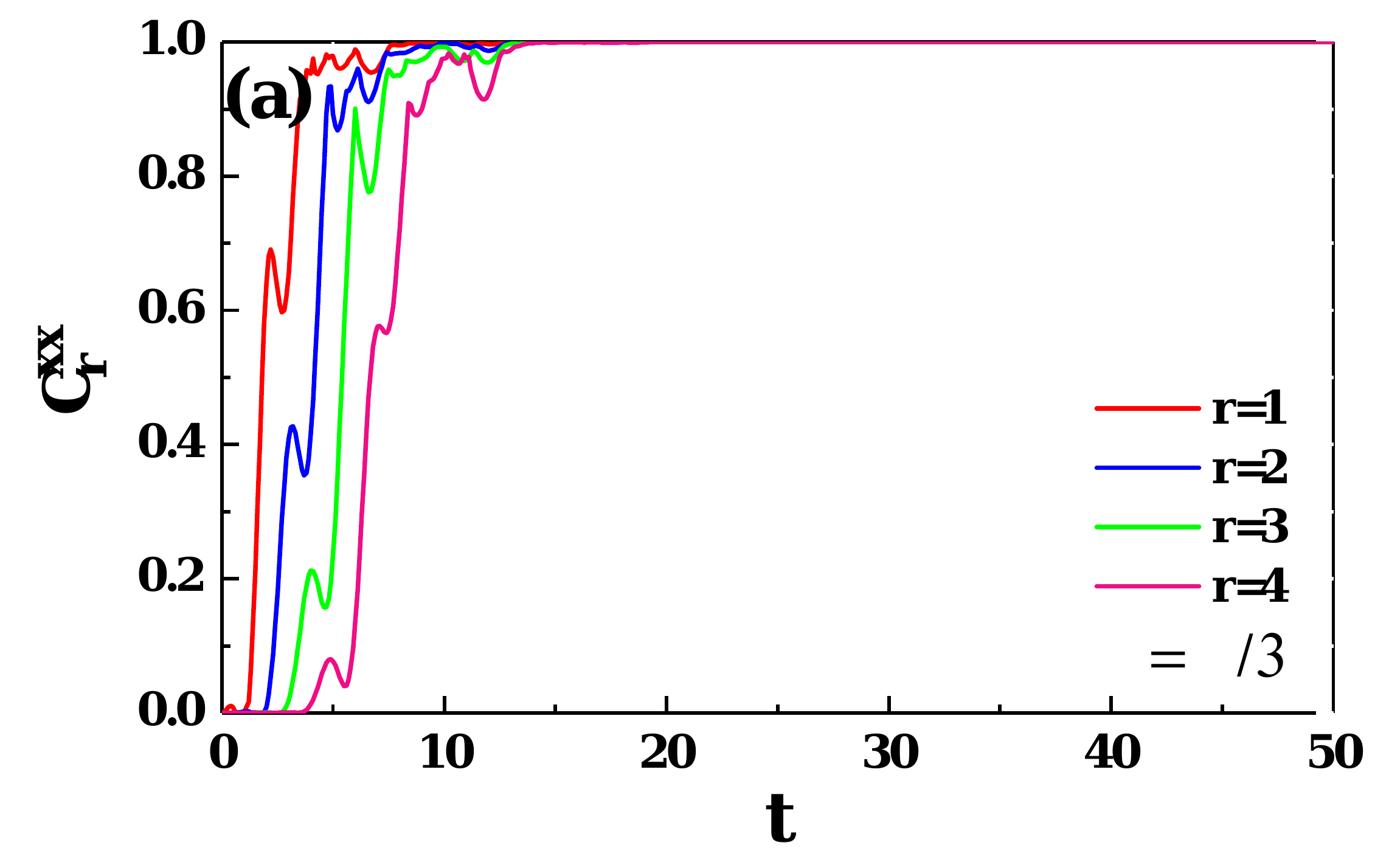}
\includegraphics[width=0.32\linewidth]{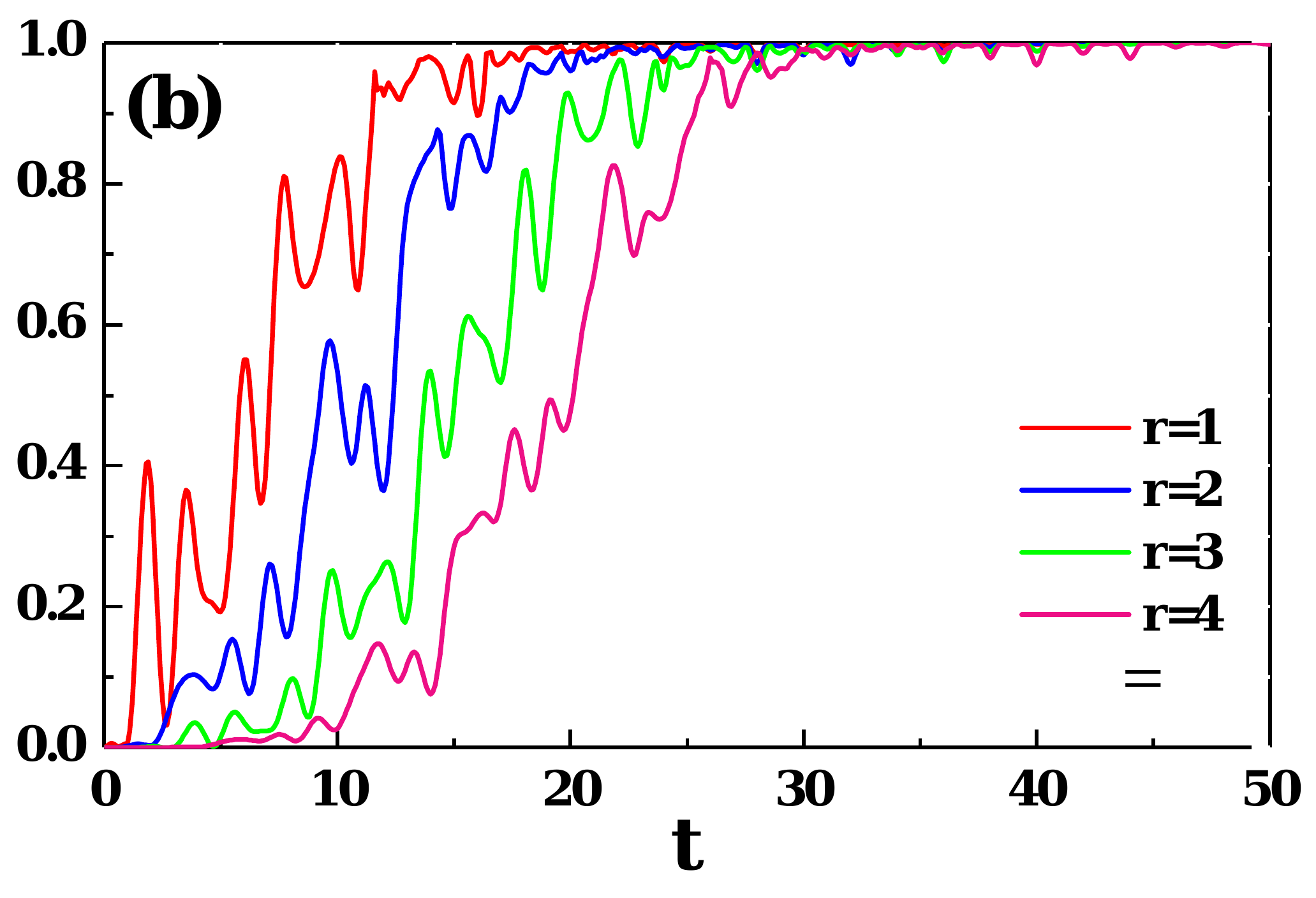}
\includegraphics[width=0.32\linewidth]{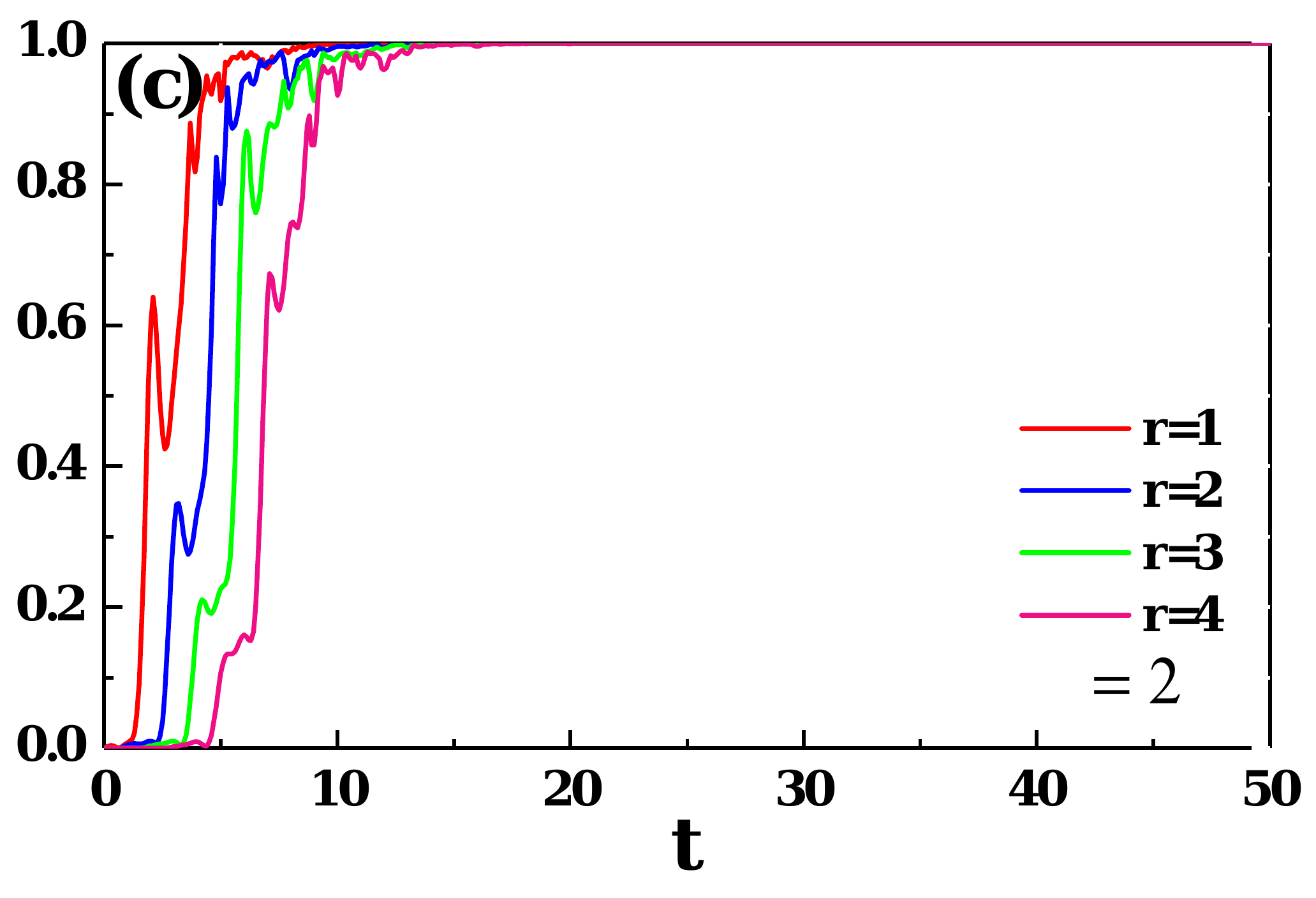}}
\centering
\end{minipage}
\caption{(Color online) Time evolution of nonlocal OTOC, $C^{xx}_{r}$, versus time for fixed separations  $r=1, 2, 3, 4$,
 of Floquet $XY$ model,
with $\beta=0$ and different values of driving frequency,
(a) $\omega=\pi/3$, (b) $\omega=\pi$, (c) $\omega=2 \pi$.
The model parameters are $J=0.25\pi, h=0.5\pi, \gamma=0.5$ and $N = 100$.}
\label{Fig5}
\end{figure*}
%

\section{Floquet XY model\label{FXYM}}
The Hamiltonian of the one-dimensional periodically driven spin-$1/2$ chain, is given as follows
{\small
\begin{eqnarray}
\nonumber
{\cal H}(t)=\!\!\!\!\!\sum_{n=-N/2}^{N/2}\!\!\!\!\!\!\!\!\!&&\Big[(J-\gamma\cos(\omega t)) s_n^x s_{n+1}^x + (J+\gamma\cos(\omega t)) s_n^y s_{n+1}^y\\
\label{eq10}
&&-\gamma\sin(\omega t) \Big(s_n^x s_{n+1}^y + s_n^y s_{n+1}^x \Big)+h s_n^z \Big],
\end{eqnarray}
}
where, $N+1$ is the size of the system, $J$, $h$ and $\gamma$ are system parameters, and $\omega$ is the driving frequency. Here, $S_n^\alpha$ are the  spin-half operators at the $n$th site, i.e. $S_n^\alpha=\frac{1}{2}\sigma^{\alpha}_{n} $. In order to calculate the spin correlation functions, we should diagonalize the above Hamiltonian. The Hamiltonian can be exactly diagonalized by Jordan-Wigner transformation, which transforms spins into spinless fermions. It should be mentioned that the fermionic representation of the Hamiltonian is equivalent to the one dimensional p-wave superconductor with time dependent pairing phase (magnetic flux)~\cite{jafari2021floquet,zamani2020floquet}. The Fourier transformed fermionic Hamiltonian can be expressed as sum of independent terms ${\cal H}(t) =\sum_{k{\cal{2}} BZ} H_{k}(t)$, in which

\begin{eqnarray}
\nonumber
H_{k}(t)=&&h_{z}(k)\big(c_{k}^{\dagger}c_{k}-c_{-k}c_{-k}^{\dagger}\big)\\
-&i&h_{xy}(k)\big(e^{-i\omega t}c_{k}^{\dagger}c_{-k}^{\dagger}+e^{i\omega t}c_{k}c_{-k}\big) ,
\label{eq11}
\end{eqnarray}
where, $h_{z}(k)=J\cos(k)+h$ and $h_{xy}(k)=\gamma\sin(k)$. The eigenstates and eigenvalues of the Hamiltonian Eq. (\ref{eq11}) are obtained by solving the time dependent Schr\"{o}dinger equation~\cite{yang2019floquet,jafari2021floquet,zamani2020floquet,naji2021} (see Appendix \ref{B}).

It is straightforward to show that the exact expression of the GLA is represented by~\cite{yang2019floquet,jafari2021floquet}
%
{\small
\bea
\no
{\cal GL}(t)=\Pi_{k}{\cal GL}_{k}(t),~~
{\cal GL}_{k}(t)={\cal R}(k,t)+i\,{\cal I}(k,t)\tanh(\beta\varepsilon_{k}),
\eea
}
%
with
%
\bea
\no
{\cal R}(k,t)&=&\cos(\varepsilon_{k}t)\cos(\omega t/2)-\frac{B_{z}(k)}{\varepsilon_{k}}\sin(\varepsilon_{k}t)\sin(\omega t/2),\\
\no
{\cal I}(k,t)&=&\sin(\varepsilon_{k}t)\cos(\omega t/2)+\frac{B_{z}(k)}{\varepsilon_{k}}\cos(\varepsilon_{k}t)\sin(\omega t/2),
\eea
%
where, $B_{z}(k)=h_{z}-\omega/2$ and $\varepsilon_{k}=\sqrt{h_{xy}^{2}(k)+B_{z}^{2}(k)}$. It has been shown that the model shows FDQPTs, at any temperature, when the driving frequency ranges from $2(h-J)$ to $2(h+J)$, i.e., $2(h-J)<\omega<2(h+J)$, where the system experience adiabatic cyclic processes~\cite{yang2019floquet,zamani2020floquet}. In the following we will examine the behavior of the OTOCs in the Floquet XY model to obtain their early and long time scaling behavior.

\subsubsection{OTOC of local operators in Floquet XY model}
The OTOCs in the Floquet XY model can be obtained by lengthy and tedious calculation (see the Appendix \ref{C}).
Fig. \ref{Fig1} represents $C^{zz}_{r}(t)$ of the Floquet XY model versus time at infinite temperature, $\beta=0$,
for different values of driving frequency and $N=100$.
As seen, $C^{zz}_{r}(t)$ reveals bounded cone structure (which indicates the bound of butterfly effect) with the velocity of wavefront
$c\approx 0.66$ for $\omega= \pi/3$ and $2\pi$ (no FDQPTs regime), and  $c\approx 0.28$ for $\omega= \pi$ (FDQPTs regime).
The numerical value of velocities is in good agreement with the maximum quasiparticle group velocities ($\partial\varepsilon_{k}/\partial k$)
of the effective time-independent Floquet Hamiltonian Eq. (\ref{eq:B7}) at fixed frequencies.
The maximum quasiparticle group velocity of the effective time-independent Floquet Hamiltonian has been plotted in Fig. \ref{Fig2} versus $\omega$.
As seen, the maximum quasiparticle group velocity gets a minimum at the mid-frequency of the region,
where FDQPT occurs, namely $\omega=\pi$.
Comparing Figs. \ref{Fig1}(a)-(c) indicates that the system in FDQPTs regime (Fig. \ref{Fig1}(b)) reveals narrower light
cone with slower spreading of local operator which expresses slower information propagating, witnessed by Fig. \ref{Fig2}.
This can be expected as the system evolves adiabatically
in FDQPTs regime~\cite{zamani2020floquet}, while the system experiences non-adiabatic cyclic process in no-FDQPTs regime.
To examine the behavior of $C^{zz}_{r}(t)$ accurately for small value of separations $r$, $C^{zz}_{r}(t)$ has been depicted versus
time in Figs. \ref{Fig1}(d)-(f). As seen, $C^{zz}_{r}(t)$ typically enhances in a short time from zero to its maximum value
and then decreases to vanishing at long time in periodic manner. Indeed, we observe that the OTOC composed with local
operators show no sign of scrambling, namely $\lim_{t\rightarrow\infty}C^{zz}_{r}(t)=0$ (which is the same
as the value at $t=0$).

In addition, as is clear, the maximum value of $C^{zz}_{r}(t)$ decreases by increasing the separation $r$.
So, it is important to probe how OTOC behaves at the early and the long times with fixed sites.
%
\begin{figure*}
\begin{minipage}{\linewidth}
\centerline{\includegraphics[width=0.32\linewidth]{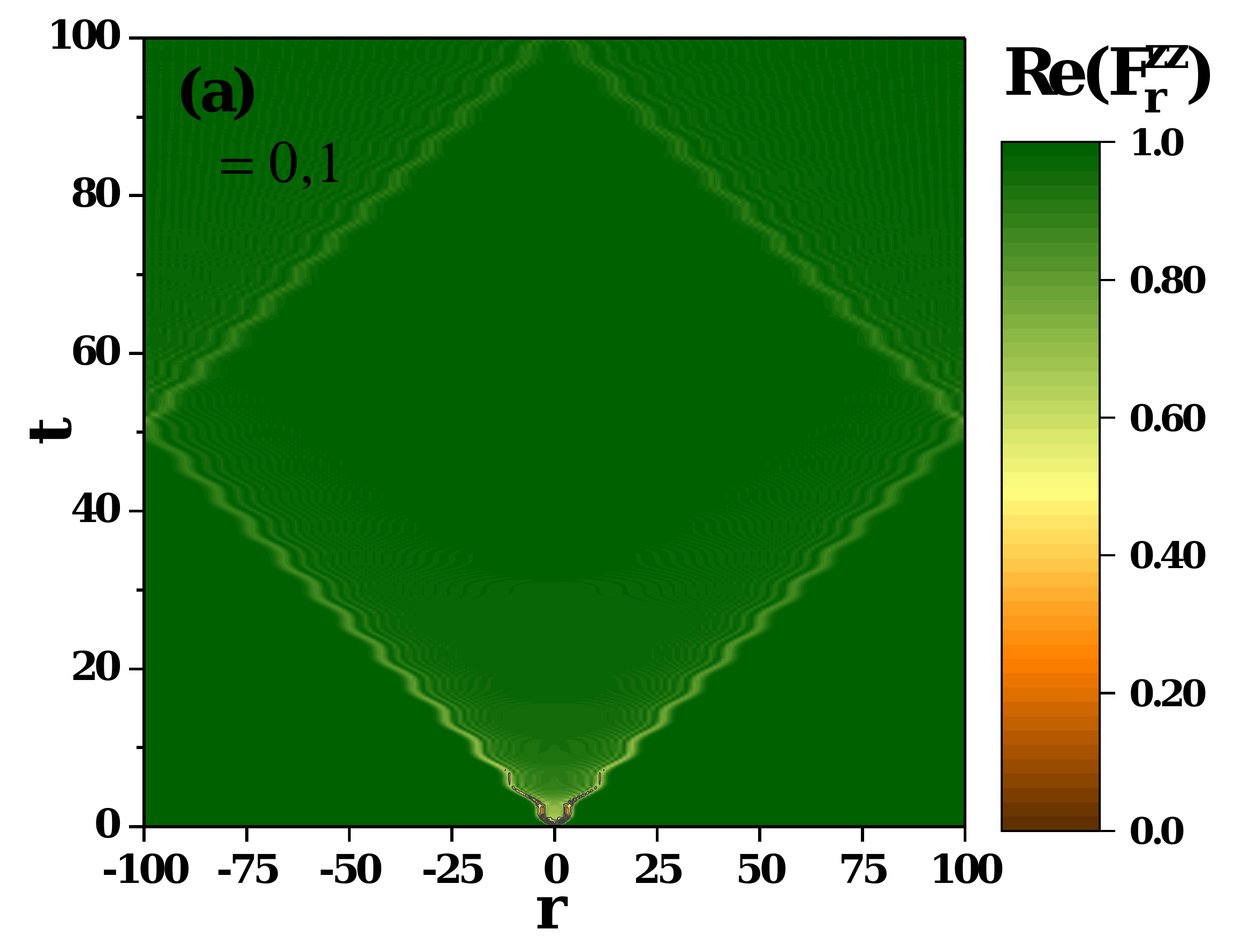}
\includegraphics[width=0.33\linewidth]{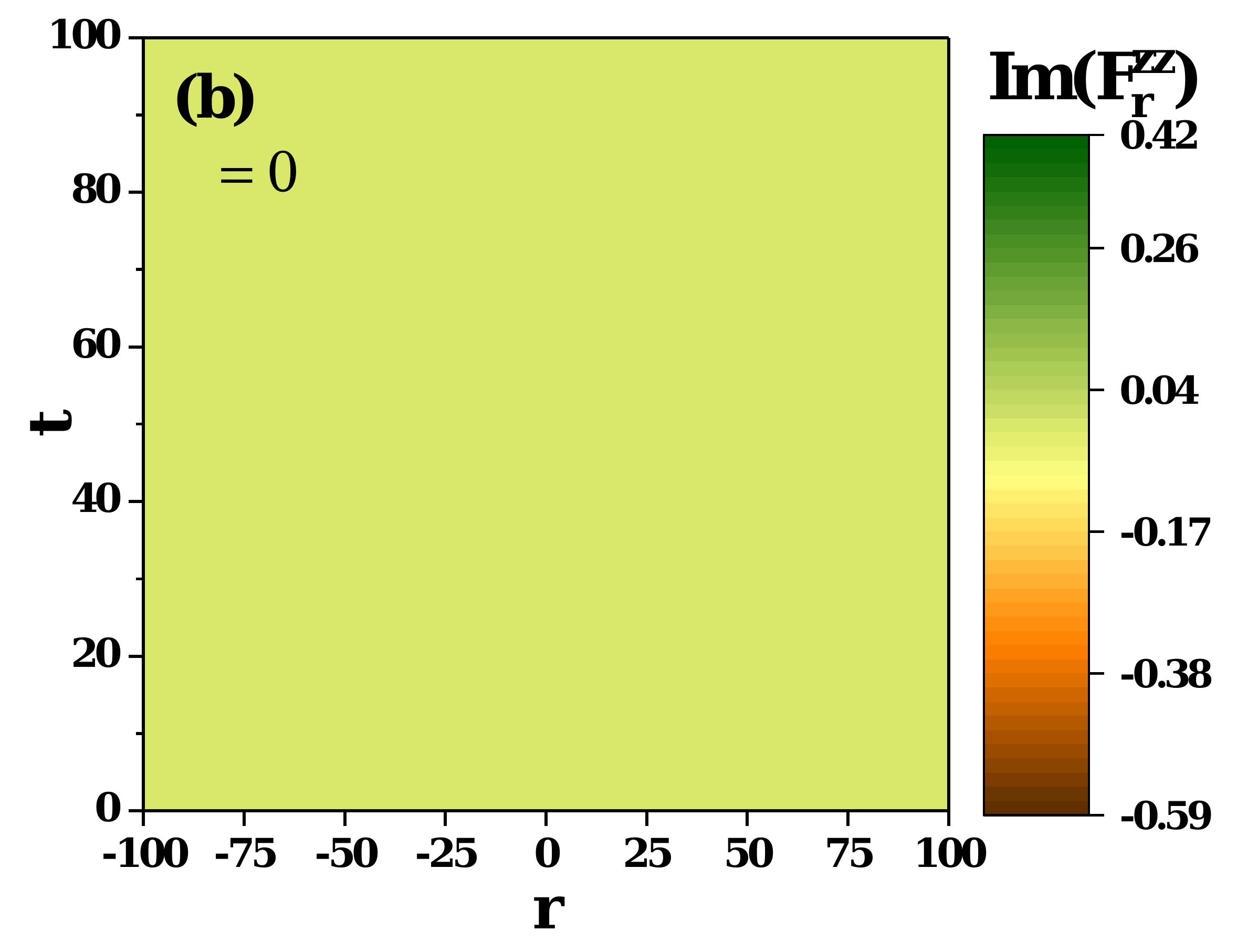}
\includegraphics[width=0.33\linewidth]{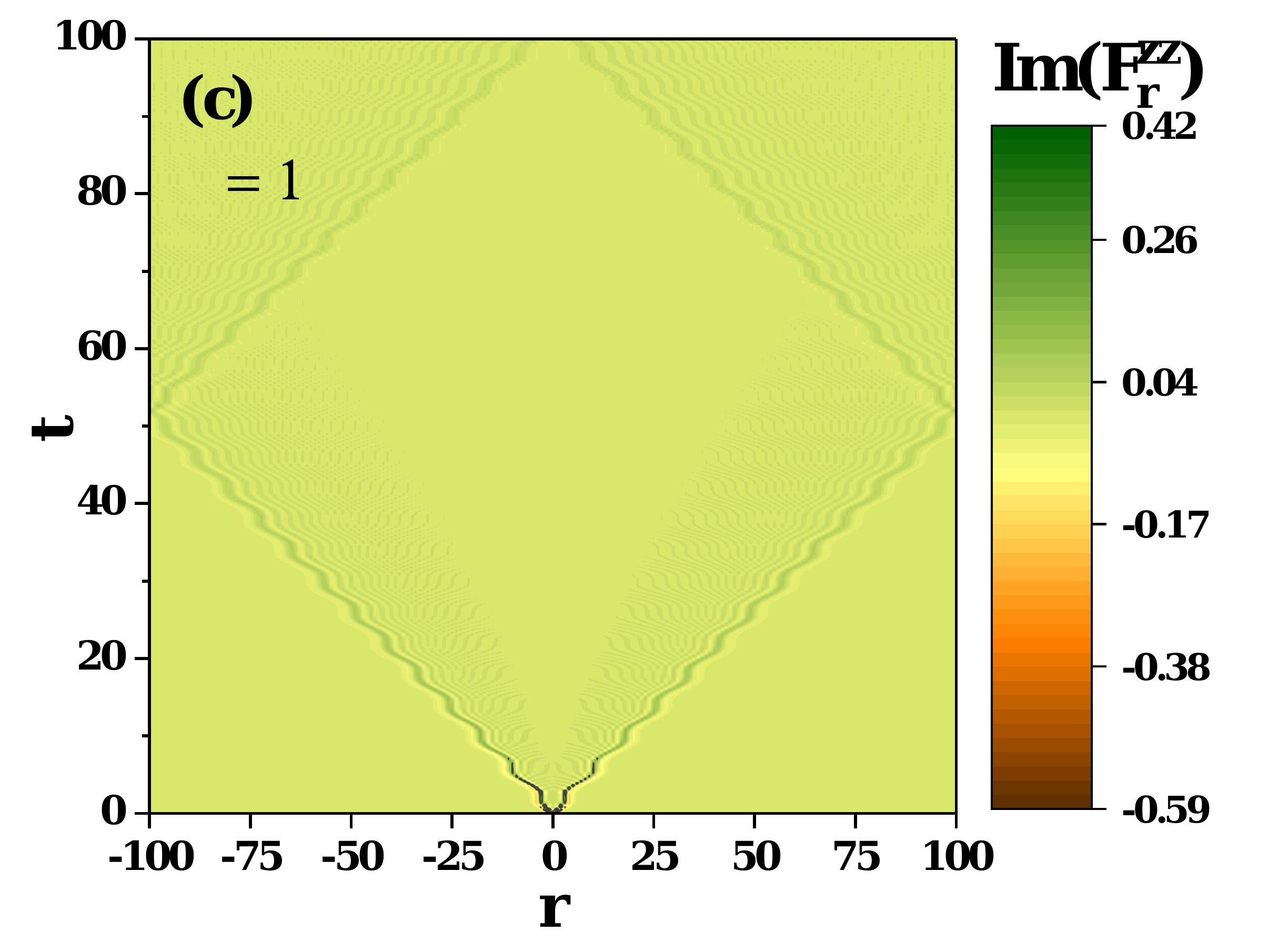}}
\centering
\end{minipage}
\caption{(Color online) Density plot of $F^{zz}_{r}(t)$ versus time and separation,
for synchronized Ising model at the strong coupling $\lambda=1$, in the presence of $h(t)=h_{0}+h_{1}\cos(\omega t)$.
(a) Real part of $F^{zz}_{r}(t)$ for both $\beta=0, 1$.
Imaginary part of $F^{zz}_{r}(t)$ for  (b) infinite temperature $\beta=0$ and (c) finite temperature $\beta=1$.
We consider $N = 200$ and Hamiltonian parameters are $h_{0}=1, h_{1}=1, \omega=\pi/2$.}
\label{Fig6}
\end{figure*}
%
%
\begin{figure*}
\begin{minipage}{\linewidth}
\centerline{\includegraphics[width=0.31\linewidth]{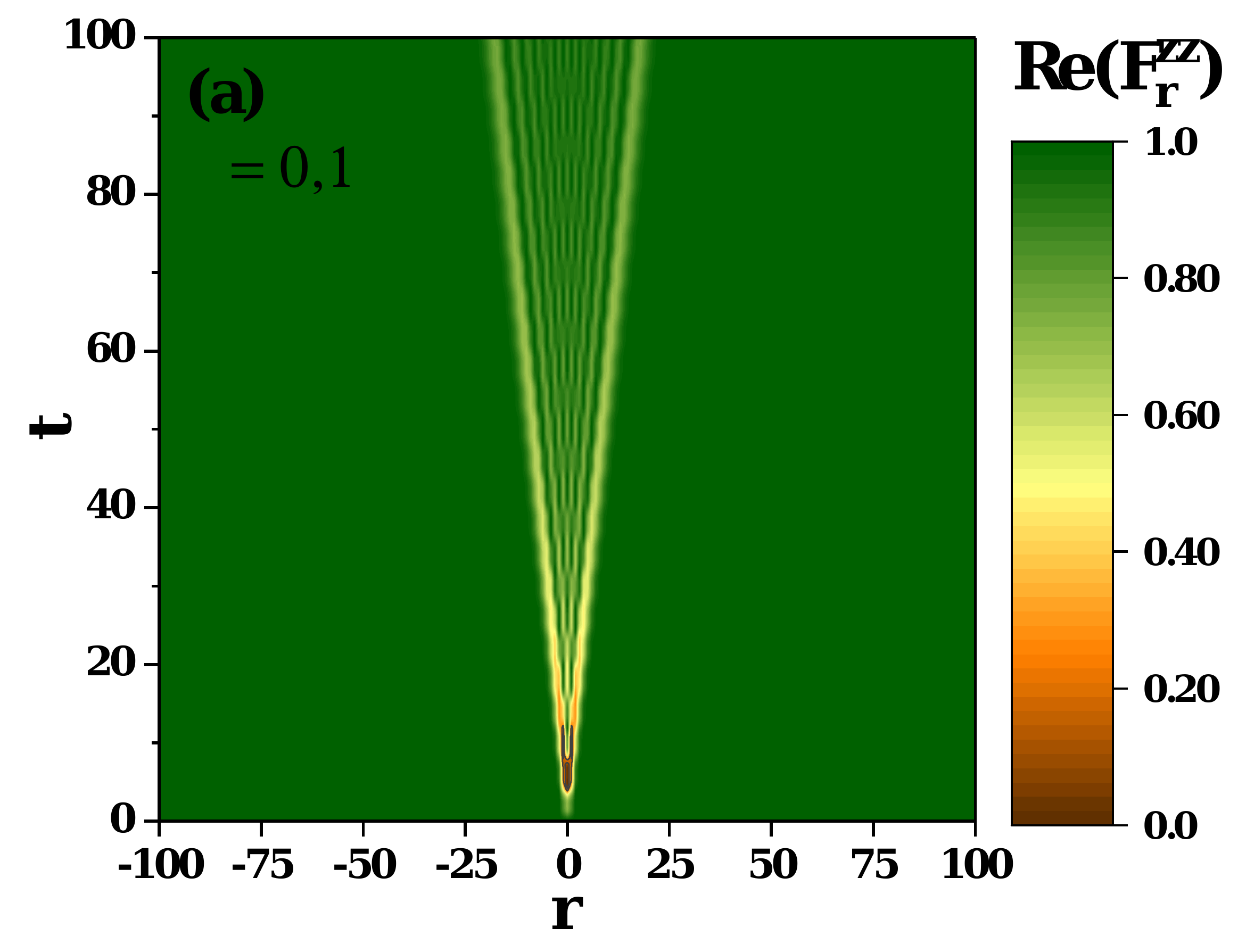}
\includegraphics[width=0.33\linewidth]{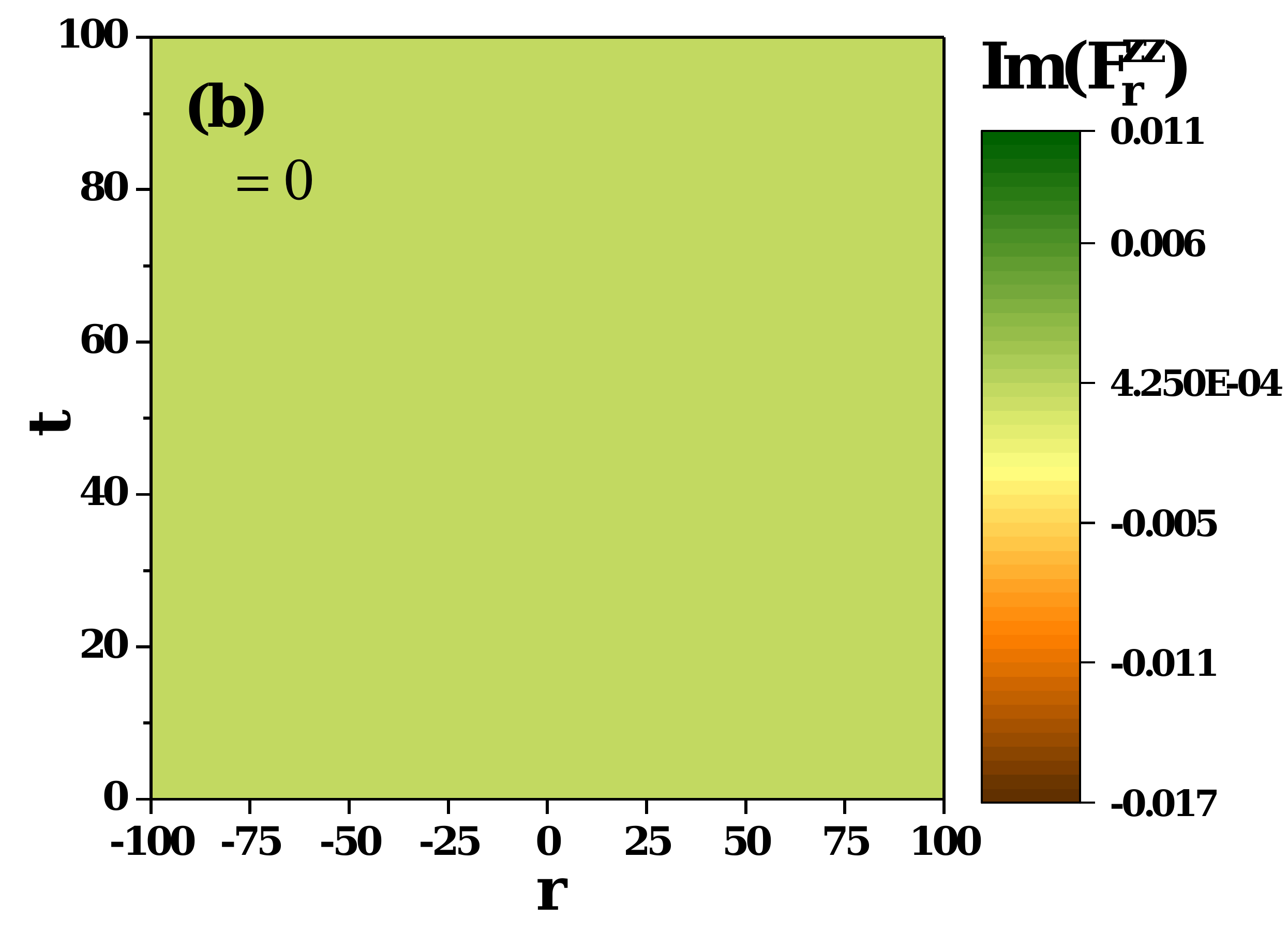}
\includegraphics[width=0.33\linewidth]{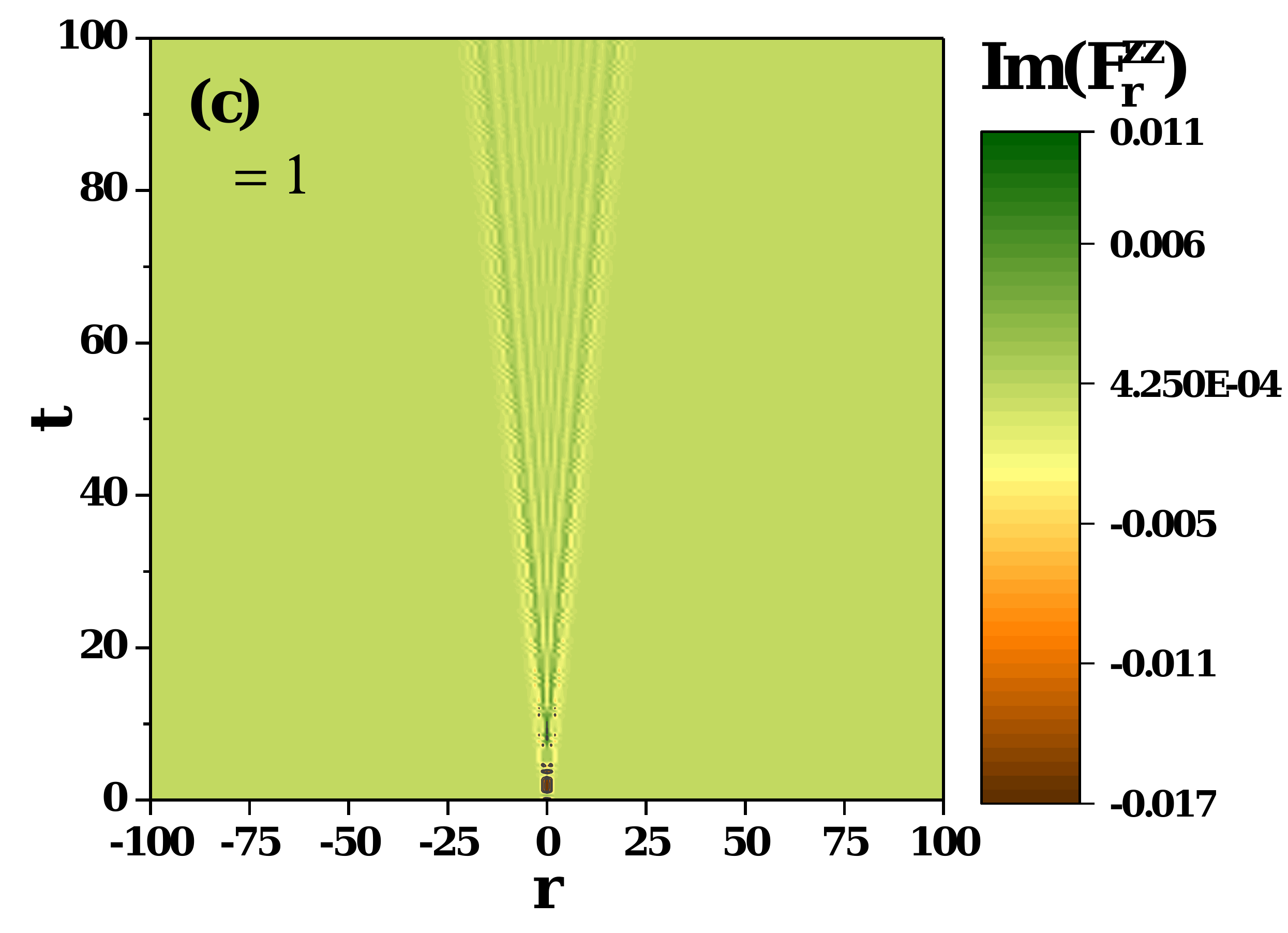}}
\centering
\end{minipage}
\caption{(Color online)
Density plot of real and imaginary parts of $F^{zz}_{r}(t)$ for weak coupling $\lambda=0.1$ while
other parameters are the same as in Fig. \ref{Fig6}.}
\label{Fig7}
\end{figure*}
%
The numerical simulation of $C^{zz}_{r}(t)$ is illustrated in Fig. \ref{Fig3} with sites $r=1, 2, 3, 4$, and we can see clearly
that the early time behavior is growing power law, $t^{2r}$, for any values of driven frequency. However, the long time behavior represents $t^{-1}$ decay
in no-FDQPTs regime ($\omega= \pi/3$, $2\pi$), independent of $r$ and $\beta$.
In FDQPTs regime ($\omega= \pi$) the long time decaying behaviour of $C^{zz}_{r}(t)$ is $t^{-2.6}$ and approximately disordered.
Consequently, we expect that $C^{zz}_{r}(t)$ could show signatures to detect the range of driving frequency over which the Floquet DQPT occurs.
For this purpose we have calculated the infinite-temperature time average of $C^{zz}_{r}(t)$ as a function of frequency $\bar{C}^{zz}_{r}=\frac{1}{T}\int_{0}^{T}C^{zz}_{r}(t^{\prime})(t^{\prime})dt^{\prime}$,
with $T=\frac{2\pi}{\omega}$. The numerical results has been illustrated in Fig. \ref{Fig4} for different $r$.
As indicated, $\bar{C}^{zz}_{r}$ is roughly constant in no-FDQPTs regime while in the FDQPTs regime its experiences large variation and
its global minimums signals the boundary values of the window frequency over which the system shows FDQPTs.
So, the time average of local OTOC can serve as a dynamical order parameter that dynamically detect
the range of driven frequency over which FDQPTs occur.
Moreover, the different long time decaying behaviour of $C^{zz}_{r}(t)$ at FDQPTs and no FDQPTs regimes can be interpreted as an indicator of
non-adiabatic to adiabatic topological transition~\cite{zamani2020floquet}.

\subsubsection{OTOC of nonlocal operators in Floquet XY model}
As mentioned before, for exactly solvable spin $1/2$ model by means of Jordan-Wigner transformation,
there are five kinds of OTOC of nonlocal operators. In Figs. \ref{Fig5}(a)-(c), $C^{xx}_{r}(t)$ has been depicted
for $r=1, 2, 3, 4$ in FDQPTs and no-FDQPTs regimes.
We can see that, $C^{xx}_{r}(t)$ in both FDQPTs and no-FDQPTs regimes increases rapidly at short initial
time from zero to reach its saturated value, $1$. Since nonlocal operators bear nonlocal information about operators,
the OTOC composed with nonlocal operators shows the signature of scrambling which is their main differences compared
with local ones. As can be seen from Fig. \ref{Fig5}(b), enhancement of $C^{xx}_{r}(t)$ in the FDQPTs regime
is slower than that in no-FDQPTs regime, which means delocalization of information, in FDQPTs regime, occurs more slowly in
comparison with no-FDQPTs case. Other OTOCs of nonlocal operators show similar behaviors (see Fig. \ref{Fig11} in Appendix \ref{C}).

\section{Synchronized Floquet XY model\label{FSXYM}}
The Hamiltonian of synchronized Floquet $XY$ model is given by~\cite{Jafari2021c},
\begin{equation}
\label{eq12}
H(t) =-\sum_{n=1}^{N}\Big[ \frac{J(t)}{2}\Big( (1+\gamma) \sigma_n^x
\sigma_{n+1}^x + (1-\gamma) \sigma_n^y \sigma_{n+1}^y \Big)+ h(t) \sigma_n^z\Big],
\end{equation}
%
where $J(t)=\lambda h(t)$, $h(t)=h_{0}+h_{1}\cos(\omega t)$ and $\gamma$ represents the anisotropy.
%
\begin{figure*}
\begin{minipage}{\linewidth}
\centerline{\includegraphics[width=0.34\linewidth]{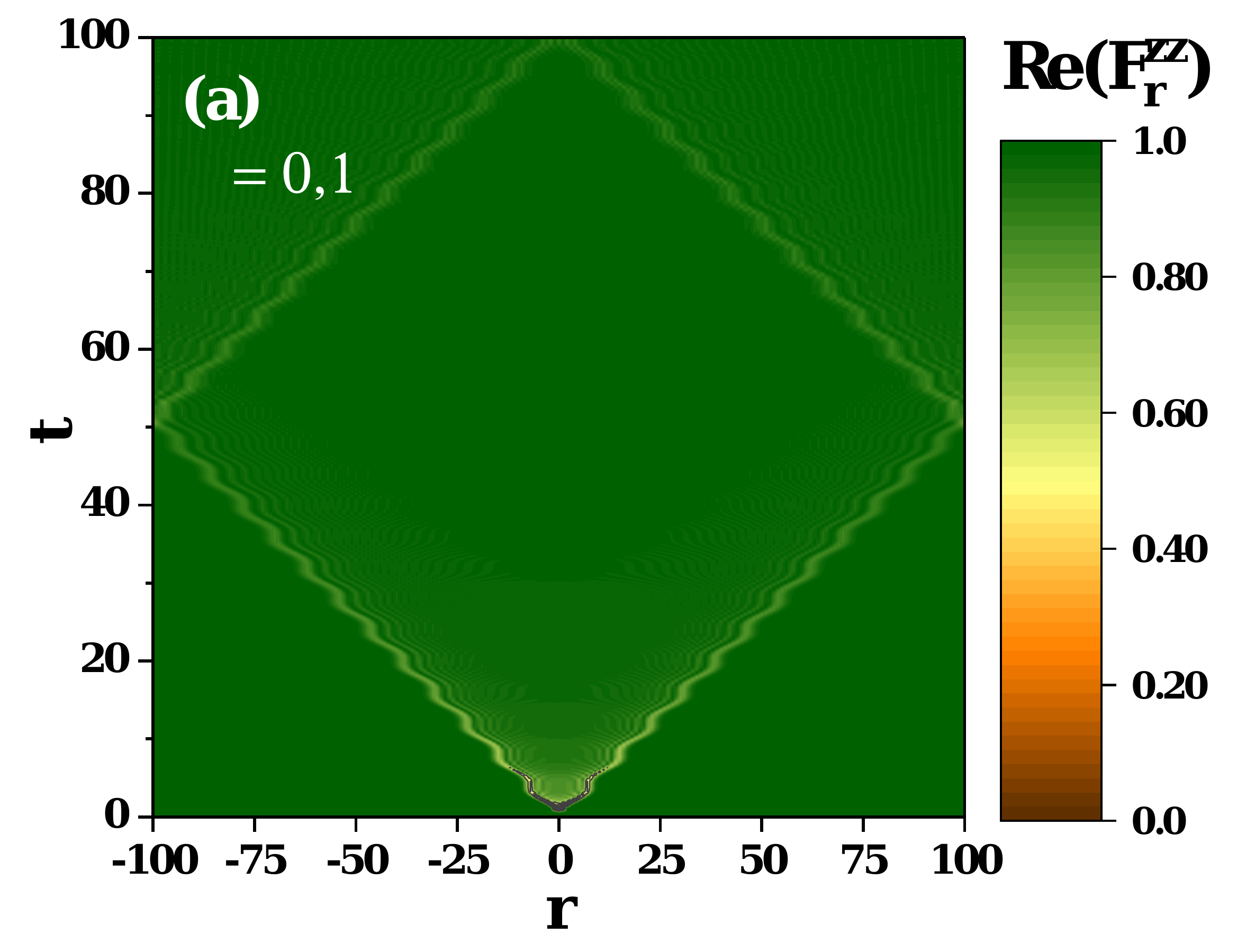}
\includegraphics[width=0.33\linewidth]{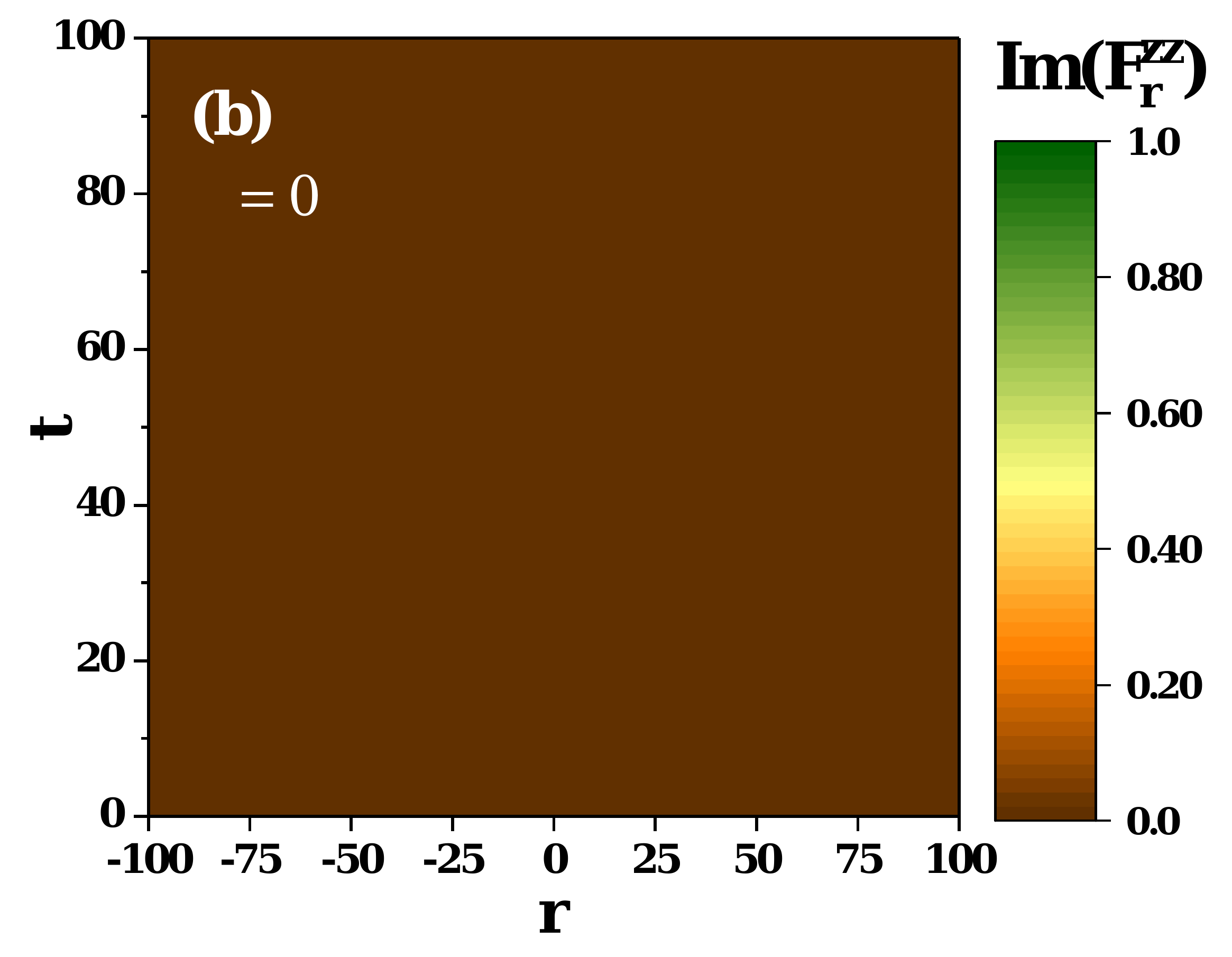}
\includegraphics[width=0.33\linewidth]{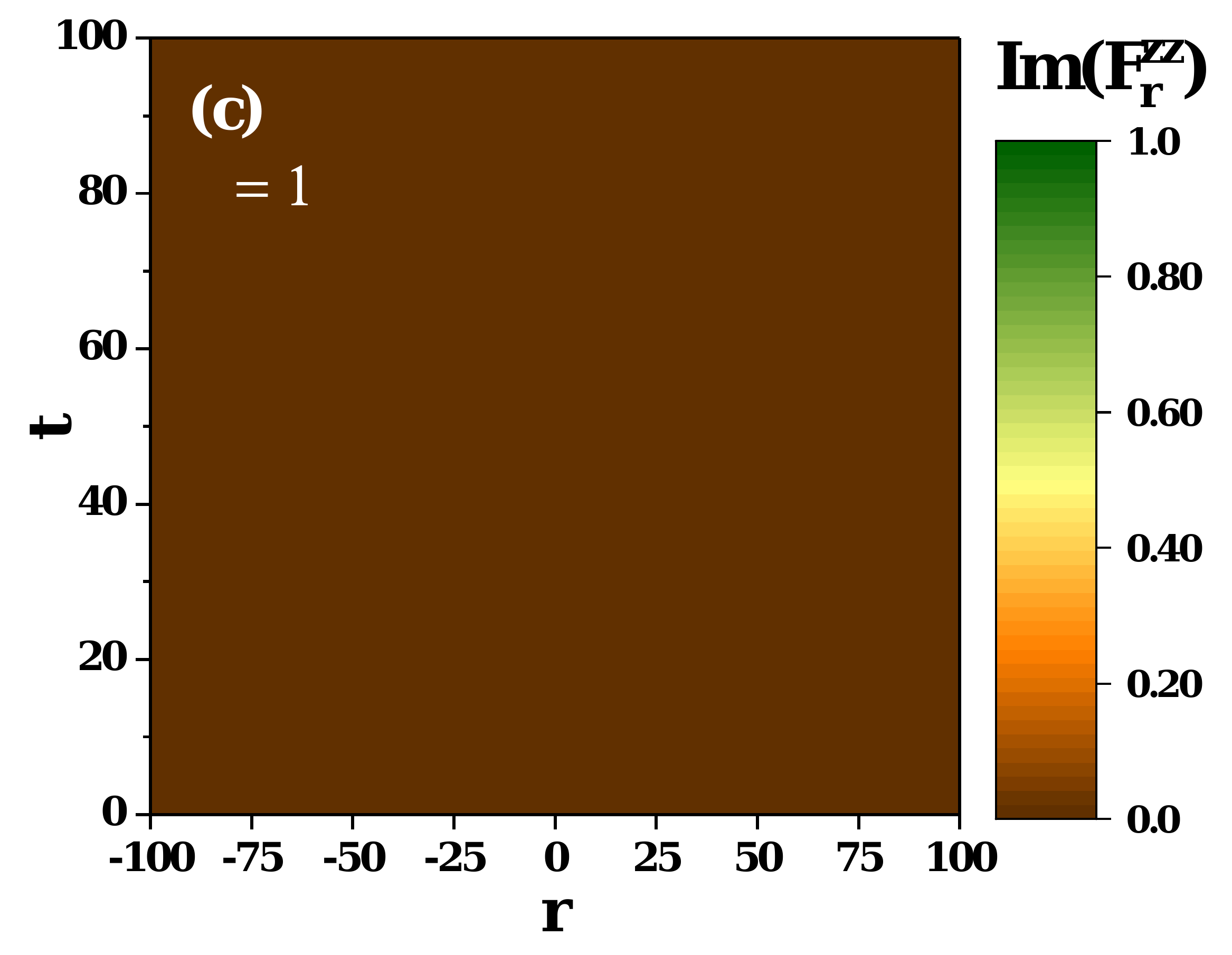}}
\centering
\end{minipage}
\caption{(Color online)
Density plot of $F^{zz}_{r}(t)$ versus time and separation,
for synchronized Ising model at the strong coupling $\lambda=1$, in the presence of $h(t)=h_{0}+h_{1}\cos(\omega t)$.
(a) Real part of $F^{zz}_{r}(t)$ for both infinite and finite temperature $\beta=0, 1$.
(b) Imaginary part of $F^{zz}_{r}(t)$ for (b) $\beta=0$ and (c) $\beta=1$.
The model is in strong coupling $\lambda=1$ and  $h_{0}=1, h_{1}=-1, \omega=\pi/2, N = 200$,
which represents FDQPT at any temperature. The imaginary part of $F^{zz}_{r}(t)$ is always zero.}
\label{Fig8}
\end{figure*}
%
%
\begin{figure*}
\begin{minipage}{\linewidth}
\centerline{\includegraphics[width=0.33\linewidth]{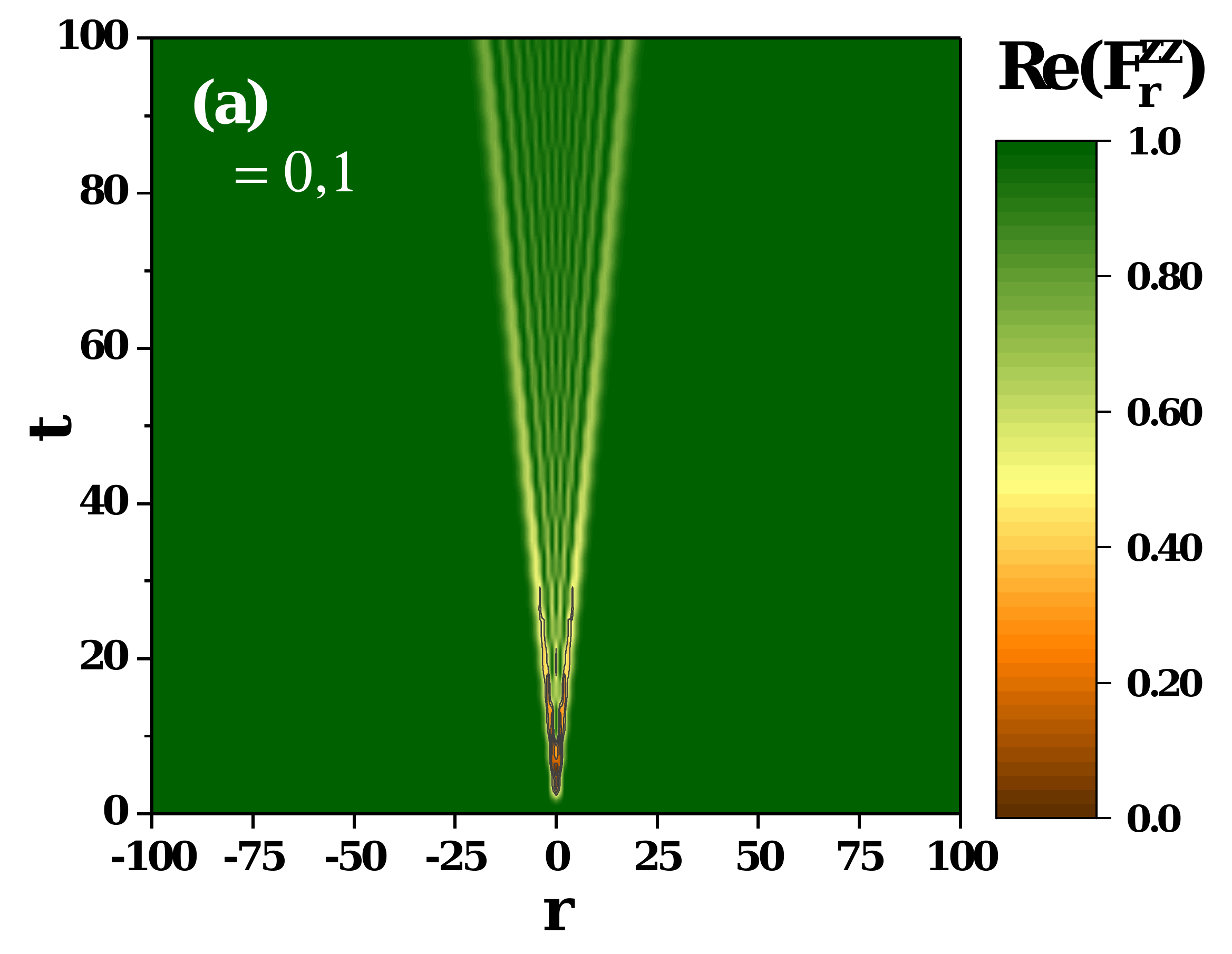}
\includegraphics[width=0.33\linewidth]{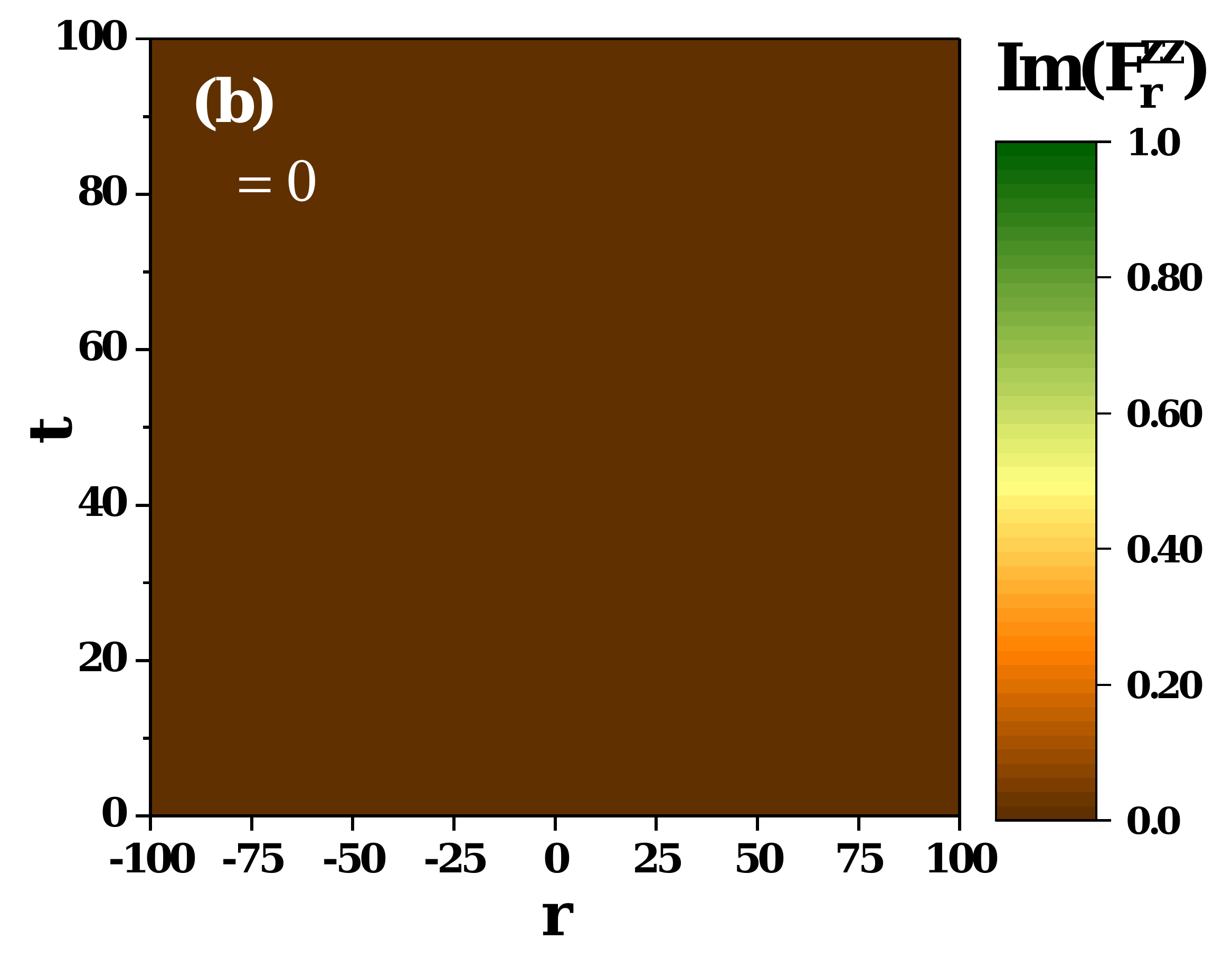}
\includegraphics[width=0.33\linewidth]{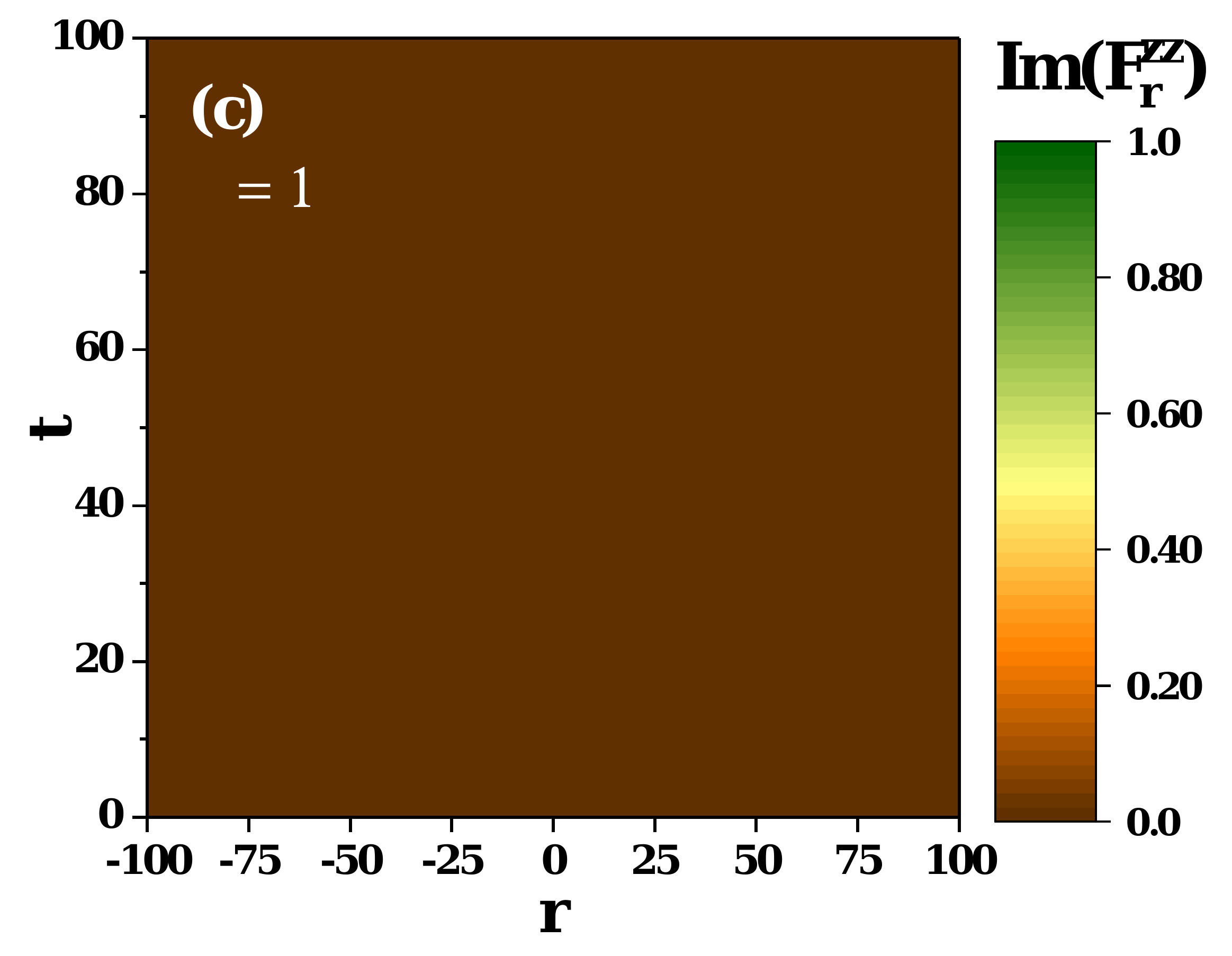}}
\centering
\end{minipage}
\caption{(Color online)
Density plot of real and imaginary parts of $F^{zz}_{r}(t)$ for weak coupling $\lambda=0.1$ while
other parameters are the same as in Fig. \ref{Fig8}, with imaginary part of $F^{zz}_{r}(t)$ is zero everywhere in (b) and (c).
}
\label{Fig9}
\end{figure*}
%
The Hamiltonian in Eq. (\ref{eq12}) is exactly solvable by means of Jordan-wigner transformation~\cite{Jafari2021c} (see Appendix \ref{D}).

It has been shown that the GLA of the synchronized Floquet XY model is obtained to be~\cite{Jafari2021c}
%
\bea
\label{eq13}
{\cal GL}_{k}(t)=\cos(\epsilon_{k}\tau)+i\sin(\epsilon_{k}\tau)\tanh(|h(0)|\epsilon_{k}\beta),
\eea
%
where $\epsilon_{k}=\sqrt{P^{2}(k)+Q^{2}(k)}$, $P(k)=2\lambda\cos(k)+2$, $Q(k)=2\lambda\gamma\sin(k)$,
$\tau=\int_{0}^{t} h(t')dt'$ and $h(0)=h_{0}+h_{1}$.
The GLA becomes zero if the temperature goes to infinity, i.e., $\beta\longrightarrow0$, at time
instances $\tau^{\ast}=(2n+1)\pi/2\epsilon_{k}, (n=0, 1, 2, \cdots)$.
In addition, the FDQPTs occur for all temperatures if $h(0)=h_{0}+h_{1}$ becomes zero, i,e,. $h_{0}=-h_{1}$. In other words,
FDQPT in synchronized Floquet system depends on the initial conditions,  which occurs
for all range of driving frequency and at any finite or infinite temperature.
For simplicity and without loss of generality we consider the isotropic case $\gamma=1$,
which corresponds to the synchronized Floquet Ising model.

\subsubsection{OTOC of local operators in the synchronized Floquet XY model}
Firstly, we investigate the case, where our model shows FDQPTs at infinite temperature, i.e.,
initial magnetic field is nonzero $h(0)\neq0$.
The local operator spreading in the synchronized Floquet $XY$ model probed by
analyzing $F^{zz}_{r}(t)$, where its vanishing at long-time limit signals the information scrambling.
Fig.  \ref{Fig6} shows numerical simulations of real and imaginary parts of
$F^{zz}_{r}(t)$ versus time and spin separation $r$ for synchronized Ising model $\gamma=1$,
at infinite ($\beta=0$) and finite ($\beta=1$) temperature with system size $N=200$ and the strong synchronized coupling $\lambda=1$.  The real part of $F^{zz}_{r}(t)$, Fig.  \ref{Fig6}(a), reveals the bounded cone structure with the
velocity of wavefront $c=2$.
The situation for weak synchronized coupling  $\lambda=0.1$ is shown in Fig. \ref{Fig7}, where the parameters of
the model are the same as those in Fig.  \ref{Fig6}.
As seen in Fig. \ref{Fig7}, the diagrams of the real part of $F^{zz}_{r}$ exhibit narrower cone structure, representing slower spreading of local operators with the velocity $c=0.2$. It indicates that the speed of operator spreading
depends monotonically on the synchronized coupling strength.

It should be mentioned that, since the synchronized Floquet XY model can not be transformed to the time-independent effective Floquet
Hamiltonian (unlike the Floquet XY model), the quasiparticle group velocity can not be defined here. So, the velocity of wavefront in the synchronized system can not be related to the quasiparticle group velocity of the model. Moreover, it is clear that in both  Figs. \ref{Fig6}(b)-\ref{Fig7}(b) the imaginary part of $F^{zz}_{r}(t)$ is zero at infinite temperature (FDQPT case).
%
\begin{figure*}
\begin{minipage}{\linewidth}
\centerline{\includegraphics[width=0.265\linewidth]{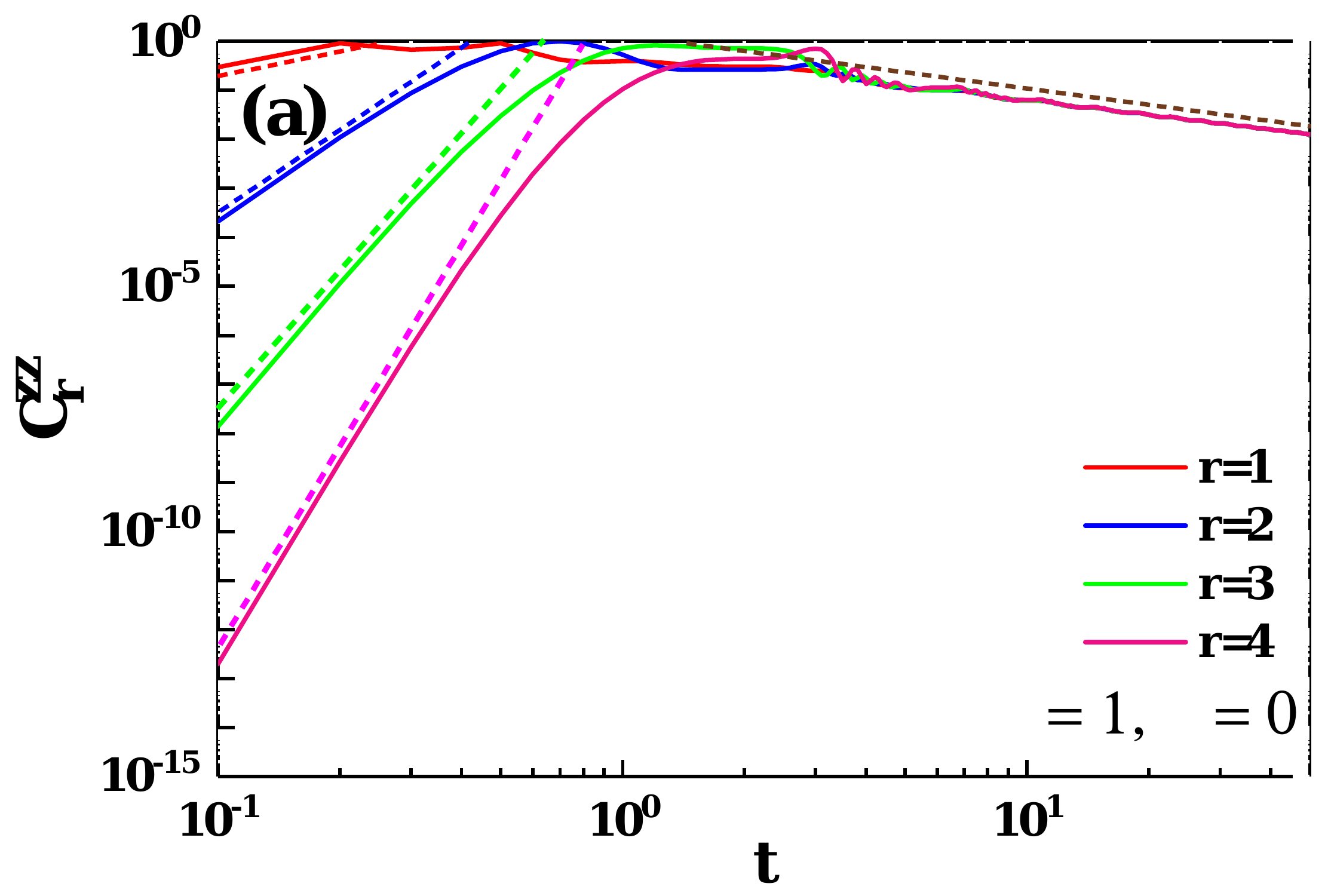}
\includegraphics[width=0.245\linewidth]{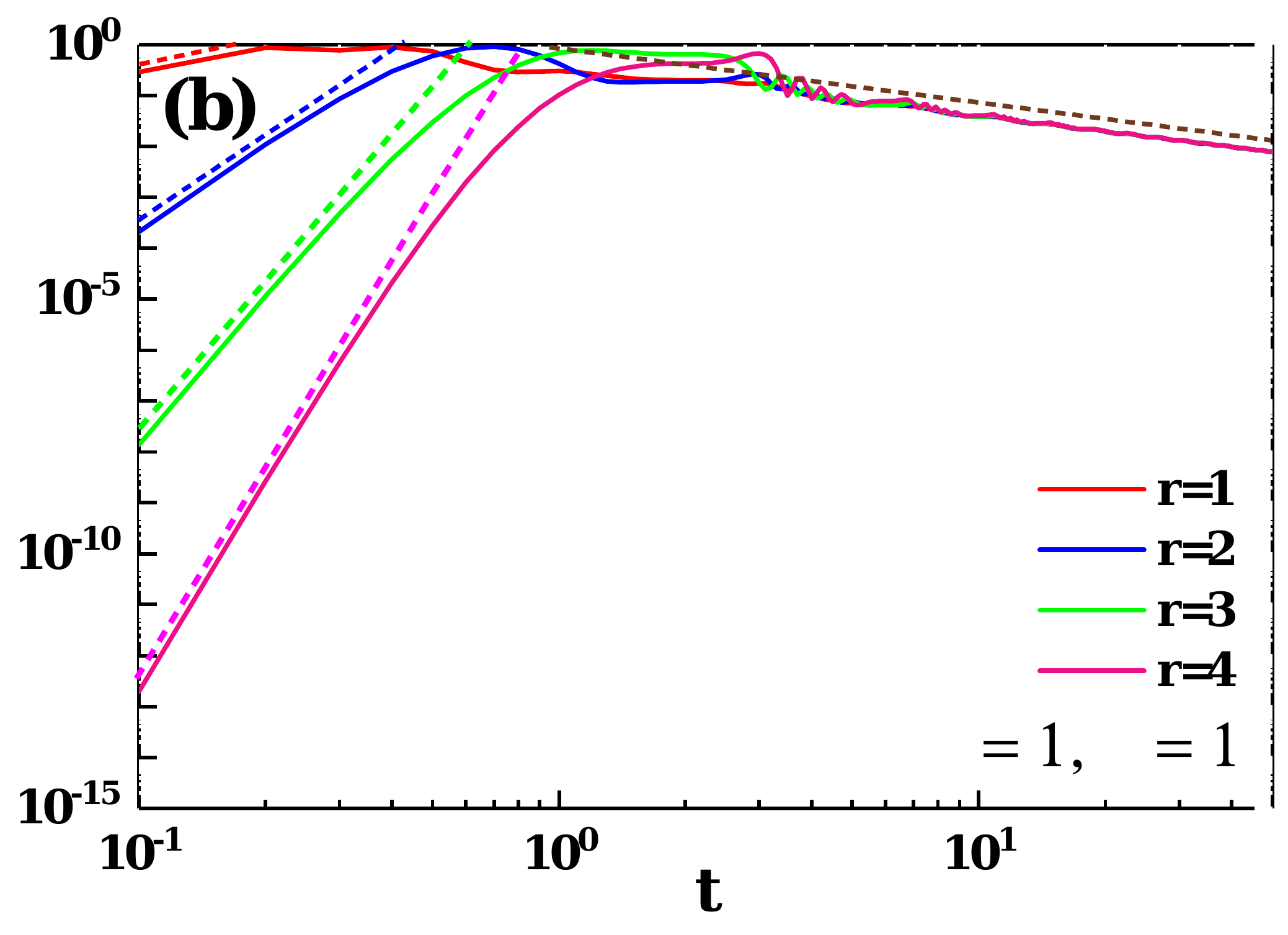}
\includegraphics[width=0.245\linewidth]{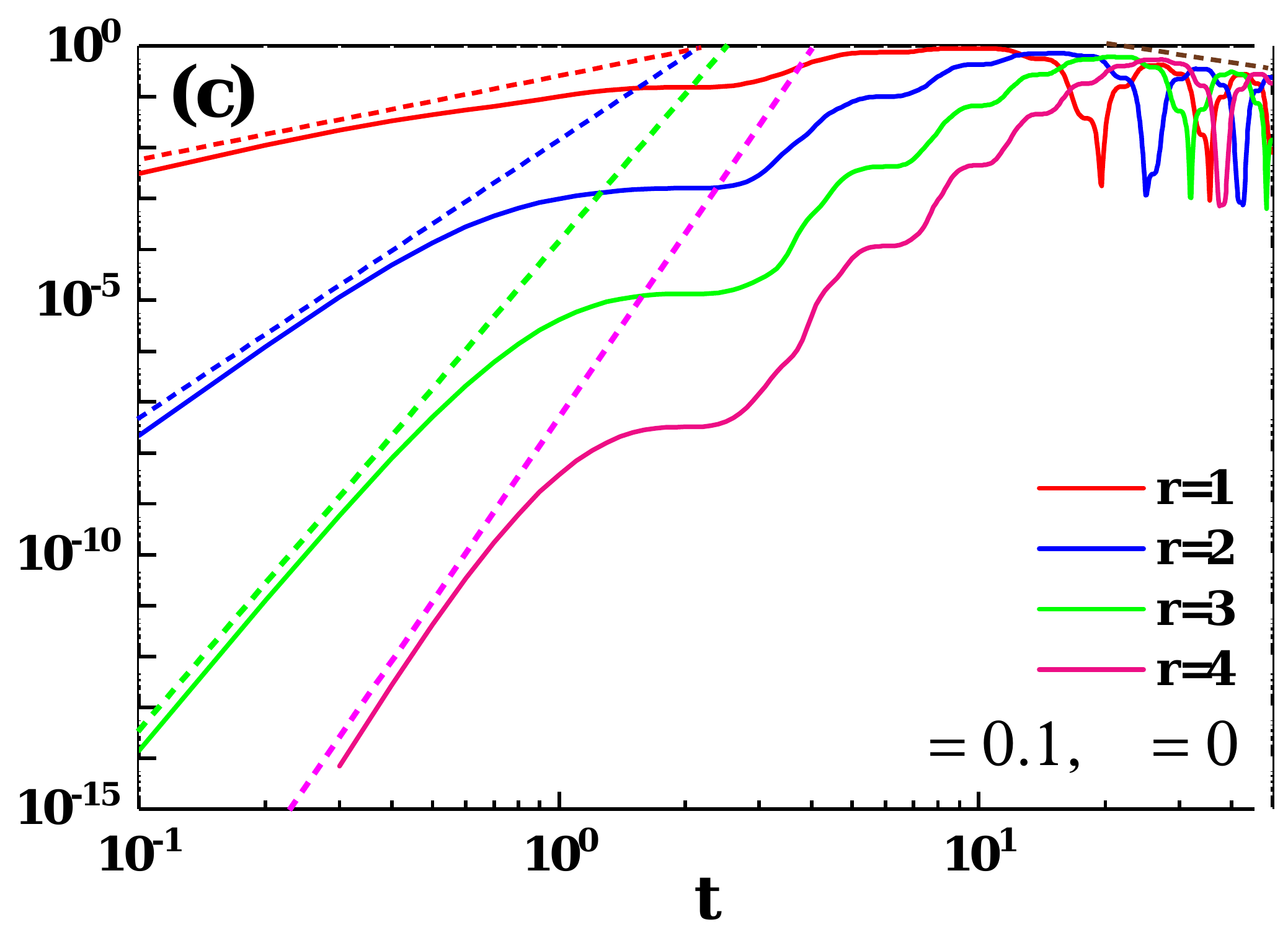}
\includegraphics[width=0.245\linewidth]{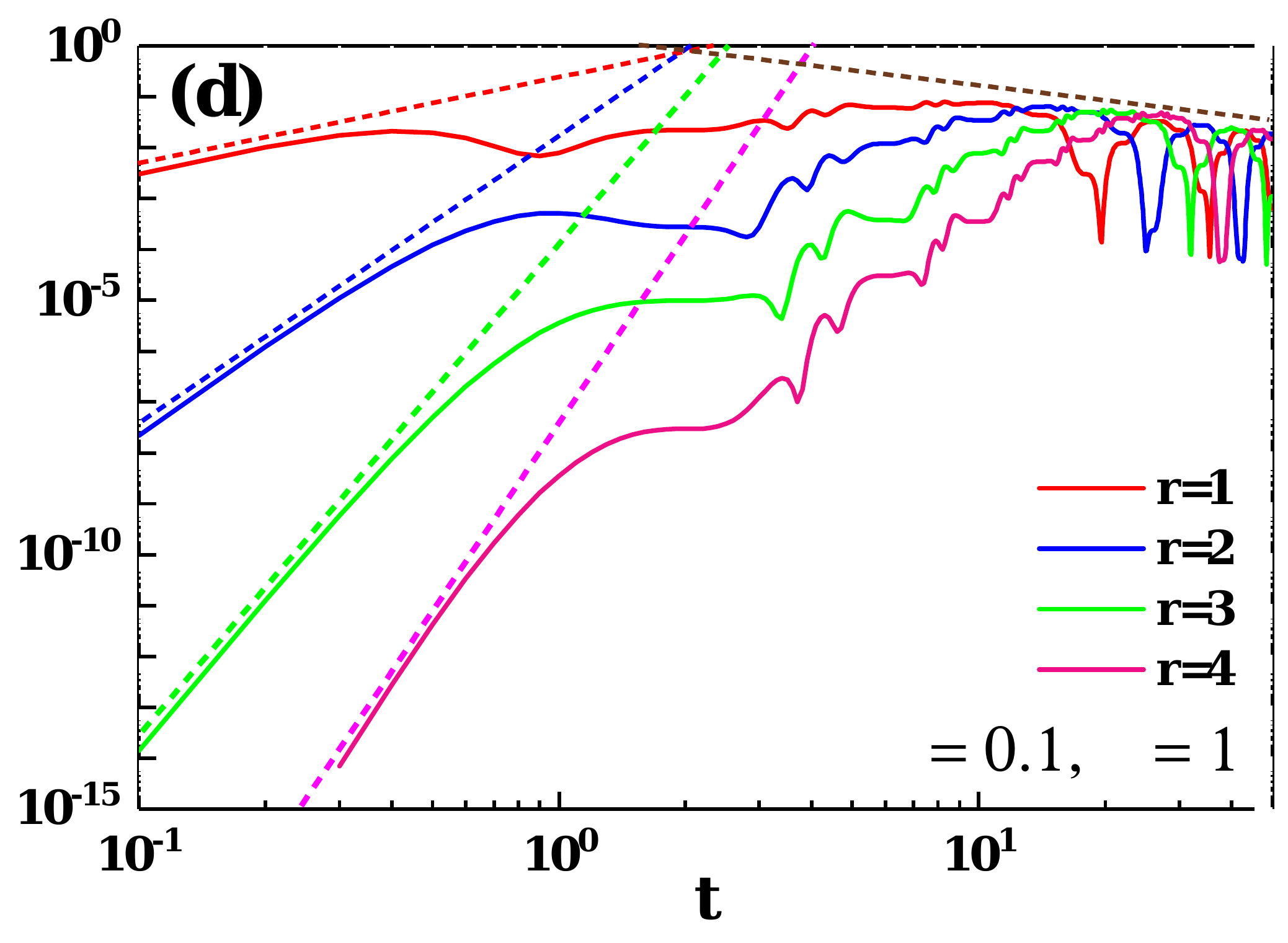}}
\centering
\end{minipage}
\begin{minipage}{\linewidth}
\centerline{\includegraphics[width=0.265\linewidth]{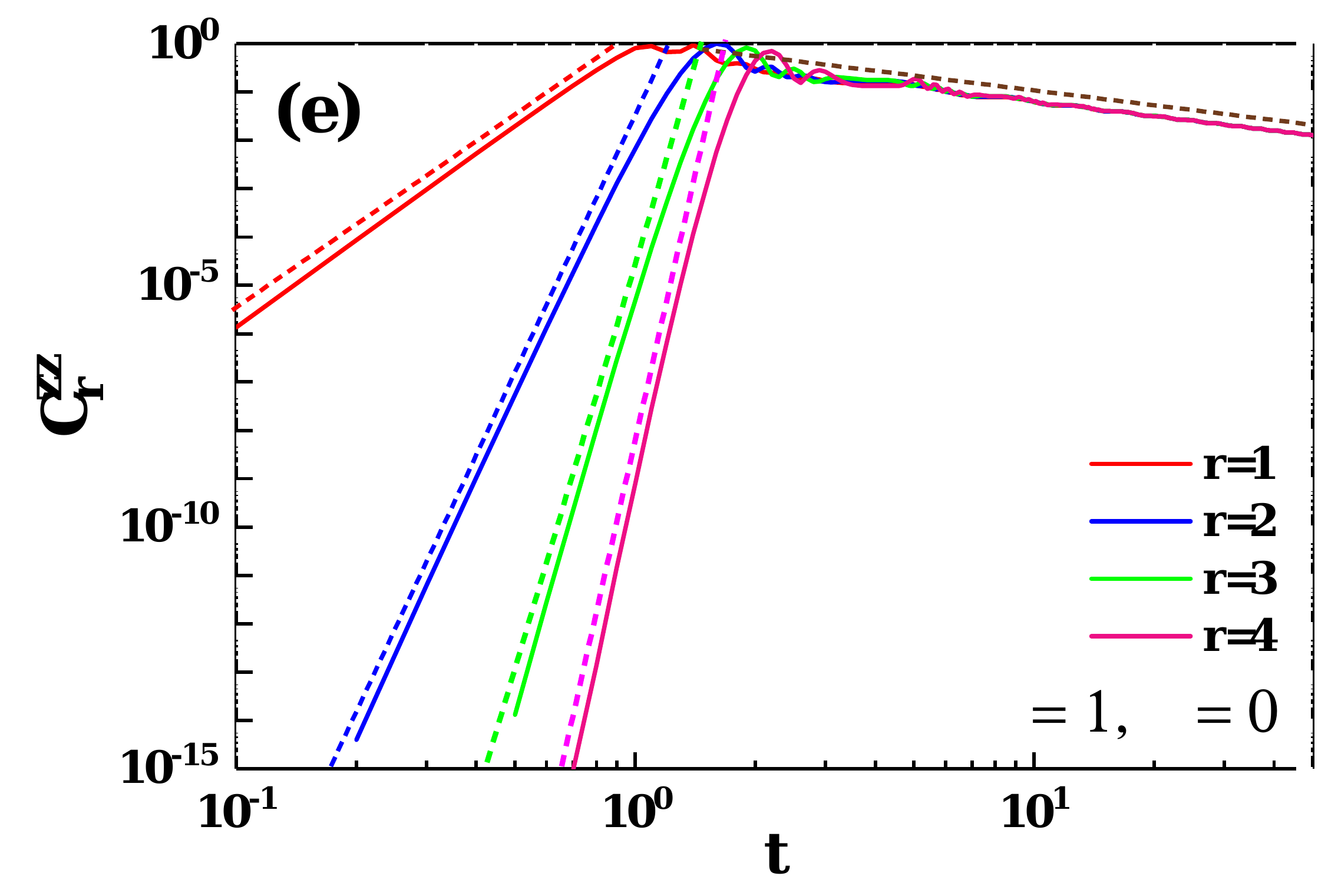}
\includegraphics[width=0.245\linewidth]{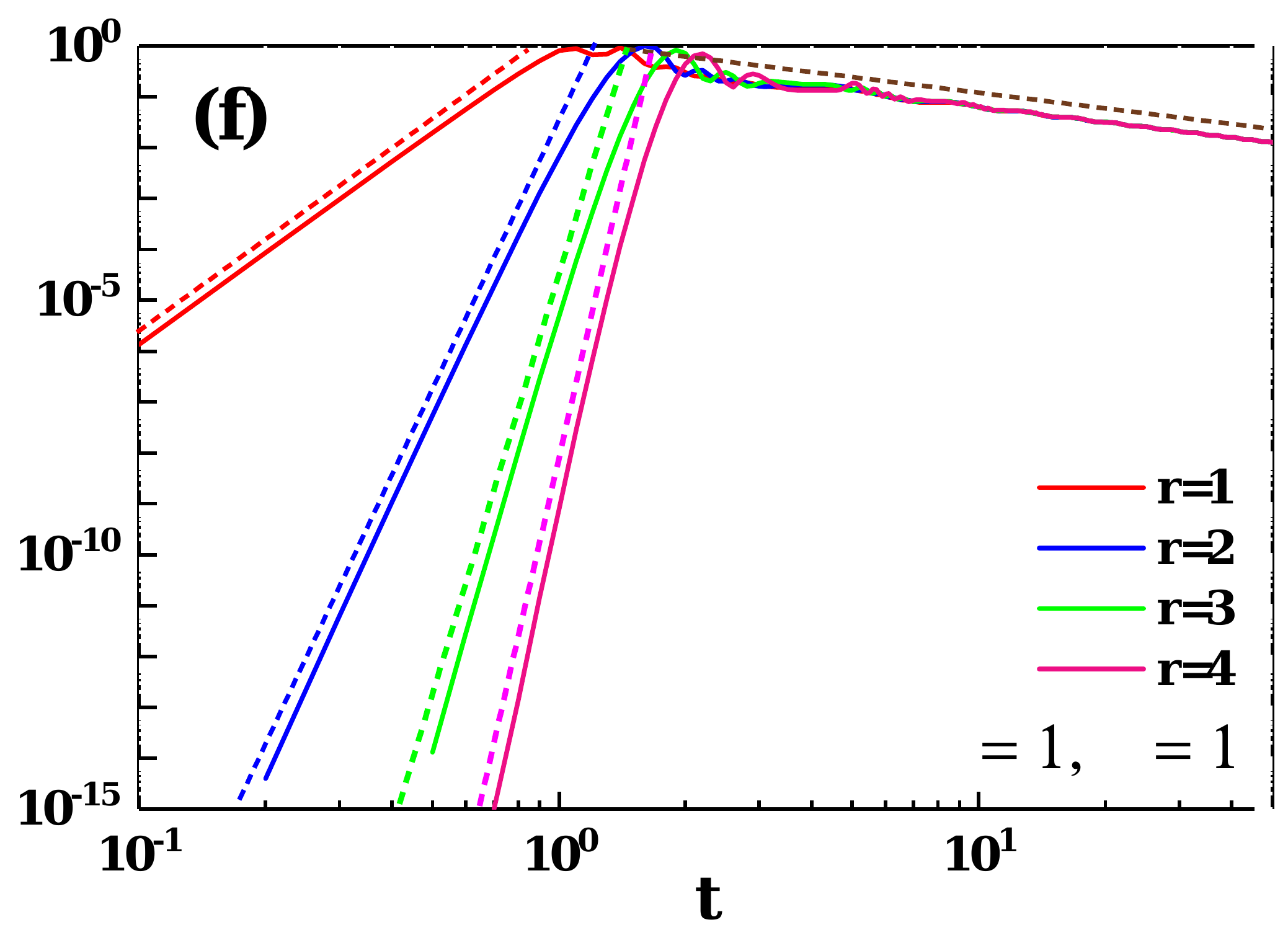}
\includegraphics[width=0.245\linewidth]{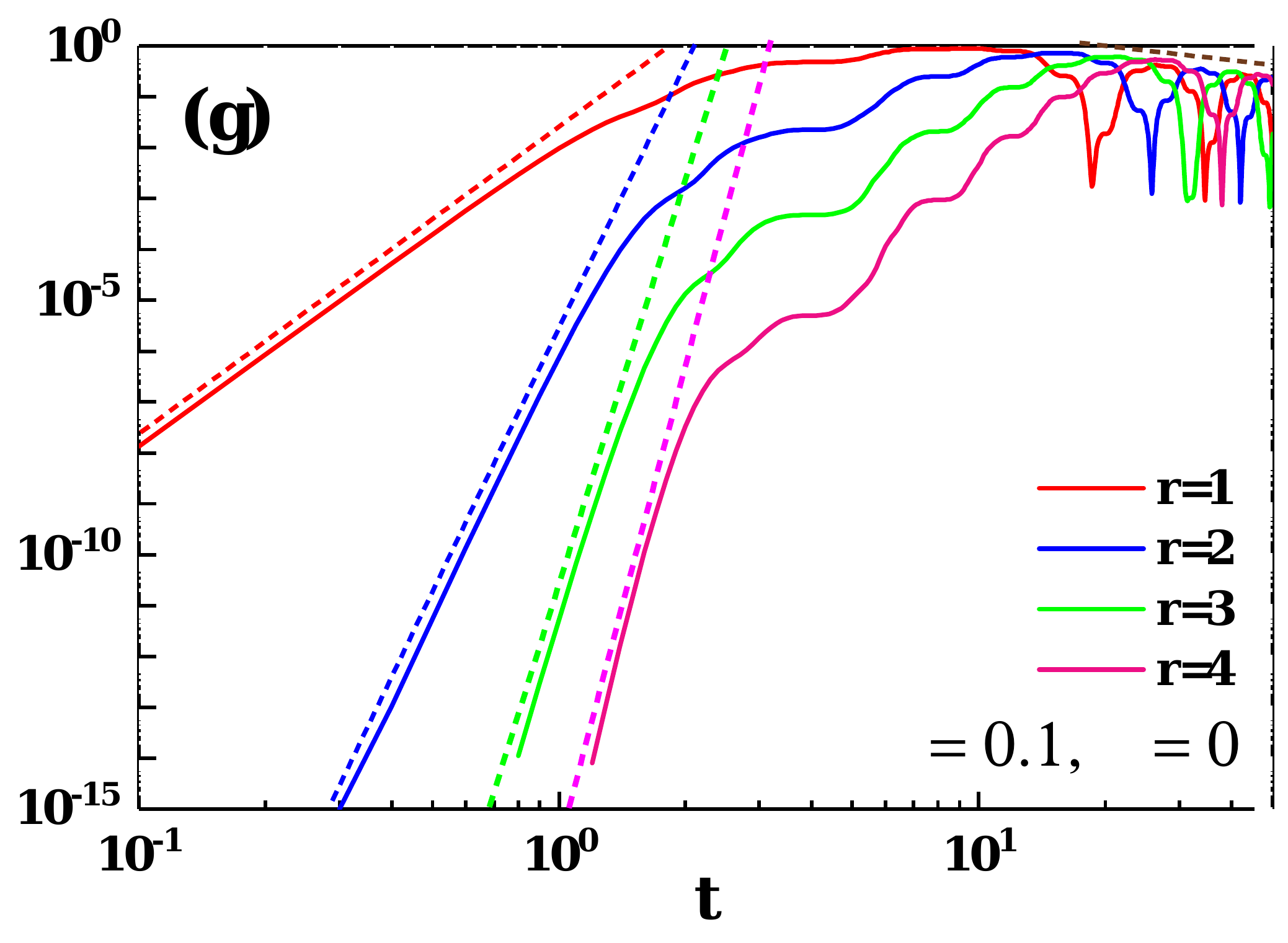}
\includegraphics[width=0.245\linewidth]{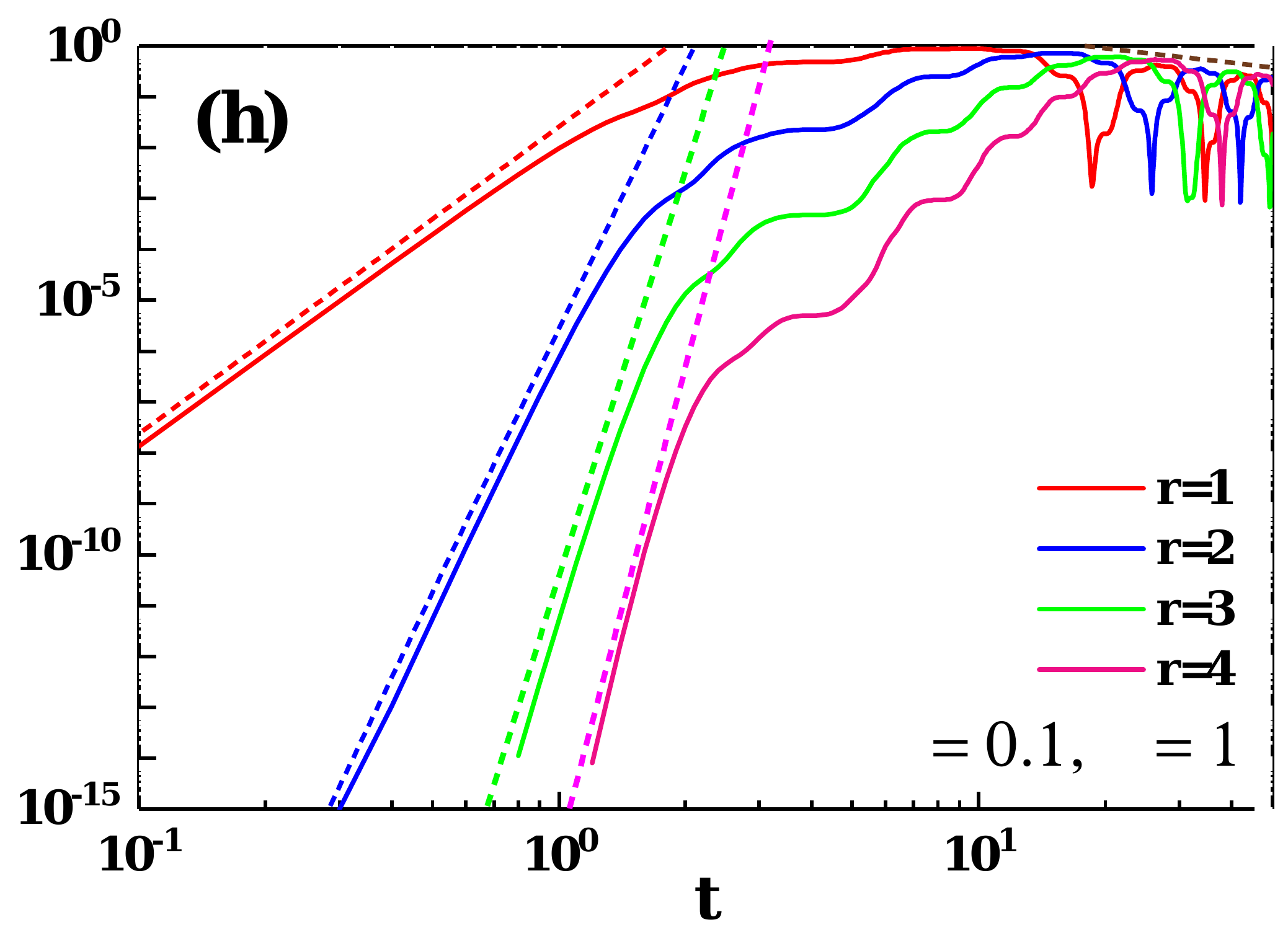}}
\centering
\end{minipage}
\caption{(Color online) Scaling behavior of $C^{zz}_{r}$
with the universal form for several fixed separations of the synchronized
Ising model, in the presence of $h(t)=h_{0}+h_{1}\cos(\omega t)$.
The upper plots indicate the results of $h(0)\neq0$
and the lower plots are for $h(0)=0$. The dashed lines are used for power
law fitting. We see approximately $t^{4r-3}$ ($t^{10r-3}$) power-law
fashion at short times and $t^{-1}$ decay at long times for the case
of $h(0)\neq0$ ($h(0)=0$). We set system size $N = 200$, frequency $\omega=\pi/2$,
and Hamiltonian parameters $h_{0}=1, h_{1}=\pm1$.
The strong/weak coupling and infinite/finite temperature cases are
(a) $\lambda=1, \beta=0$, (b) $\lambda=1, \beta=1$, (c) $\lambda=0.1, \beta=0$, (d) $\lambda=0.1, \beta=1$,
(e) $\lambda=1, \beta=0$, (f) $\lambda=1, \beta=1$, (g) $\lambda=0.1, \beta=0$, (h) $\lambda=0.1, \beta=1$.
}
\label{Fig10}
\end{figure*}
%

The numerical results have also shown that, $\lim_{t\rightarrow\infty}Re\{F^{zz}_{r}(t)\}=1$, indicating no scrambling in OTOCs of local operator,
analogous to that of Floquet XY model. Although, the qualitative behavior of $F^{zz}_{r}(t)$ at finite temperature (no-FDQPTs case) approximately is similar to that at infinite temperature, the imaginary part of $F^{zz}_{r}$ becomes non-zero at finite temperature independent of
the synchronized coupling value (Figs. \ref{Fig6}(c)-\ref{Fig7}(c)).

Furthermore, analysing the OTOCs of nonlocal operators have shown that the system is scrambled at infinite and
finite temperature, which is expected from nonlocal nature of inherited information (see Appendix \ref{E}).
At infinite temperature (FDQPTs case), the imaginary part of OTOCs of nonlocal operators are also zero, while in no-FDQPTs 
case (finite temperature) the imaginary part of OTOCs of nonlocal operators become nonzero.

As the second case, we consider $h(0)=0$, where FDQPTs occur at any temperature for any values of driven frequency.
The density plot of real and imaginary parts of $F^{zz}_{r}$, are shown in Figs. \ref{Fig8}-\ref{Fig9} for strong and weak synchronized coupling $\lambda=1$ and $\lambda=0.1$, respectively. The numerical analysis exhibits that, the behavior of real part of OTOCs of both local and nonlocal operators at infinite and finite temperature are the same. However, the imaginary part of OTOCs of both local and nonlocal operators vanish at any temperatures.
Consequently, we come to conclusion that the OTOCs with both local and non local operators can be considered as
a diagnostic tool to dynamically detect the FDQPTs in the synchronized Floquet XY model. In other words, the imaginary part of OTOCs becomes zero when the system undergoes the FDQPT.

Finally, to exactly assess how a local operator behaves dynamically and verify its universal form, the
evolution of $C^{zz}_{r}$ for some fixed separations at finite and infinite temperature, has been depicted in Fig. \ref{Fig10}.
Since the interactions of Hamiltonian are local, we expect the power-law growth of $C^{zz}_{r}$, similar to previous studies~\cite{bao2020out,lin2018out}.
As is clear, the short-time behavior of $C^{zz}_{r}$, in the case of $h(0)\neq0$ ($h(0)= 0$) at any temperature, reveals the power-law
trend $t^{n}$ with position-dependent power $n\approx 4r-3$ ($n\approx 10r-3$), which has been extracted from the numerical results.
Moreover, $C^{zz}_{r}$ approaches its limiting value at long times, in a slow power law  $t^{-1}$,
independent of the value of separations and temperature.

\section{Conclusion}
In this paper, we have studied the dynamical quantum phase transition of two periodically time driven Hamiltonian, the Floquet XY model and synchronized Floquet XY model, via analyzing the behavior of out-of-time-order correlation. Our results indicate that out-of-time-order correlation is a proper diagnostic tool for studying the dynamical characteristics of quantum systems and can represent features of dynamical behavior. We discovered that out-of-time-order correlation of local operators, could precisely detect the dynamical quantum phase transition. In the Floquet $XY$ chain, the infinite-temperature time averaged out-of-time-order correlation of local operators can serve as a dynamical order parameter that dynamically detect the range of driven frequency over which FDQPTs occur. The aforementioned time averaged gets a jump with a peak at the boundary of FDQPT.
Moreover, the speed of wave front of information spreading in the system becomes minimum in the region,
which shows FDQPT. In the synchronized Floquet $XY$ chain, it was indicated that vanishing of the imaginary part of OTOC signals the occurrence of dynamical quantum phase transition. In addition, the temperature dependence of the generalized Loshmidt echo comes from its imaginary part,
which suggests that there is a connection between the real and imaginary parts of the generalized
Loschmidt echo and that of out-of-time-order correlation. Further investigations would be interesting to establish a precise relation between
the real and imaginary parts of the generalized Loschmidt echo and that of out-of-time-order correlation.


\appendix

\section{OTOC of nonlocal operators\label{A}}

Using Eqs. (\ref{eq3}) and (\ref{eq7}), the expressions of $\Gamma^{xx}_{r}(t)$, $\Gamma^{xy}_{r}(t)$ and $\Gamma^{xz}_{r}(t)$ are written in the following forms:
%
\begin{eqnarray}
\Gamma^{xx}_{r}(t)&&=\langle (\sigma_{\frac{N}{2}}^{x}(t) \sigma_{N-r}^{x}(t) \sigma_{0}^{x} \sigma_{\frac{N}{2}-r}^{x})^2 \rangle
\nonumber\\
&&=\langle\Big[(\prod_{l=N/2}^{N-r-1}B_{l}(t)A_{l+1}(t))(\prod_{l=0}^{N/2-r-1}B_{l}A_{l+1})\Big]^{2}\rangle
\nonumber\\
\Gamma^{xz}_{r}(t)&&=\langle (\sigma_{\frac{N}{2}}^{x}(t) \sigma_{N-r}^{x}(t) \sigma_{0}^{z} \sigma_{\frac{N}{2}-r}^{z})^2 \rangle
\nonumber\\
&&=\langle\Big[(\prod_{l=N/2}^{N-r-1}B_{l}(t)A_{l+1}(t))(A_{0}A_{N/2-r}B_{0}B_{N/2-r})\Big]^{2}\rangle
\nonumber\\
\Gamma^{xy}_{r}(t)&&=\langle (\sigma_{\frac{N}{2}}^{x}(t) \sigma_{N-r}^{x}(t) \sigma_{0}^{y} \sigma_{\frac{N}{2}-r}^{y})^2 \rangle
\nonumber\\
&&=\langle\Big[(\prod_{l=N/2}^{N-r-1}B_{l}(t)A_{l+1}(t))(\prod_{l=0}^{N/2-r-1}A_{l}B_{l+1})\Big]^{2}\rangle.
\label{eq:A1}
\end{eqnarray}
%

\section{Exact solution of the Floquet XY chain \label{B}}

Considering the identity $\sum_{k{\cal{2}} BZ}\cos(k)=0$, one can rewrite Eq. (\ref{eq11}) as follows:
%
\begin{eqnarray}
	{\cal H}_{k}(t)&&= h_{z}(k)(c_{k}^{\dagger}c_{k}+c_{-k}^{\dagger}c_{-k})
	\nonumber\\
	&& - ih_{xy}(k)(e^{-i\omega t}c_{k}^{\dagger}c_{-k}^{\dagger}+e^{i\omega t}c_{k}c_{-k})-h .
	\label{eq:B1}
\end{eqnarray}
%

It is convenient to use the following basis for the $k$-th subspace, which are defined in Heisenberg picture
%
\begin{equation}
	| 0 \rangle, ~~c_{k}^{\dagger}| 0 \rangle, ~~c_{-k}^{\dagger}| 0 \rangle, ~~c_{k}^{\dagger}c_{-k}^{\dagger}| 0 \rangle
	\label{eq:B2}.
\end{equation}
%
In this representation, the Hamiltonian ${\cal H}_{k}(t)$ can be expressed as
%
{\small
\begin{eqnarray}
	{\cal H}_{k}(t)=
	\left(
	\begin{array}{cccc}
		-h & ih_{xy}(k)e^{i\omega t} & 0 & 0 \\
		-ih_{xy}(k)e^{-i\omega t}  & 2h_{z}(k)-h & 0 & 0 \\
		0 & 0 & h_{z}(k)-h & 0 \\
		0 & 0 & 0 & h_{z}(k)-h \\
	\end{array}
	\right). \nonumber \\
	\label{eq:B3}
\end{eqnarray}
}
%

By solving the time-dependent Schr\"{o}dinger equation, we obtain the eigenvalues and eigenvectors of Hamiltonian ${\cal H}_{k}(t)$
%
\begin{equation}
	i\frac{d}{dt}|\psi_{k}^{\pm}(t)\rangle={\cal H}_{k}(t)|\psi_{k}^{\pm}(t)\rangle .
	\label{eq:B4}
\end{equation}
%

The exact solution of the Schr\"{o}dinger equation is found by going to the rotating frame given by the periodic unitary transformation
%
\begin{eqnarray}
	U_{R}(t)=
	\left(
	\begin{array}{cccc}
		1 & 0 & 0 & 0 \\
		0 & e^{-i\omega t} & 0 & 0 \\
		0 & 0 & 1 & 0 \\
		0 & 0 & 0 & 1 \\
	\end{array}
	\right).
	\label{eq:B5}
\end{eqnarray}
%

In the rotating frame the eigenstate is given by $|\chi_{k}^{\pm}(t)\rangle=U_{R}^{\dagger}(t)|\psi^{\pm}(k,t)\rangle$. Substituting the transformed eigenstate into the Schr\"{o}dinger equation, the time dependent Hamiltonian is transformed to its time-independent form ${\cal H}_{k}|\chi_{k}^{\pm}(t)\rangle=E^{\pm}|\chi_{k}^{\pm}(t)\rangle$ where
%
{\small
\begin{eqnarray}
	{\cal H}_{k}=
	\left(
	\begin{array}{cccc}
		-h & ih_{xy}(k) & 0 & 0 \\
		-ih_{xy}(k) & 2h_{z}(k)-h-\omega & 0 & 0 \\
		0 & 0 & h_{z}(k)-h & 0 \\
		0 & 0 & 0 & h_{z}(k)-h \\
	\end{array}
	\right).
	\label{eq:B6}
\end{eqnarray}
}
%

The Hamiltonian ${\cal H}_{k}$ is in block-diagonal form,  which leads to the following eigenvalues and eigenvectors:
%
\begin{eqnarray}
	&& E^{1,2}_{k}= (h_{z}(k)-h-\frac{\omega}{2})\pm \varepsilon_{k}
	\nonumber\\
	&& E^{3,4}_{k}= h_{z}(k)-h
	\label{eq:B8},
\end{eqnarray}
where $\varepsilon_{k}=\sqrt{(h_{xy}(k))^2+(h_{z}(k)-\frac{\omega}{2})^2}$ and
%
\begin{eqnarray}
	&&|\chi^{1}_{k}\rangle =\left(
	\begin{array}{c}
		\cos(\gamma_{k}/2) \\
		\sin(\gamma_{k}/2) \\
	\end{array}
	\right),
	\nonumber\\
	&&|\chi^{2}_{k}\rangle =\left(
	\begin{array}{c}
		\sin(\gamma_{k}/2) \\
		-\cos(\gamma_{k}/2) \\
	\end{array}
	\right),
	\label{eq:B9}
\end{eqnarray}
%
in which
%
\begin{eqnarray}
	\gamma_{k}=2 \arctan\Big[\frac{h_{z}(k)-\frac{\omega}{2}-\varepsilon_{k}}{h_{xy}(k)}\Big].
	\label{eq:B10}
\end{eqnarray}
%

%
\begin{figure*}
\begin{minipage}{\linewidth}
\centerline{\includegraphics[width=0.35\linewidth]{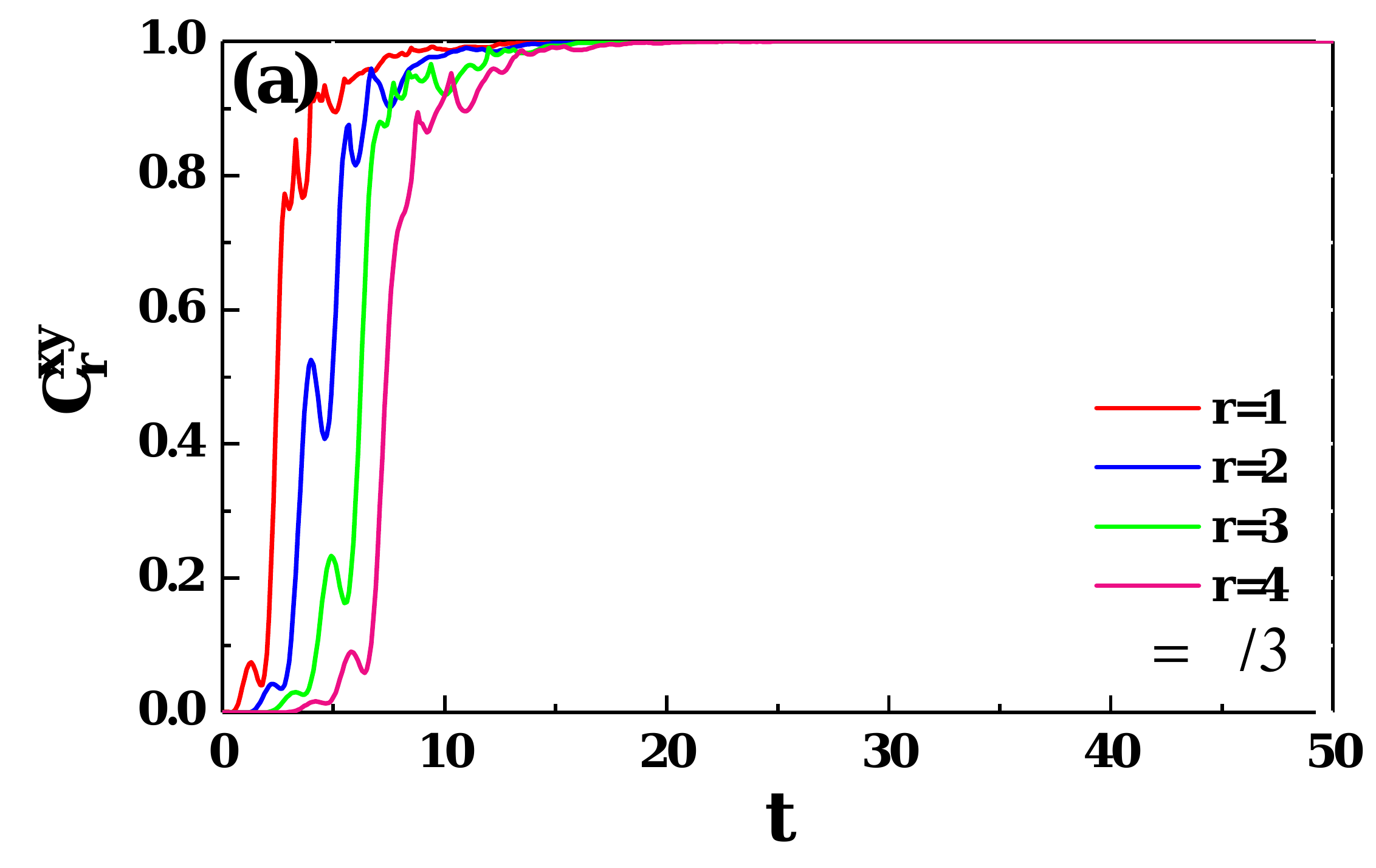}
\includegraphics[width=0.32\linewidth]{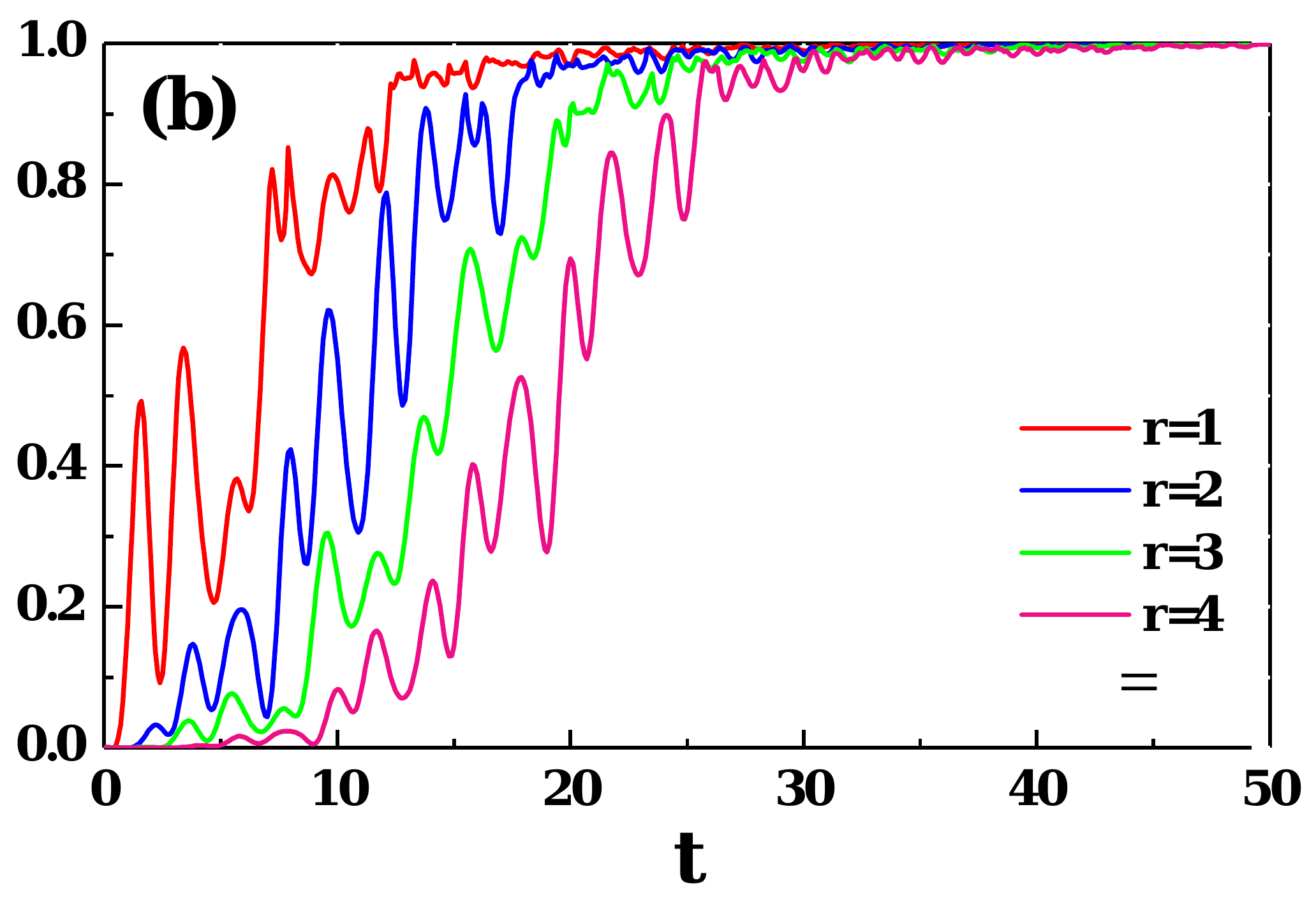}
\includegraphics[width=0.32\linewidth]{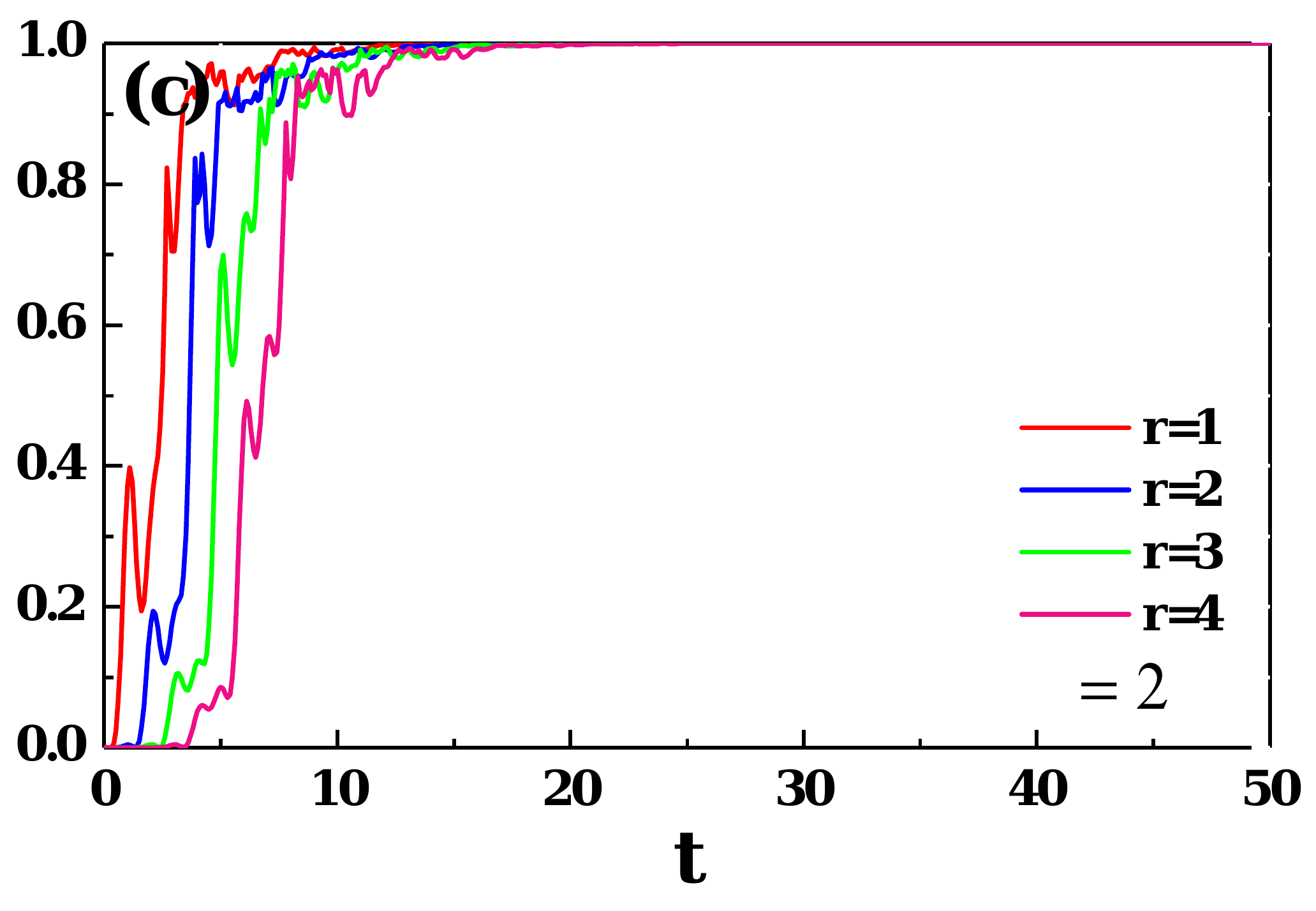}}
\centering
\end{minipage}
\begin{minipage}{\linewidth}
\centerline{\includegraphics[width=0.35\linewidth]{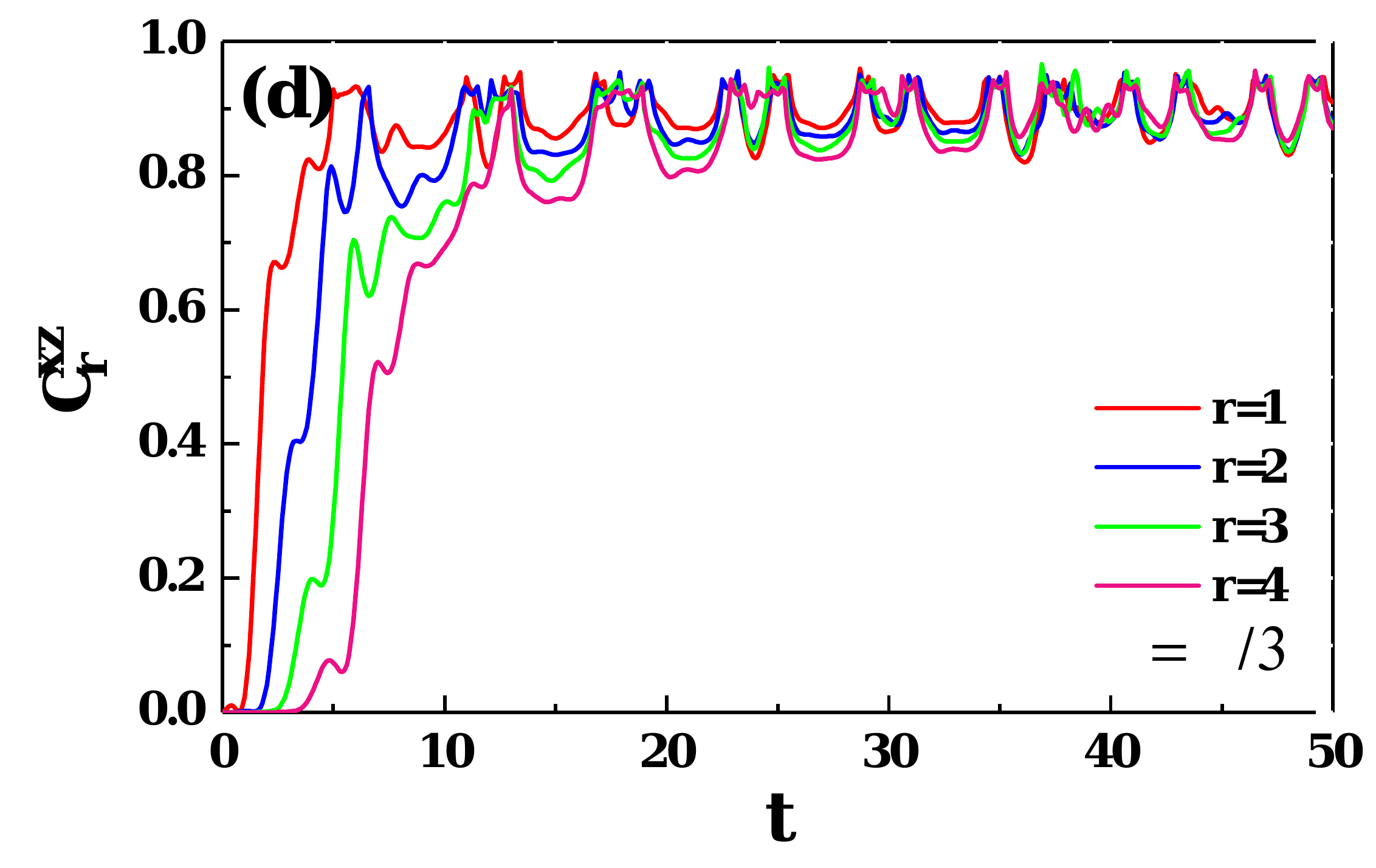}
\includegraphics[width=0.32\linewidth]{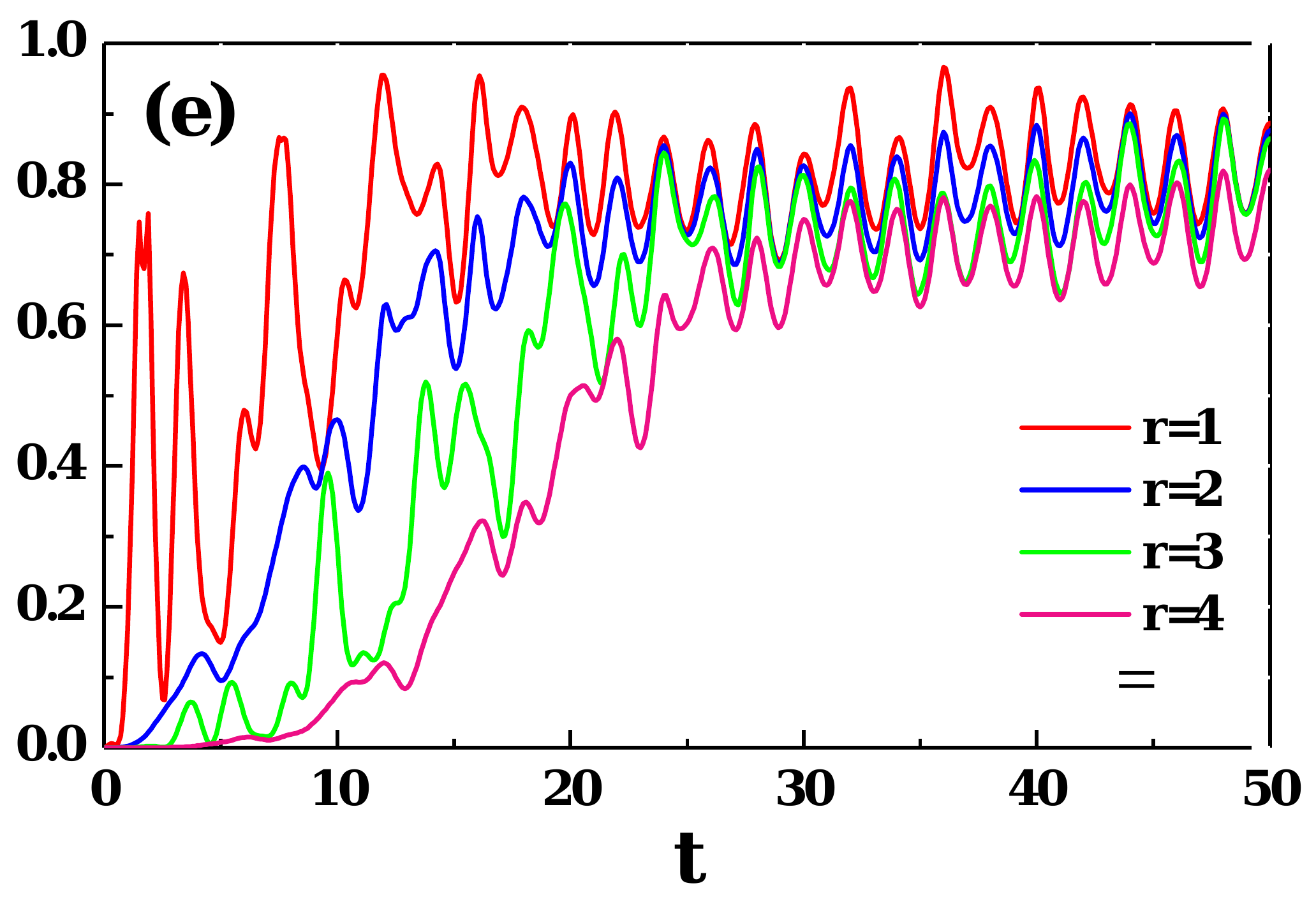}
\includegraphics[width=0.32\linewidth]{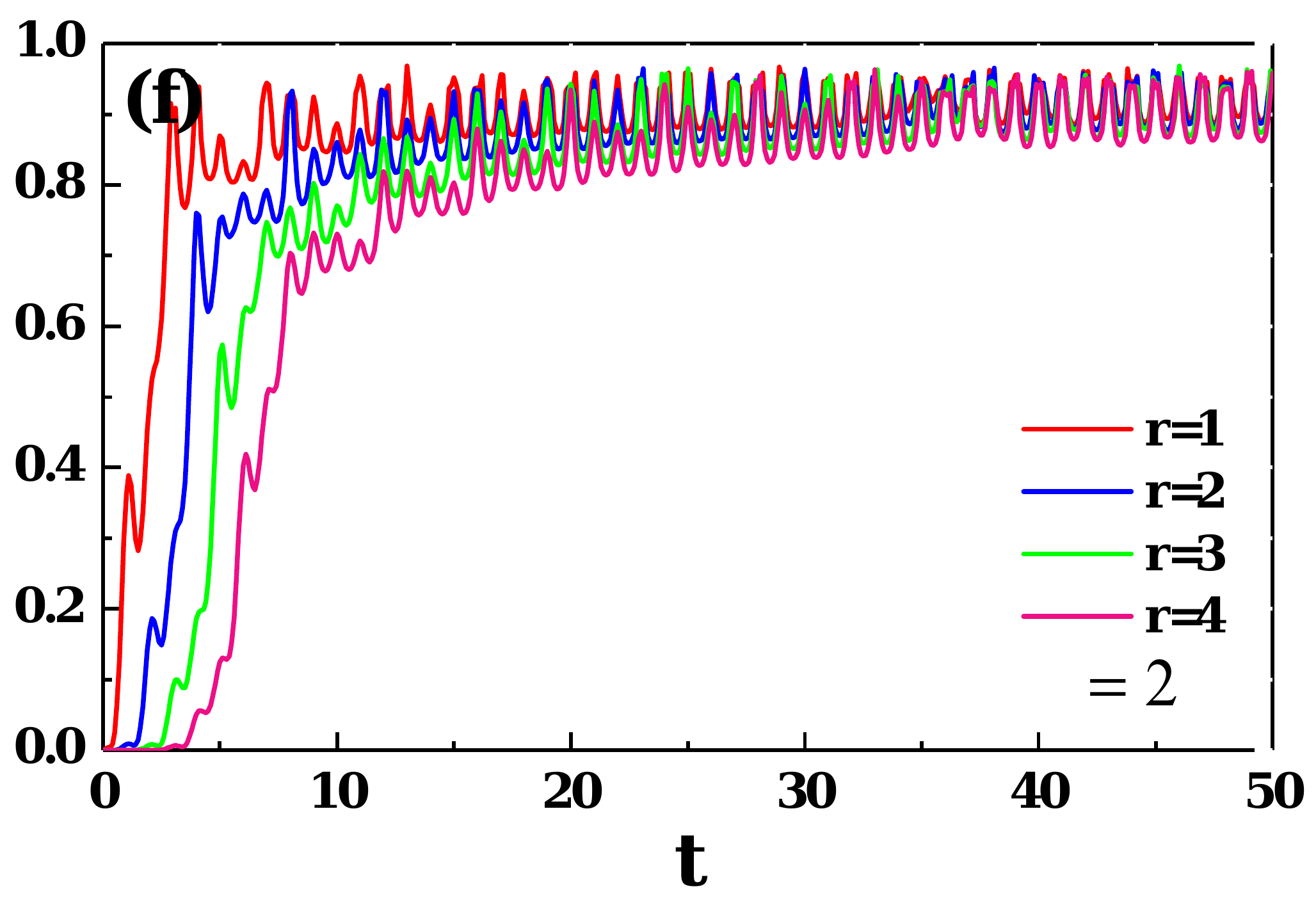}}
\centering
\end{minipage}
\caption{(Color online) Time evolution of $C^{xy}_{r}$
and $C^{xz}_{r}$ for several fixed separations in the Floquet $XY$ model,
with $\beta=0$ and different values of $\omega$. We set system size $N = 100$, and the
Hamiltonian parameters are $J=0.25\pi, h=0.5\pi, \gamma=0.5$.}
\label{Fig11}
\end{figure*}

\section{Calculation the Floquet OTOC \label{C}}

To obtain the time evolution operator of Floquet Hamiltonian, $U_{k}(t)=U_{R}(t)U_{F}(t)$, we need to calculate
$U_{F}(t)=e^{-i{\cal H}_{k}t}$. The upper-left block of ${\cal H}_{k}$ is given by ${\cal H}^{\prime}_{k}$
\begin{eqnarray}
{\cal H}^{\prime}_{k}&&=
	\left(
\begin{array}{cc}
	-h & ih_{xy}(k) \\
	-ih_{xy}(k) & 2h_{z}(k)-h-\omega \\
	\end{array}
	\right)
	\nonumber\\
	&& = \big(h_{z}(k)-\frac{\omega}{2}-h\big) \mathbb{1} + \varepsilon_{k}{\hat h_{l}}(k)\cdot\vec{\sigma}.
	\label{eq:B7}
\end{eqnarray}
At first, we calculate $e^{-i{\cal H}_{k}^{\prime}t}$,
\begin{widetext}
%
\begin{eqnarray}
	e^{-i{\cal H}_{k}^{\prime}t}&&= e^{-it(-h+h_{z}(k)-\frac{\omega}{2})} e^{-it\varepsilon_{k}{\hat h_{l}}(k)\cdot\vec{\sigma}}
	\nonumber\\
	&& = e^{-it(-h+h_{z}(k)-\frac{\omega}{2})}\Big[\cos(\varepsilon_{k} t)-i\sin(\varepsilon_{k} t)({\hat h_{l}}(k)\cdot\vec{\sigma})\Big]
	\nonumber\\
	&& = e^{-it(-h+h_{z}(k)-\frac{\omega}{2})}
	\left(
	\begin{array}{cc}
		\cos(\varepsilon_{k} t)+i\frac{h_{z}(k)-\omega/2}{\varepsilon_{k}}\sin(\varepsilon_{k} t) & \frac{h_{xy}(k)}{\varepsilon_{k}}\sin(\varepsilon_{k} t) \\
		-\frac{h_{xy}(k)}{\varepsilon_{k}}\sin(\varepsilon_{k} t) & \cos(\varepsilon_{k} t)-i\frac{h_{z}(k)-\omega/2}{\varepsilon_{k}}\sin(\varepsilon_{k} t) \\
	\end{array}
	\right) .
	\label{eq:C1}
\end{eqnarray}
%
Using the above equation, we arrive at
%
\begin{eqnarray}
	U_{F}(t)= e^{-it(h_{z}(k)-h)}
	\left(
	\begin{array}{cccc}
		\Big(\cos(\varepsilon_{k} t)+i\frac{h_{z}(k)-\omega/2}{\varepsilon_{k}}\sin(\varepsilon_{k} t)\Big)e^{\frac{i\omega t}{2}} & \frac{h_{xy}(k)}{\varepsilon_{k}}\sin(\varepsilon_{k} t)e^{\frac{i\omega t}{2}} & 0 & 0 \\
		-\frac{h_{xy}(k)}{\varepsilon_{k}}\sin(\varepsilon_{k} t)e^{\frac{i\omega t}{2}} & \Big(\cos(\varepsilon_{k} t)-i\frac{h_{z}(k)-\omega/2}{\varepsilon_{k}}\sin(\varepsilon_{k} t)\Big)e^{\frac{i\omega t}{2}} & 0 & 0 \\
		0 & 0 & 1 & 0 \\
		0 & 0 & 0 & 1 \\
	\end{array}
	\right)
	\label{eq:C2},
\end{eqnarray}
and the time evolution operator is given by
\begin{eqnarray}
	U_{k}(t)= e^{-it(h_{z}(k)-h)}
	\left(
	\begin{array}{cccc}
		\Big(\cos(\varepsilon_{k} t)+i\frac{h_{2z}(k)-\omega/2}{\varepsilon_{k}}\sin(\varepsilon_{k} t)\Big)e^{\frac{i\omega t}{2}} & \frac{h_{xy}(k)}{\varepsilon_{k}}\sin(\varepsilon_{k} t)e^{\frac{i\omega t}{2}} & 0 & 0 \\
		-\frac{h_{xy}(k)}{\varepsilon_{k}}\sin(\varepsilon_{k} t)e^{-\frac{i\omega t}{2}} & \Big(\cos(\varepsilon_{k} t)-i\frac{h_{z}(k)-\omega/2}{\varepsilon_{k}}\sin(\varepsilon_{k} t)\Big)e^{-\frac{i\omega t}{2}} & 0 & 0 \\
		0 & 0 & 1 & 0 \\
		0 & 0 & 0 & 1 \\
	\end{array}
	\right) .
	\label{eq:C3}
\end{eqnarray}
%

Similarly, the initial mixed state density matrix of Floquet system in thermal equilibrium with a heat bath, corresponding to ${\cal H}_{k}$ is
%
\begin{eqnarray}
	\rho_{k}(0)= &&\frac{e^{-\beta{\cal H}_{k}}}{{\rm Tr}(e^{-\beta{\cal H}_{k}})}= \frac{1}{2(\cosh(\beta\varepsilon_{k})e^{\beta\omega/2}+1)}\times
	\nonumber\\
	&&\left(
	\begin{array}{cccc}
		\Big(\cosh(\beta\varepsilon_{k})+\frac{h_{z}(k)-\omega/2}{\varepsilon_{k}}\sinh(\beta\varepsilon_{k})\Big)e^{\frac{\beta\omega}{2}} & -i\frac{h_{xy}(k)}{\varepsilon_{k}}\sinh(\beta\varepsilon_{k})e^{\frac{\beta\omega}{2}} & 0 & 0 \\
		i\frac{h_{xy}(k)}{\varepsilon_{k}}\sinh(\beta\varepsilon_{k})e^{\frac{\beta\omega}{2}} & \Big(\cosh(\beta\varepsilon_{k})-\frac{h_{z}-\omega/2}{\varepsilon_{k}}\sinh(\beta\varepsilon_{k})\Big)e^{\frac{\beta\omega}{2}} & 0 & 0 \\
		0 & 0 & 1 & 0 \\
		0 & 0 & 0 & 1 \\
	\end{array}
	\right) .
	\label{eq:C4}
\end{eqnarray}
%
\end{widetext}

Since the Hamiltonian is decomposable, one can find the density matrix at time $t$ for $k$-th subspace, by solving the
Liouville equation. Using Eqs. (\ref{eq:C3})-(\ref{eq:C4}) and following the relation $\rho_{k}(t)=U_{k}(t) \rho_{k}(0) U_{k}^{\dagger}(t)$, one can conclude the density matrix at time $t$. To compute the OTOC we must first calculate $c_{k}^{\dagger}\pm c_{-k}$, $c_{-k}^{\dagger}\pm c_{k}$ as follows
%
{\small
\begin{eqnarray}
	&& c_{k}^{\dagger}+c_{-k}=
	\left(
	\begin{array}{cccc}
		0 & 0 & 0 & 1 \\
		0 & 0 & 0 & 1 \\
		1 & -1 & 0 & 0 \\
		0 & 0 & 0 & 0 \\
	\end{array}
	\right),
	c_{k}^{\dagger}-c_{-k}=
	\left(
	\begin{array}{cccc}
		0 & 0 & 0 & -1 \\
		0 & 0 & 0 & -1 \\
		1 & 1 & 0 & 0 \\
		0 & 0 & 0 & 0 \\
	\end{array}
	\right),
	\nonumber\\
	&&c_{-k}^{\dagger}+c_{k}=
	\left(
	\begin{array}{cccc}
		0 & 0 & 1 & 0 \\
		0 & 0 & -1 & 0 \\
		0 & 0 & 0 & 0 \\
		1 & 1 & 0 & 0 \\
	\end{array}
	\right),
	c_{-k}^{\dagger}-c_{k}=
	\left(
	\begin{array}{cccc}
		0 & 0 & -1 & 0 \\
		0 & 0 & -1 & 0 \\
		0 & 0 & 0 & 0 \\
		1 & -1 & 0 & 0 \\
	\end{array}
	\right). \nonumber
	\label{eq:C5}
\end{eqnarray}
}
%
Then, by considering Eq. (\ref{eq4}), and the above equations, time dependent Majorana correlation functions are obtained. Finally, following the procedure of Pffafian method (sections II.B and II.C), one would compute the local and nonlocal OTOCs.

\section{Exact solution of the synchronized Floquet XY model \label{D}}
Applying Jordan-Wigner as well as Fourier transformations on Eq. (\ref{eq12}), and within the anti-periodic boundary condition used to minimize boundary effects, the Hamiltonian in terms of fermionic creation and annihilation operators is identical to
%
\begin{eqnarray}
H(t)&&=\sum_{k>0} \Big[2(-J(t)\cos(k)-h(t)) (c_{k}^{\dagger}c_{k}-c_{-k}c_{-k}^{\dagger})
\nonumber\\
&& -2 (iJ(t)\gamma\sin(k)) (c_{k}^{\dagger}c_{-k}^{\dagger}+c_{k}c_{-k})\Big] .
\label{eq:D1}
\end{eqnarray}
%
The resulting Hamiltonian can be written as $H(t) =\sum_{k>0} H_{k}(t)$, where the local Hamiltonian reads
%
\begin{equation}
H_{k}(t)=h_{z}(k,t) (c_{k}^{\dagger}c_{k}+c_{-k}^{\dagger}c_{-k}) - ih_{xy}(k,t) (c_{k}^{\dagger}c_{-k}^{\dagger}+c_{k}c_{-k})
\label{eq:D2},
\end{equation}
where, $h_{z}(k,t)=-2J(t)\cos(k)-2h(t)$, $h_{xy}(k,t)=2J(t)\gamma\sin(k)$ and the wave number $k$ is equal to
$k=(2p-1)\pi/N$ and the integer $p$ runs from $-N/2+1$ to $N/2$, where $N$ is the total number of spins (sites) in the chain. Hence, $H_{k}(t)$ can be diagonalized using the procedure of Bogoliubov transformation, which is given by:
%
\begin{eqnarray}
&& c_{k}=u_{k} \gamma_{k}+ i v_{k} \gamma_{-k}^{\dagger},
\nonumber\\
&& c_{-k}=u_{k} \gamma_{-k}- i v_{k} \gamma_{k}^{\dagger},
\nonumber\\
&& c_{k}^{\dagger}=u_{k} \gamma_{k}^{\dagger}- i v_{k} \gamma_{-k},
\nonumber\\
&& c_{-k}^{\dagger}=u_{k} \gamma_{-k}^{\dagger}+ i v_{k} \gamma_{k} .
\label{eq:D3}
\end{eqnarray}
%
The Bogoliubov transformation completes the diagonalization of Hamiltonian as
%
\begin{equation}
H_{k}(t)=\sum_{k>0}\Delta_{k}(t)(\gamma_{k}^{\dagger}\gamma_{k}-\frac{1}{2})
\label{eq:D4},
\end{equation}
%
where $\Delta_{k}(t)=((h_{z}(k,t))^{2}+(h_{xy}(k,t))^{2})^{\frac{1}{2}}$ is the dispersion of elementary excitations and by considering $u_{k}(t)=\cos(\theta_{k}(t)/2)$ and $v_{k}(t)=\sin(\theta_{k}(t)/2)$, the Bogoliubov angle $\theta_{k}(t)$ satisfies the relation $\tan(\theta_{k}(t))=-\frac{h_{xy}(k,t)}{h_{z}(k,t)}$. The ground state (Bogoliubov vacuum), the state that is annihilated by $\gamma_{k}$, and the excited state of the above Hamiltonian, for anti-periodic boundary conditions, are given by
%
\begin{eqnarray}
&&|\Psi_{0}^{ap}\rangle=\Pi_{k} (\cos(\theta_{k}(t)/2)+i\sin(\theta_{k}(t)/2)c_{k}^{\dagger}c_{-k}^{\dagger})|0\rangle
\nonumber\\
&& |\Psi_{1}^{ap}\rangle=\Pi_{k} (i\sin(\theta_{k}(t)/2)+\cos(\theta_{k}(t)/2)c_{k}^{\dagger}c_{-k}^{\dagger})|0\rangle .
\nonumber\\
\label{eq:D5} .
\end{eqnarray}
%
%
\begin{figure*}
\begin{minipage}{\linewidth}
\centerline{\includegraphics[width=0.25\linewidth]{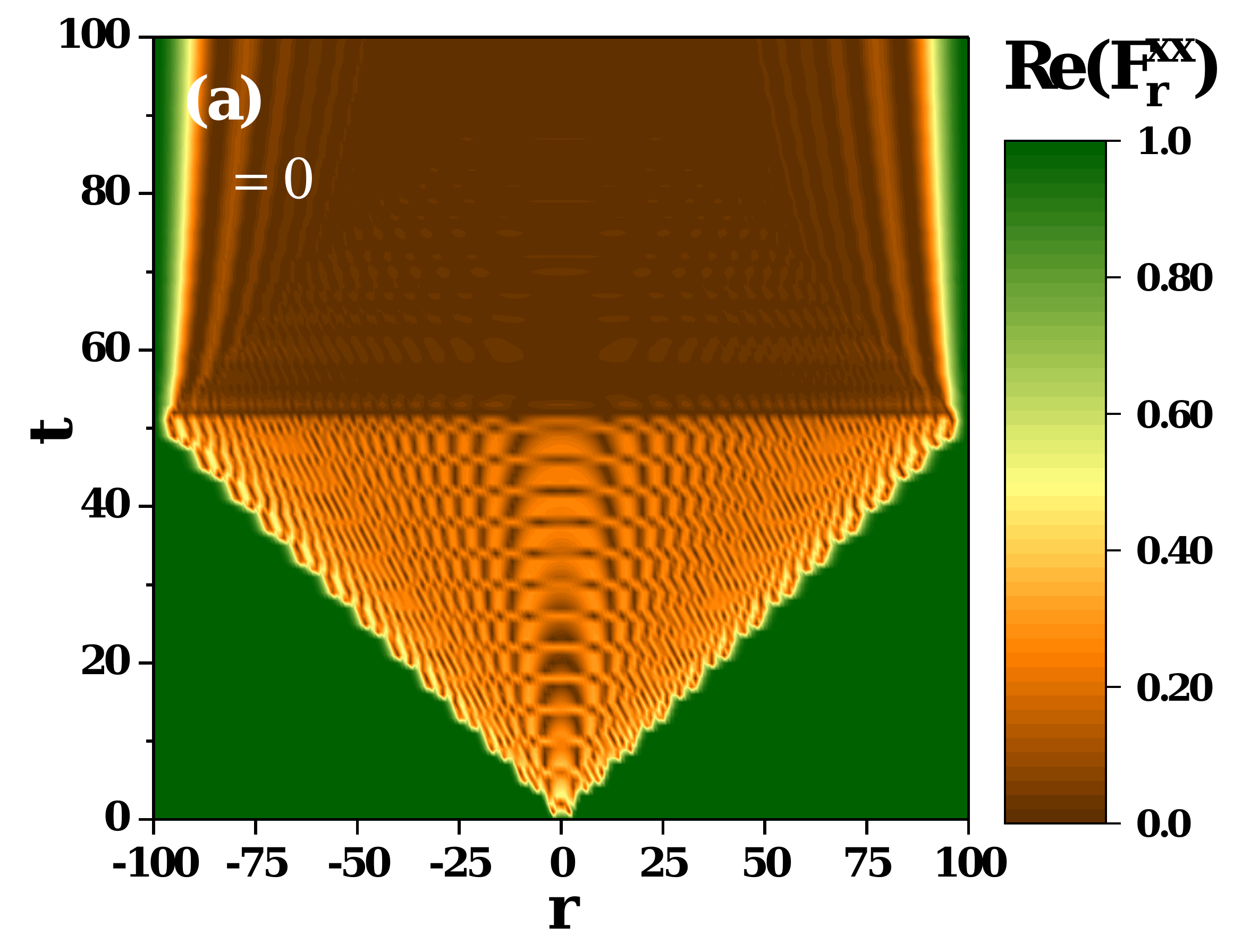}
\includegraphics[width=0.25\linewidth]{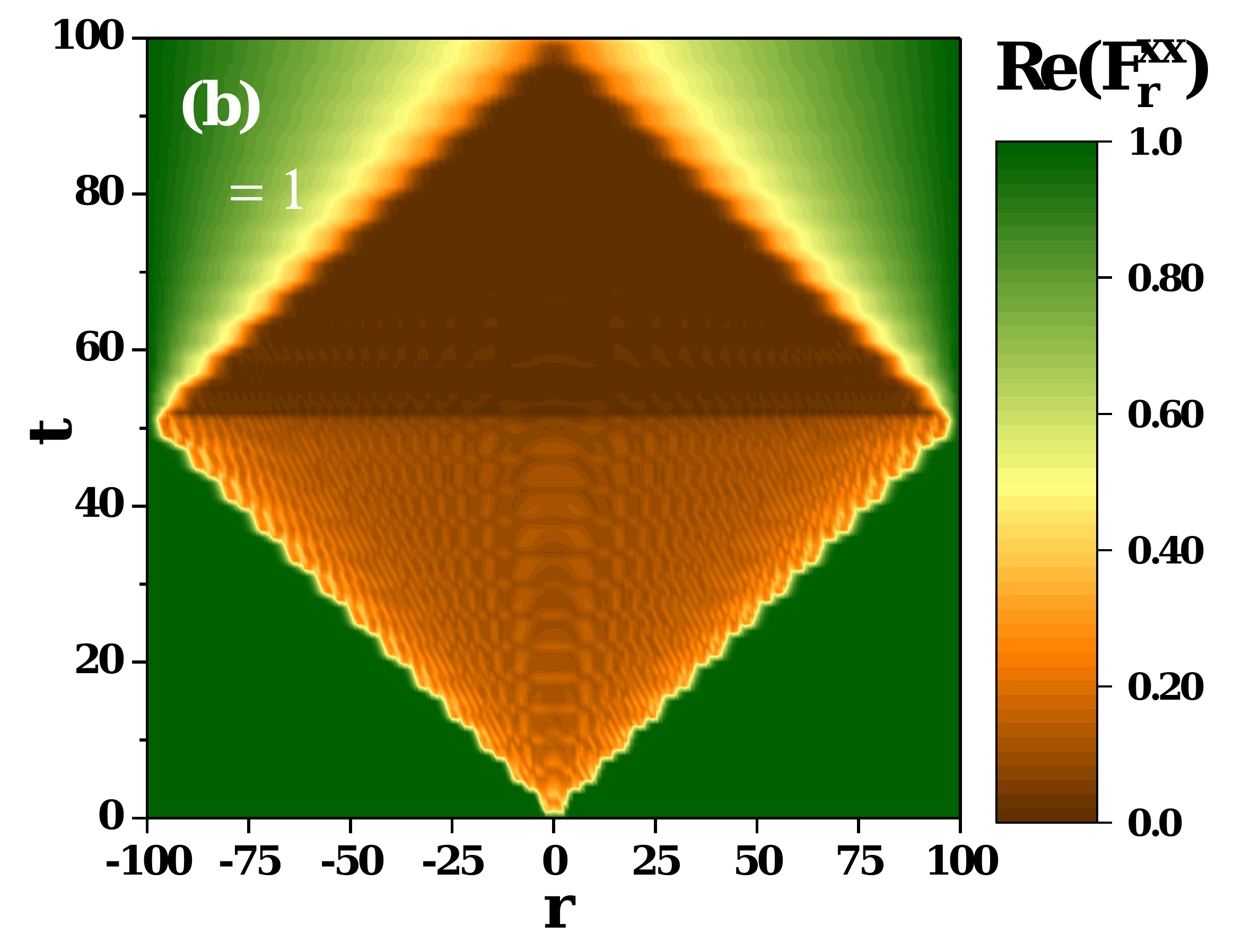}
\includegraphics[width=0.26\linewidth]{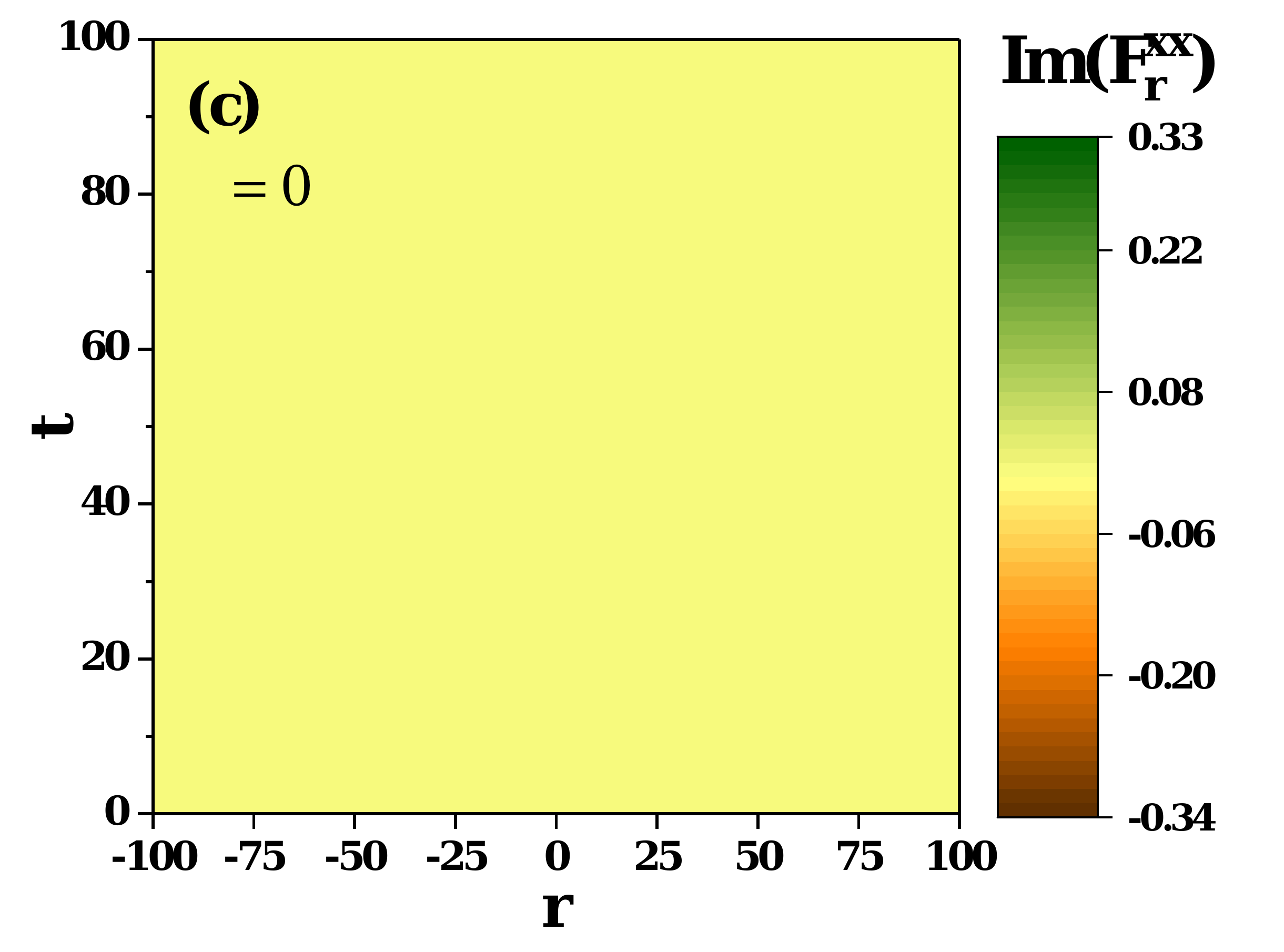}
\includegraphics[width=0.26\linewidth]{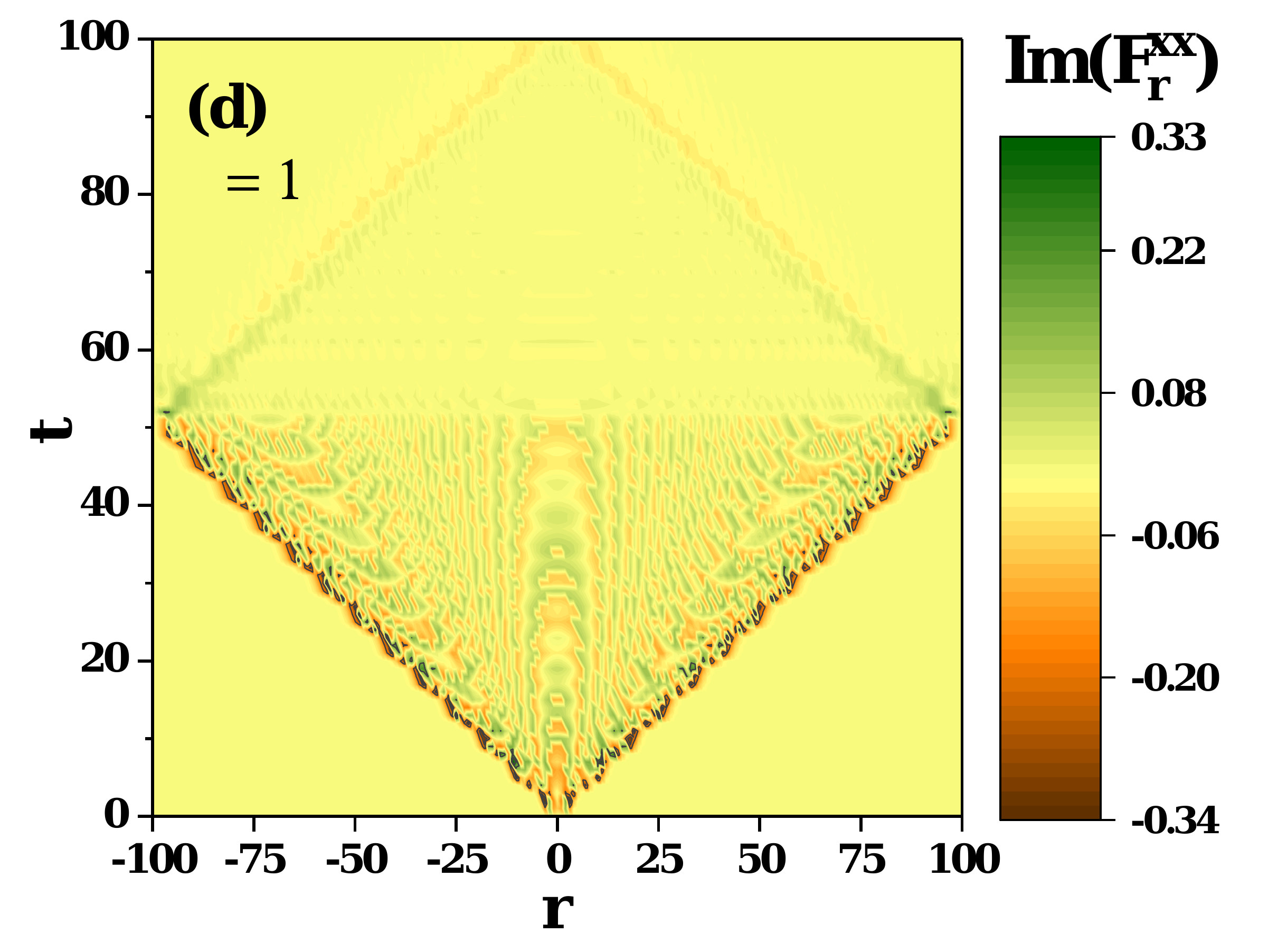}}
\centering
\end{minipage}
\caption{(Color online) Density plot of the real and
imaginary parts of $F^{xx}_{r}$ versus time and separation,
for the  synchronized Ising model, in the presence of $h(t)=h_{0}+h_{1}\cos(\omega t)$
for inverse temperature $\beta=0$ and $\beta=1$. We set system size
$N = 200$ and strong coupling  $\lambda=1$, other parameters are $h_{0}=1, h_{1}=1, \omega=\pi/2$.}
\label{Fig12}
\end{figure*}
%
%
\begin{figure*}
\begin{minipage}{\linewidth}
	\centerline{\includegraphics[width=0.25\linewidth]{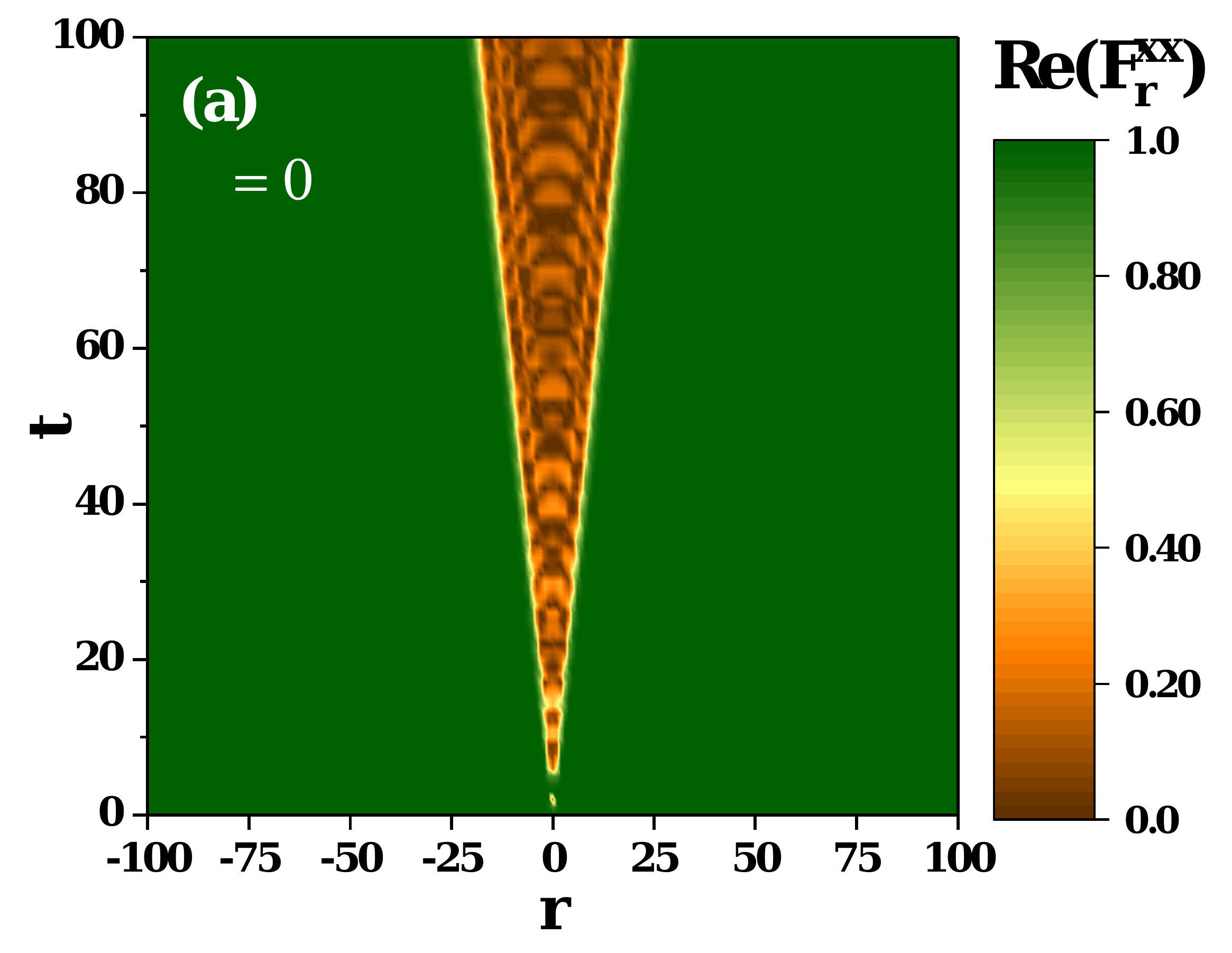}
		\includegraphics[width=0.25\linewidth]{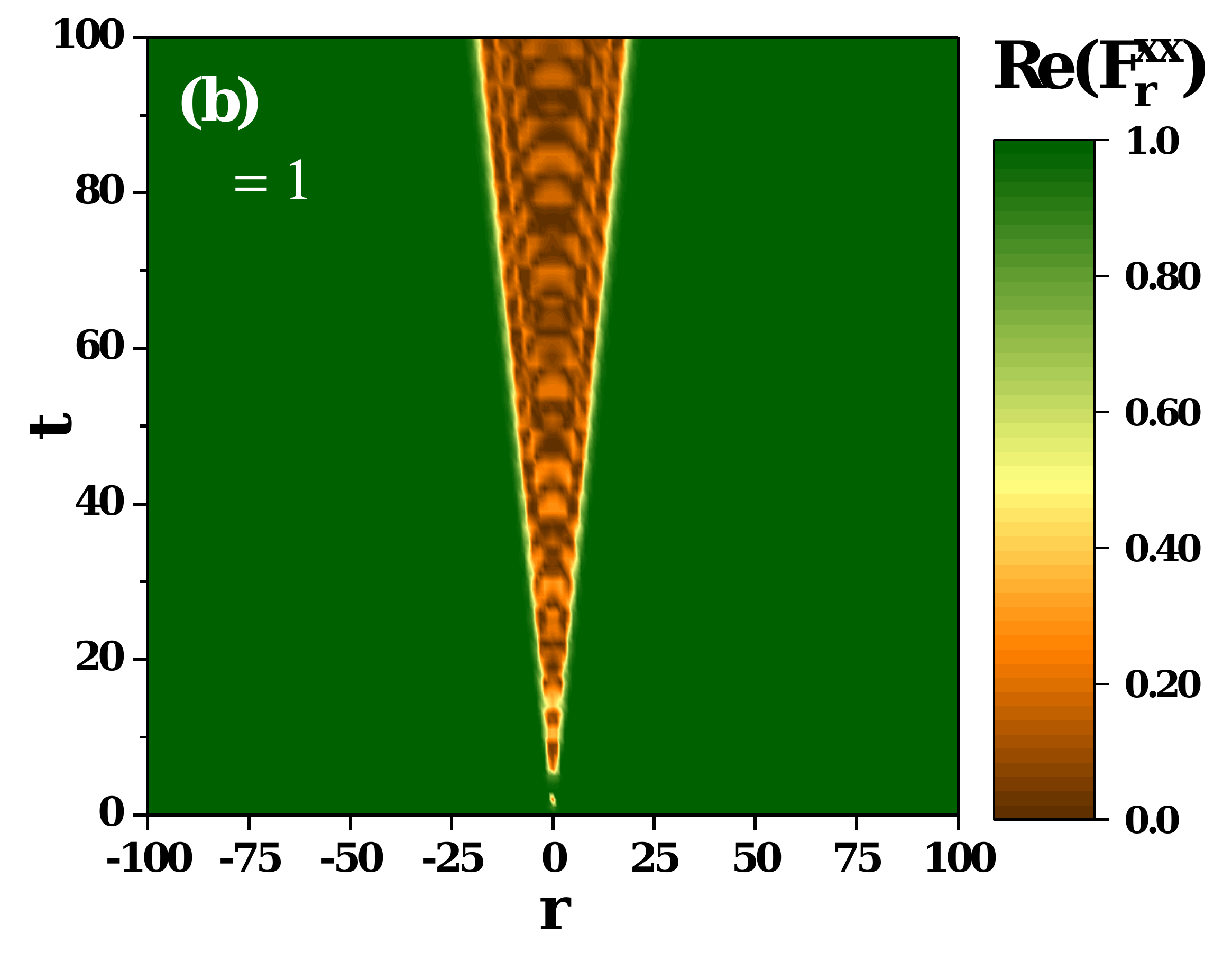}
		\includegraphics[width=0.26\linewidth]{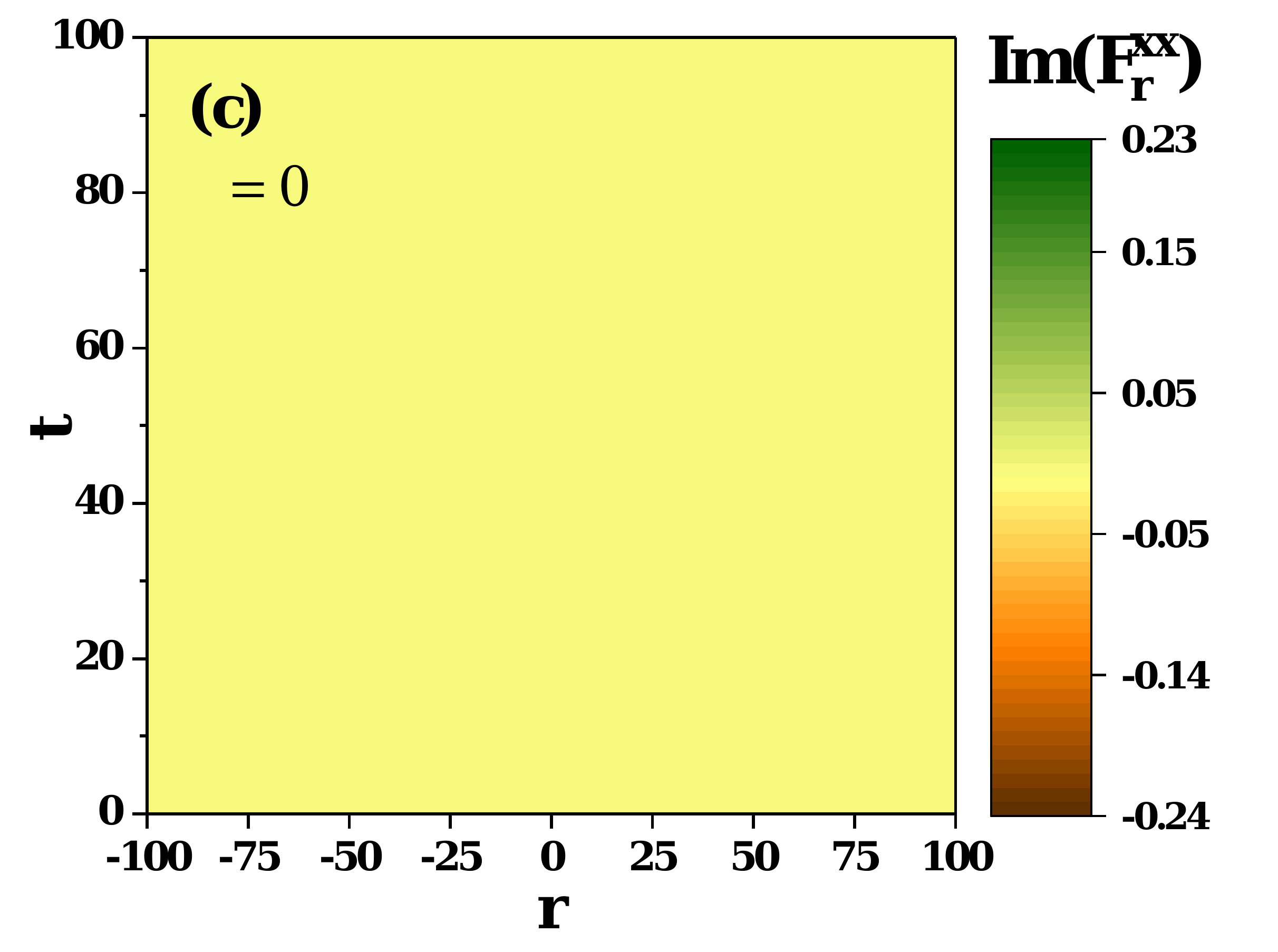}
		\includegraphics[width=0.26\linewidth]{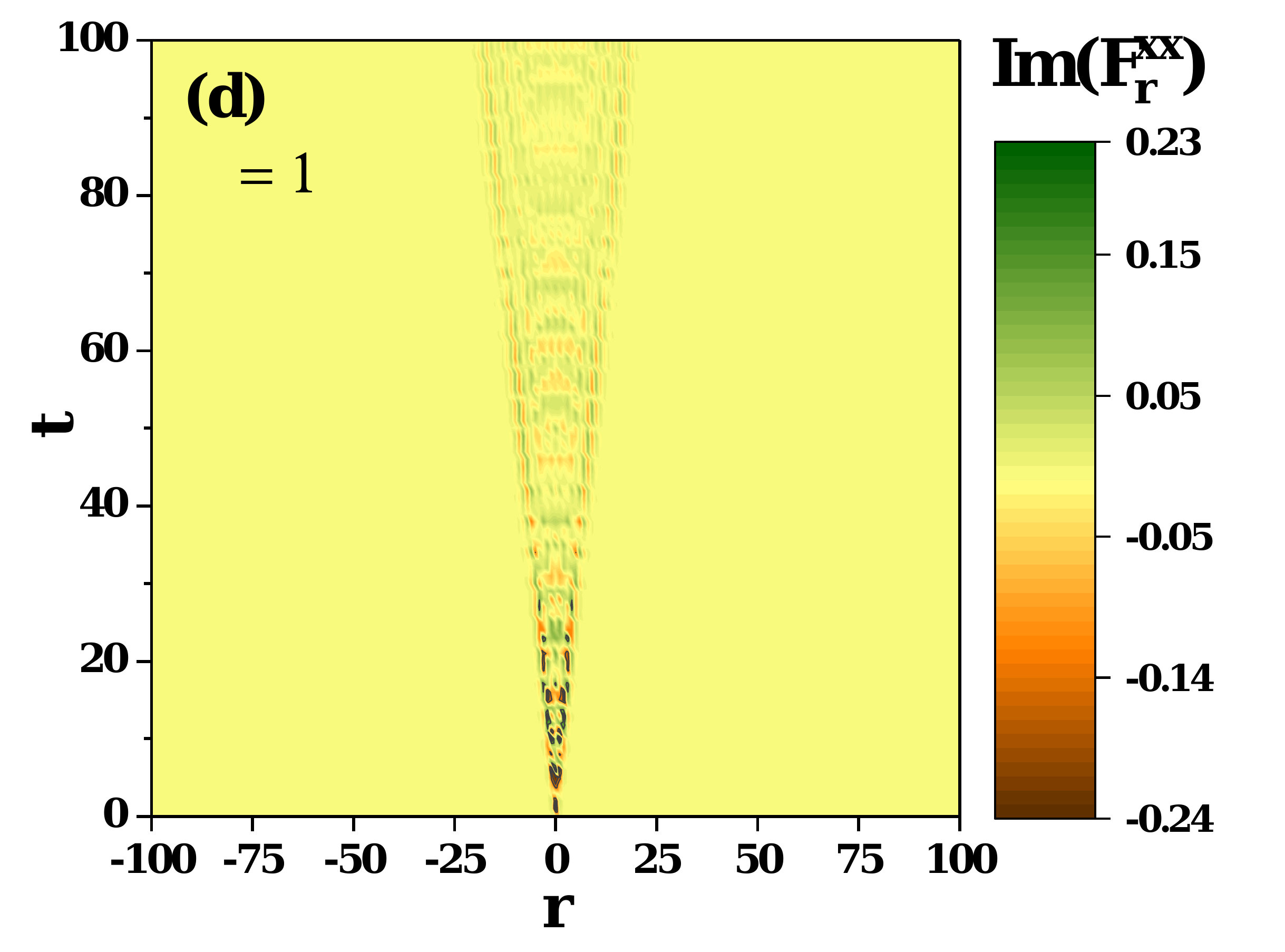}}
	\centering
\end{minipage}
\caption{(Color online)
Density plot of the real and
imaginary parts of $F^{xx}_{r}$ versus time and separation,
for the  synchronized Ising model, in the presence of $h(t)=h_{0}+h_{1}\cos(\omega t)$
for inverse temperature $\beta=0$ and $\beta=1$. We set system size
$N = 200$ at weak coupling  $\lambda=0.1$, other parameters are $h_{0}=1, h_{1}=1, \omega=\pi/2$.
}
\label{Fig13}
\end{figure*}
%
where $|0\rangle$ is the vacuum of system. For the synchronized model ($J(t)=\lambda h(t)$), that we consider afterward, we have $h_{z}(k,t)=-h(t) P(k)$, $h_{xy}(k,t)=h(t) Q(k)$, $\Delta_{k}(t)=h(t) \epsilon_{k}$ and $\tan(\theta^{\prime}_{k}(t))=\frac{Q(k)}{P(k)}$, in which $P(k)=2\lambda\cos(k)-2$, $Q(k)=2\lambda\gamma\sin(k)$ and $\epsilon_{k}=((P(k))^2+(Q(k))^2)^{\frac{1}{2}}$ are time independent. Moreover, we focus on the case of harmonically time dependent magnetic field $h(t)=h_{0}+h_{1}\cos(\omega t)$.

In the Bogoliubov basis (Eq. (\ref{eq:D5})), the Hamiltonian $H_{k}(t)$, density matrix $\rho_{k}(0)$ and time evolution operator $U_{k}(t)$ are expressed as
%
\begin{eqnarray}
	\label{eq:D6}
	H_{k}(t)= h(t)
	\left(
	\begin{array}{cc}
		-\frac{\epsilon_{k}}{2} & 0 \\
		0 & \frac{\epsilon_{k}}{2} \\
	\end{array}
	\right),
\end{eqnarray}
%
%
\begin{eqnarray}
	\label{eq:D7}
	U_{k}(t)=
	\left(
	\begin{array}{cc}
		e^{i \frac{\epsilon_{k}}{2} \tau} & 0 \\
		0 & e^{-i \frac{\epsilon_{k}}{2} \tau} \\
	\end{array}
	\right),
\end{eqnarray}
%
%
\begin{eqnarray}
	\label{eq:D8}
	\rho_{k}(0)= \frac{1}{2 \cosh(\beta h(0)\frac{\epsilon_{k}}{2})}
	\left(
	\begin{array}{cc}
		e^{\beta h(0)\frac{\epsilon_{k}}{2}} & 0 \\
		0 & e^{-\beta h(0)\frac{\epsilon_{k}}{2}} \\
	\end{array}
	\right),
\end{eqnarray}
%
where $\tau=\int_{0}^{t} h(t')dt'$, and the density matrix at time $t$ is obtained to be $\rho_{k}(t)=U_{k}(t) \rho_{k}(0) U_{k}(t)$.

\section{OTOC in synchronized Ising chain \label{E}}
Considering Eq. (\ref{eq:D3}), we obtain
%
\begin{eqnarray}
&& c_{k}^{\dagger}+c_{-k}= e^{-i\theta^{\prime}_k/2}(\gamma_{k}^{\dagger}+\gamma_{-k}),
\nonumber\\
&& c_{k}^{\dagger}-c_{-k}= e^{i\theta^{\prime}_k/2}(\gamma_{k}^{\dagger}-\gamma_{-k}),
\nonumber\\
&& c_{-k}^{\dagger}+c_{k}= e^{i\theta^{\prime}_k/2}(\gamma_{-k}^{\dagger}+\gamma_{k}),
\nonumber\\
&& c_{-k}^{\dagger}-c_{k}= e^{-i\theta^{\prime}_k/2}(\gamma_{-k}^{\dagger}-\gamma_{k}).
\label{eq:E1}
\end{eqnarray}
%
Then using Eq. (\ref{eq4}) we get
%
\begin{eqnarray}
\langle A_p(t) A_q\rangle &&=\frac{1}{N} \sum_{k} e^{ik(p-q)} \times
\nonumber\\
&&\langle U^{\dagger}_{1k}(t) (\gamma_{k}^{\dagger}+\gamma_{-k})U_{1k}(t)(\gamma_{-k}^{\dagger}+\gamma_{k})\rangle,
\nonumber\\
\langle A_p(t) B_q\rangle &&=\frac{1}{N} \sum_{k} e^{ik(p-q)} e^{-i\theta^{\prime}_k} \times
\nonumber\\
&&\langle U^{\dagger}_{1k}(t) (\gamma_{k}^{\dagger}+\gamma_{-k})U_{1k}(t)(\gamma_{-k}^{\dagger}-\gamma_{k})\rangle,
\nonumber\\
\langle B_p(t) A_q\rangle &&=\frac{1}{N} \sum_{k} e^{ik(p-q)} e^{i\theta^{\prime}_k} \times
\nonumber\\
&&\langle U^{\dagger}_{1k}(t) (\gamma_{k}^{\dagger}-\gamma_{-k})U_{1k}(t)(\gamma_{-k}^{\dagger}+\gamma_{k})\rangle,
\nonumber\\
\langle B_p(t) B_q\rangle &&=\frac{1}{N} \sum_{k} e^{ik(p-q)} \times
\nonumber\\
&&\langle U^{\dagger}_{1k}(t) (\gamma_{k}^{\dagger}-\gamma_{-k})U_{1k}(t)(\gamma_{-k}^{\dagger}-\gamma_{k})\rangle.
\label{eq:E2}
\end{eqnarray}
%
Finally, following the above equations and considering  Eqs. (\ref{eq:D6})-(\ref{eq:D8}), time dependent Majorana correlation functions for mixed state synchronized case are given by
%
\begin{eqnarray}
\langle A_p(t) A_q\rangle &&=\frac{1}{N} \sum_{k} e^{ik(p-q)} \times
\nonumber\\
&&\Big[\cos(\epsilon_{k} \tau) -i\sin(\epsilon_{k} \tau)\tanh(\beta h(0) \frac{\epsilon_{k}}{2})\Big],
\nonumber\\
\langle A_p(t) B_q\rangle &&=\frac{1}{N} \sum_{k} e^{ik(p-q)} e^{-i\theta^{\prime}_k} \times
\nonumber\\
&&\Big[\cos(\epsilon_{k} \tau) \tanh(\beta h(0) \frac{\epsilon_{k}}{2})-i\sin(\epsilon_{k} \tau)\Big],
\nonumber\\
\langle B_p(t) A_q\rangle &&=-\frac{1}{N} \sum_{k} e^{ik(p-q)} e^{i\theta^{\prime}_k} \times
\nonumber\\
&&\Big[\cos(\epsilon_{k} \tau) \tanh(\beta h(0) \frac{\epsilon_{k}}{2})-i\sin(\epsilon_{k} \tau)\Big],
\nonumber\\
\langle B_p(t) B_q\rangle &&=-\frac{1}{N} \sum_{k} e^{ik(p-q)} \times
\nonumber\\
&&\Big[\cos(\epsilon_{k} \tau) -i\sin(\epsilon_{k} \tau)\tanh(\beta h(0) \frac{\epsilon_{k}}{2})\Big].
\nonumber\\
\label{eq:E3}
\end{eqnarray}
%
%
\begin{figure*}
\begin{minipage}{\linewidth}
\centerline{\includegraphics[width=0.33\linewidth]{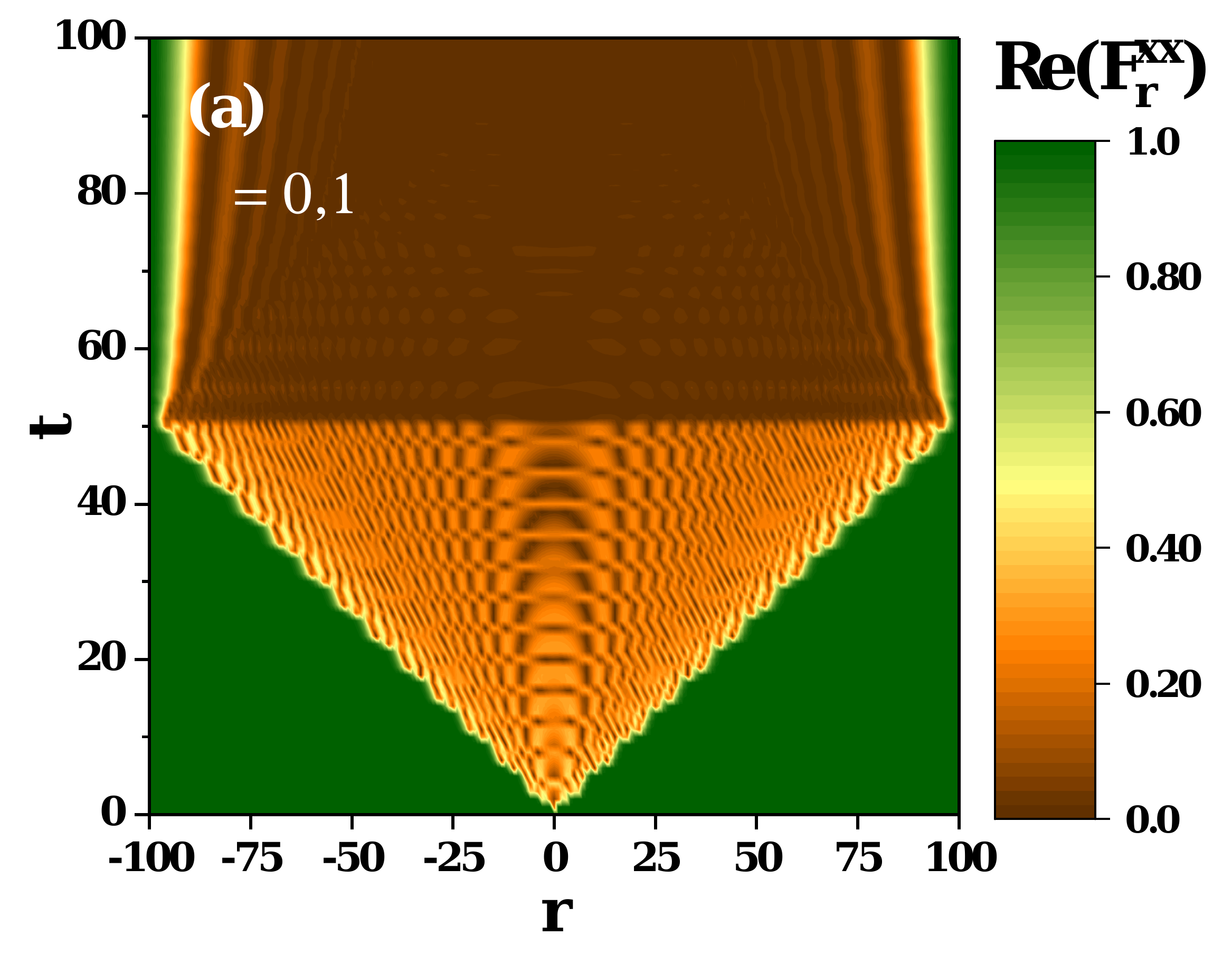}
\includegraphics[width=0.33\linewidth]{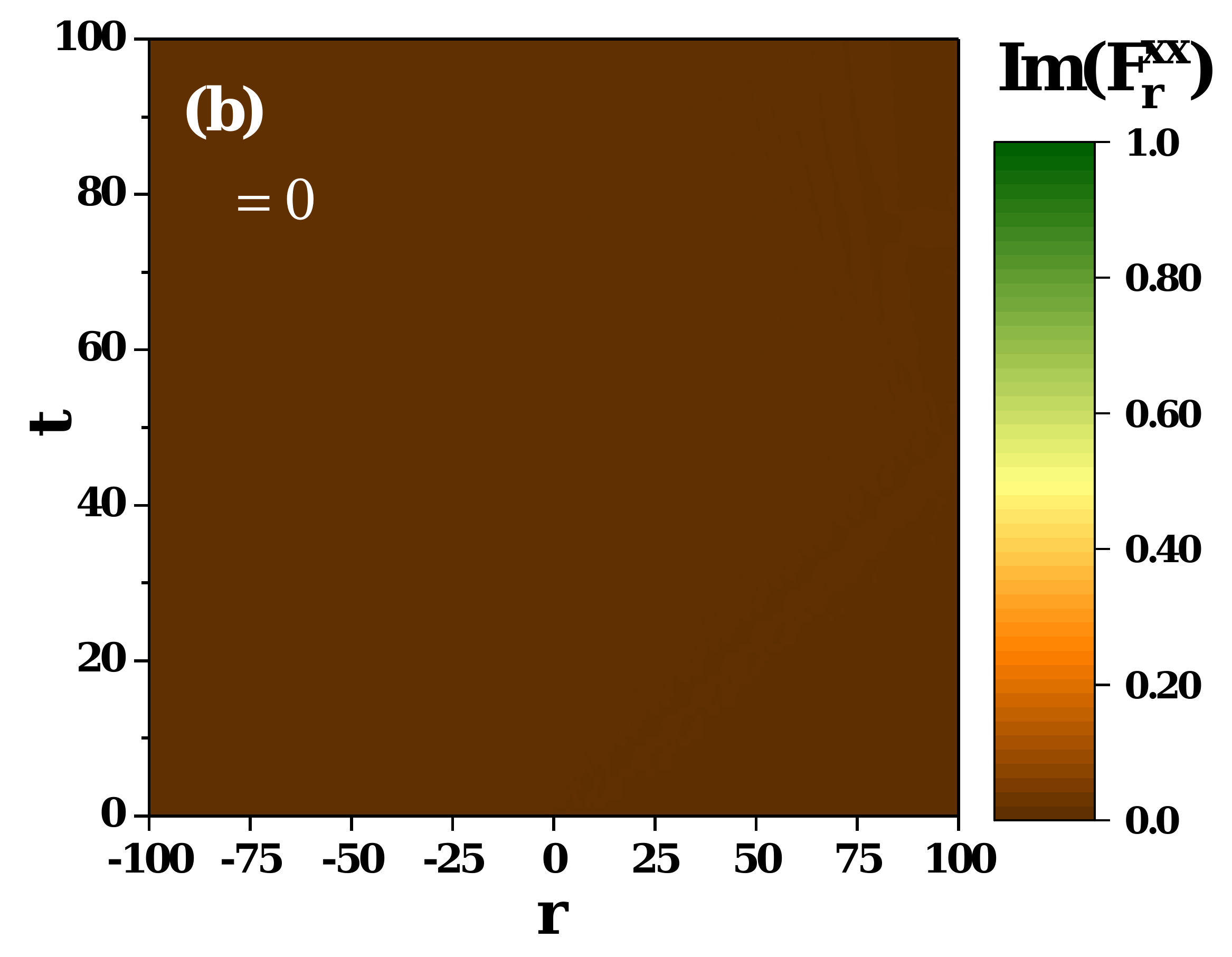}
\includegraphics[width=0.33\linewidth]{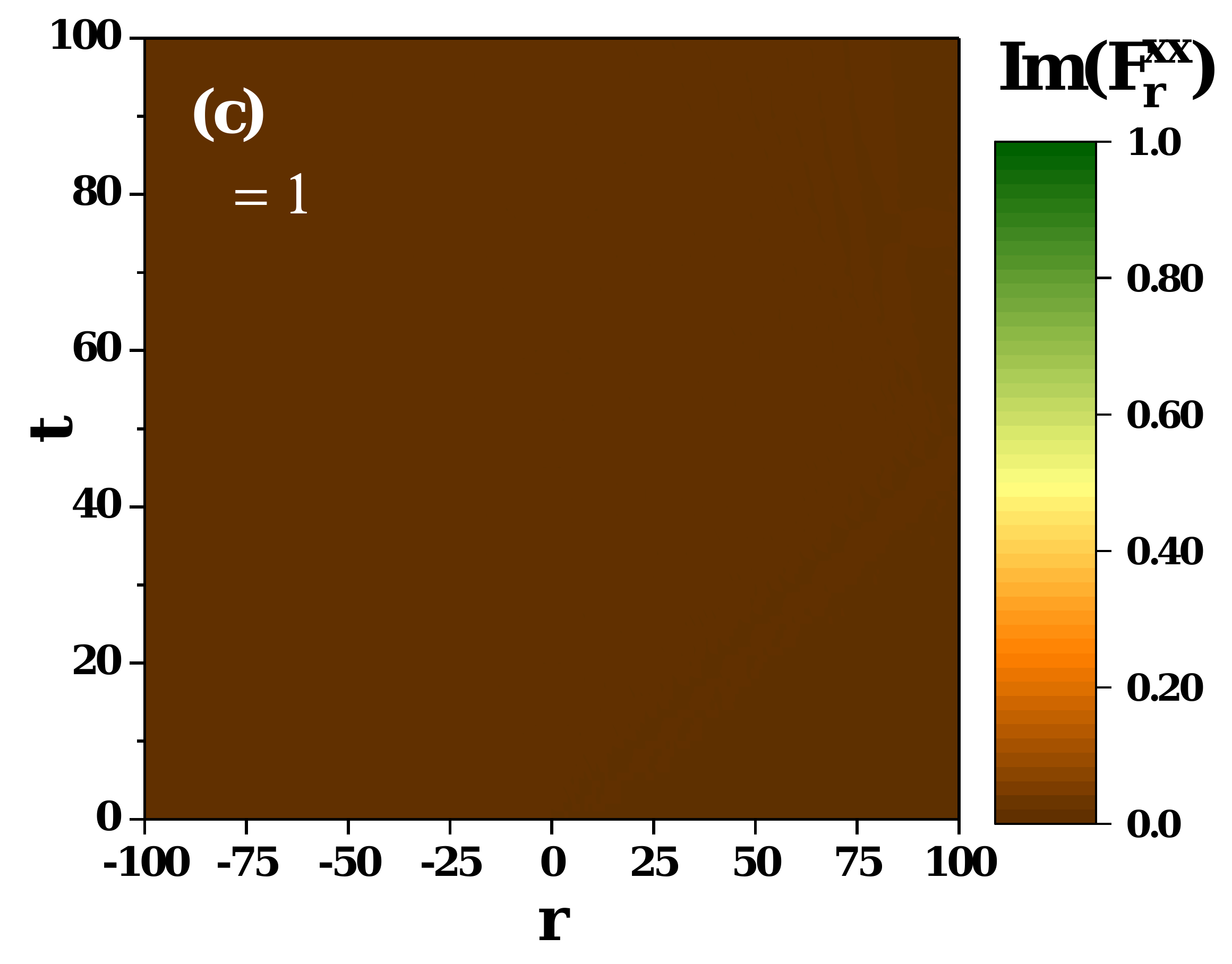}}
\centering
\end{minipage}
\caption{(Color online)
Density plot of the real and
imaginary parts of $F^{xx}_{r}$ versus time and separation,
for the  synchronized Ising model, in the presence of $h(t)=h_{0}+h_{1}\cos(\omega t)$
where $h_{0}=1, h_{1}=1, \omega=\pi/2$, i.e. $h(t=0)=0$.
(a) The real part of $F^{xx}_{r}$  is the same for $\beta=0$ and $\beta=1$.
The imaginary part of  $F^{xx}_{r}$ is plotted
for inverse temperature (b) $\beta=0$ and (c) $\beta=1$.
The system is in the strong coupling $\lambda=1$ and $N=200$.
}
\label{Fig14}
\end{figure*}
%
%
\begin{figure*}
\begin{minipage}{\linewidth}
\centerline{\includegraphics[width=0.33\linewidth]{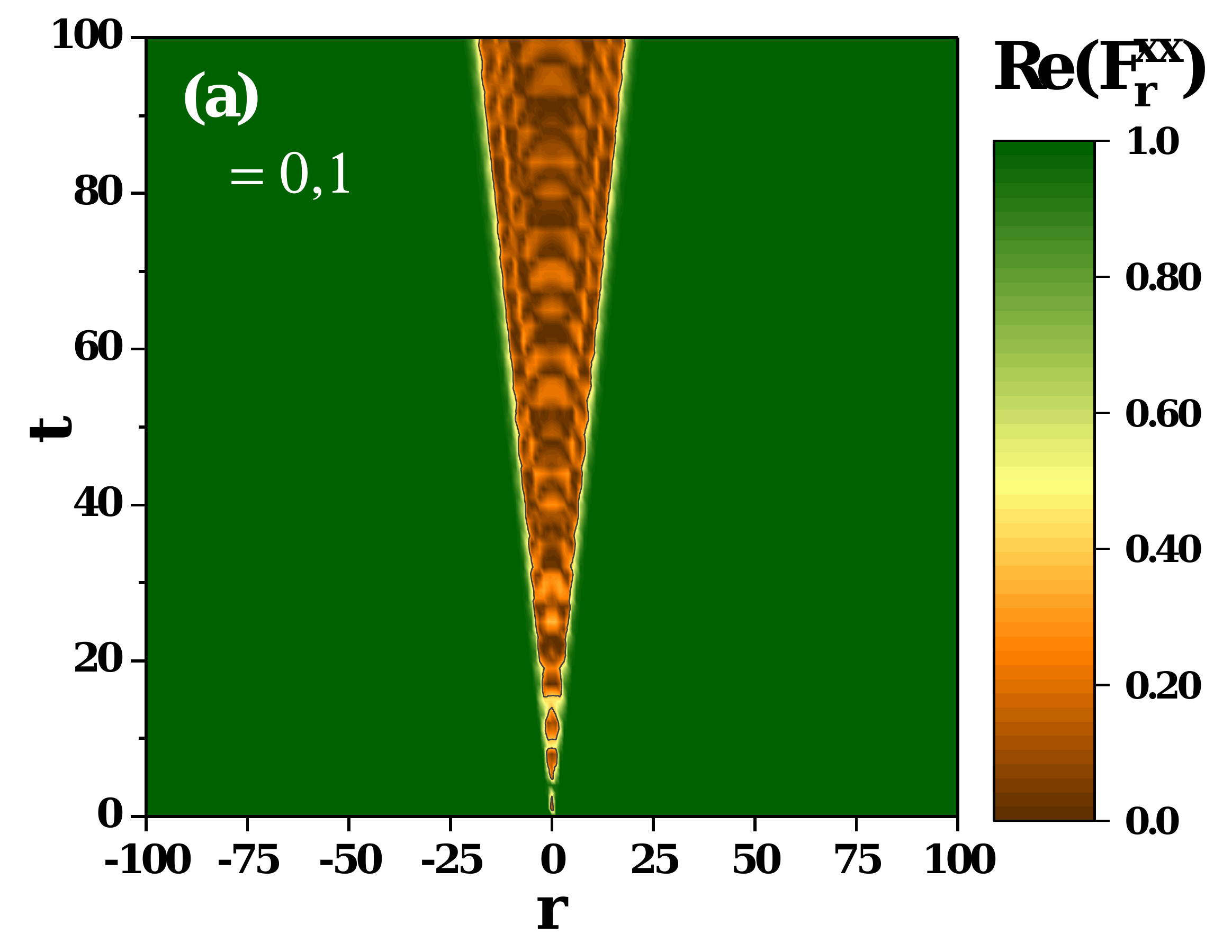}
\includegraphics[width=0.33\linewidth]{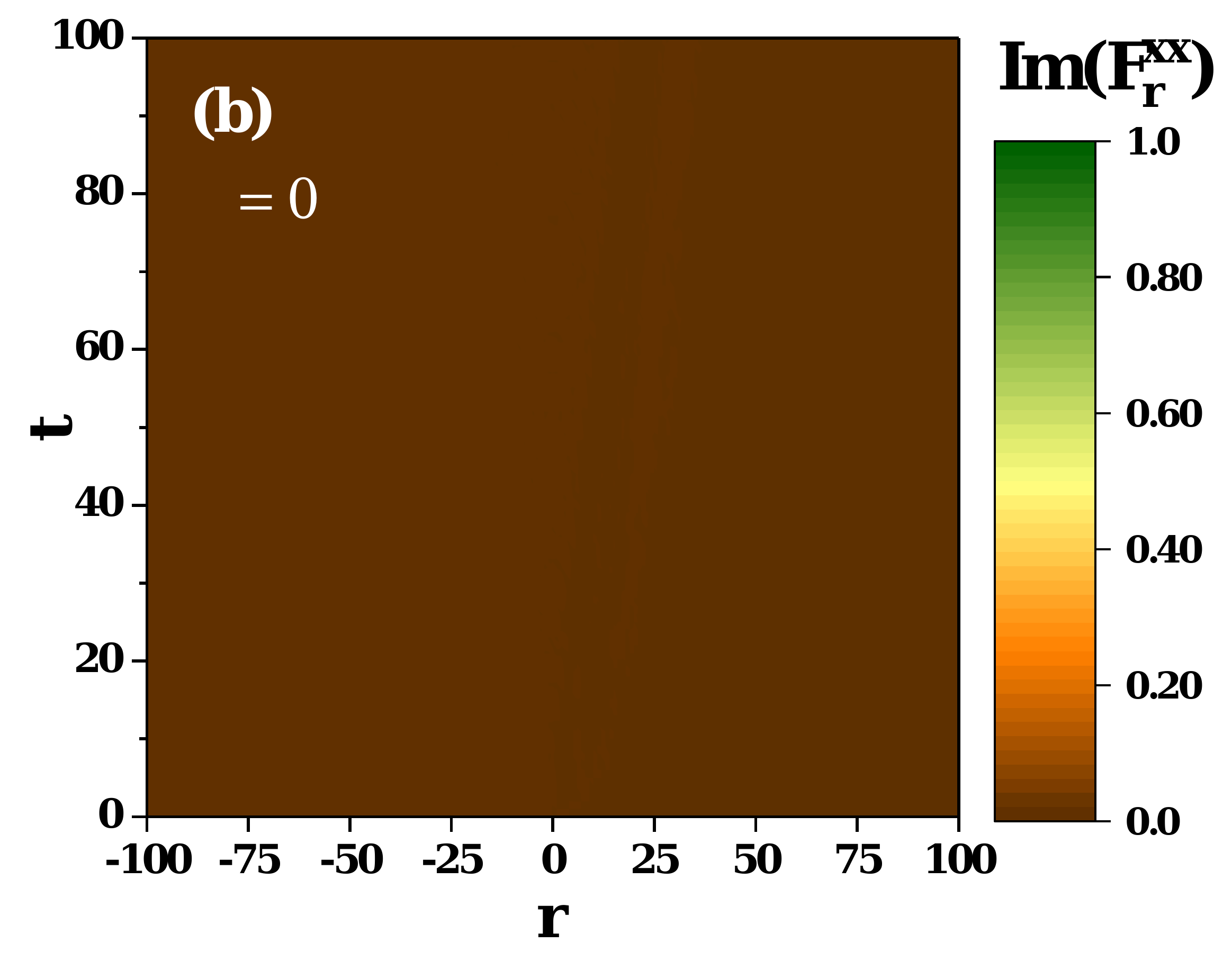}
\includegraphics[width=0.33\linewidth]{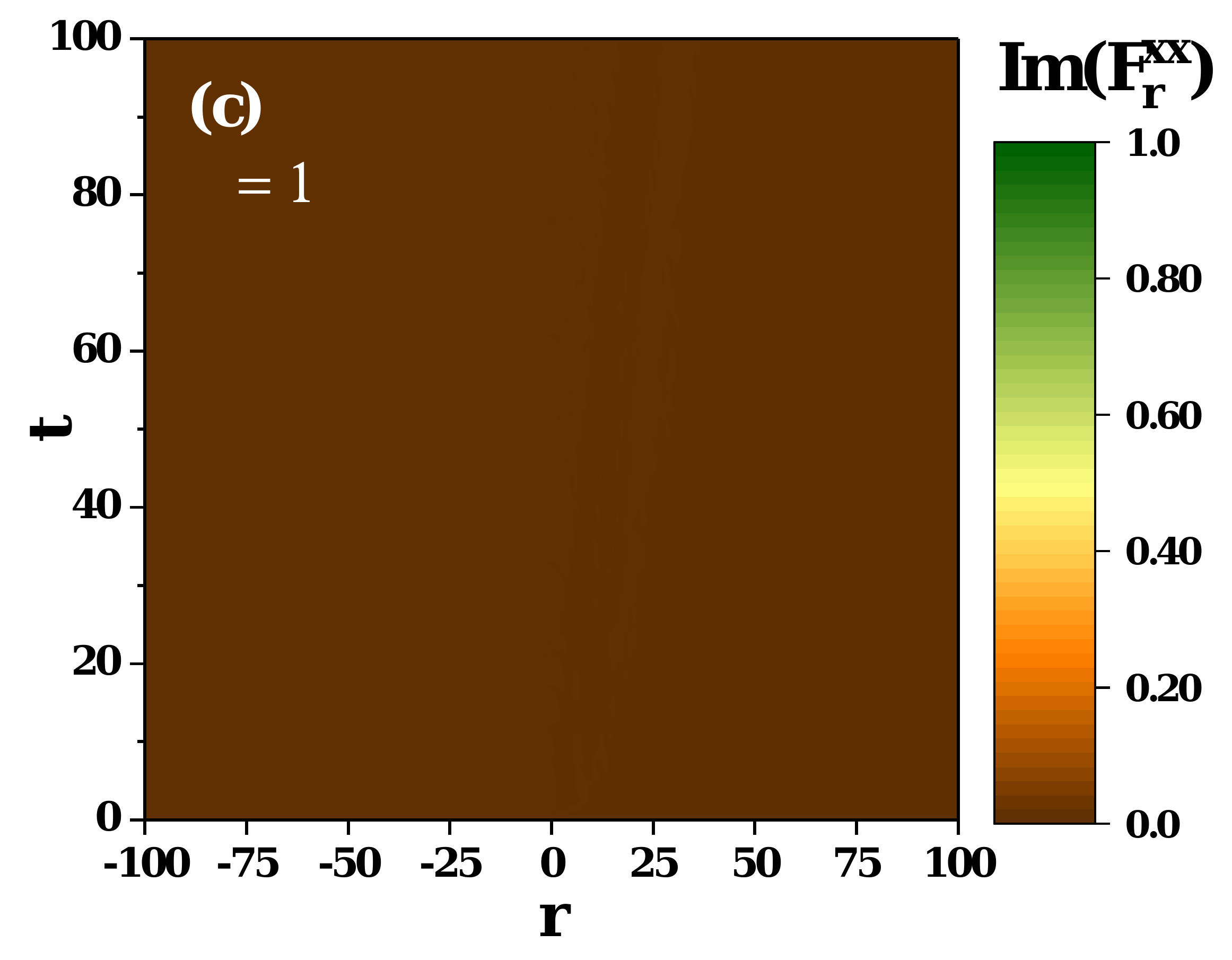}}
\centering
\end{minipage}
\caption{(Color online)
The explanation is the same as of Fig.(\ref{Fig14}) except the
coupling $\lambda=0.1$, which is weak.
}
\label{Fig15}
\end{figure*}
%
\subsubsection{OTOC of nonlocal operators in the synchronized Floquet XY model}

The general behavior of OTOC composed of nonlocal operator for the synchronized Ising model, is illustrated in Figs. \ref{Fig12}-\ref{Fig15}, using the procedure described in section \ref{OTOC}, for system size $N=200$. As seen, the $xx$ OTOC shows the signature of operator spreading, although with some differences in comparison with the  $zz$ OTOC. Figures \ref{Fig12}-\ref{Fig13} exhibit the evolution of real and imaginary parts of $F^{xx}_{r}$ in time, at high and low temperature and for $h(0)\neq0$ case.

The OTOC with nonlocal operator has been depicted in Figs. \ref{Fig14}-\ref{Fig15} for $h(0)=0$ case. As can be observed, diagrams reveal no temperature dependence and so decreasing the temperature from its infinite value, does not significantly alter the quantitative behavior of OTOC in this context. Hence, similar to the situation of $C^{zz}_{r}$, when the initial time magnetic field is zero, vanishing of the imaginary part of $F^{xx}_{r}$ signals the occurrence of FDQPT independent of temperature, that is in agreement with the results of Loschmidt amplitude analysis. So, it would be suitable to detect the mixed state FDQPT of synchronized Ising chain due to analyzing the vanishing of $Im(F^{zz}_{r})$ as well as $Im(F^{xx}_{r})$ at any temperature.

\bibliography{OTOCReferences}

\end{document}